\documentclass[aps,prd,aps,floatfix,tightenlines,superscriptaddress,
amsmath,amssymb,nofootinbib]{revtex4}
\usepackage{amssymb,amsbsy,epsfig,color,graphicx}
\usepackage{color}
\usepackage{[longtable}
\usepackage{array}
\usepackage{dcolumn}   
\usepackage{cellspace}
\usepackage{mathtools}
\usepackage{amstext}
\usepackage{amssymb}
\usepackage{stmaryrd}
\usepackage{stackrel}
\usepackage{graphicx}
\usepackage{esint}
\usepackage[utf8]{inputenc}
\usepackage{blindtext}
\usepackage{float}
\restylefloat{table}
\usepackage{booktabs}
\usepackage{enumitem}
\usepackage{esint}
\usepackage{amssymb}
\usepackage{hyperref}
\usepackage{cancel}
\usepackage{slashed}
\usepackage{soul}
\usepackage{amssymb,amsfonts}
\usepackage{latexsym}
\usepackage[utf8]{inputenc}

\usepackage{etoolbox} 
\usepackage{lipsum} 
\usepackage[capitalize]{cleveref}

\def\XXint#1#2#3{{\setbox0=\hbox{$#1{#2#3}{\int}$}
\vcenter{\hbox{$#2#3$}}\kern-.5\wd0}}

\usepackage{multirow}
\usepackage[caption=false]{subfig}
\renewcommand\[{\begin{equation}}
\renewcommand\]{\end{equation}}

\newcommand{\ba}{\begin{aligned}}
\newcommand{\ea}{\end{aligned}}
\newcommand{\LF}{\left(}
\newcommand{\RF}{\right)}
\newcommand{\LT}{\left[}
\newcommand{\RT}{\right]}

\newcommand{\cO}{{\cal O}}
\newcommand{\cF}{{\cal F}}

\newcommand{\be}{\begin{equation}}
\newcommand{\ee}{\end{equation}}

\newcommand{\bea}{\begin{eqnarray}}
\newcommand{\eea}{\end{eqnarray}}

\renewcommand{\texttt}{{}}

\def\bs{\begin{subequations}}
\def\es{\end{subequations}}

\def\cF{\mathcal{F}}

\def\cK{\mathcal{K}}
\def\cL{\mathcal{L}}

\def\cO{\mathcal{O}}

\def\Fc{\mathcal{F}}

\def\Kc{\mathcal{K}}
\def\Lc{\mathcal{L}}

\def\Oc{\mathcal{O}}

\newcommand{\tia}[1]{}
\newcommand{\beas}{\begin{eqnarray*}}
	\newcommand{\eeas}{\end{eqnarray*}}
\newcommand{\bal}{\begin{aligned}}
	\newcommand{\eal}{\end{aligned}}

\makeatletter

\appto{\appendix}{%
\@ifstar{\def\theequation@prefix{A.}}%
{}%
}
\makeatother

\begin{document}

\title{Perturbations in higher derivative gravity beyond maximally symmetric spacetimes}

\author{K. Sravan Kumar}
\email{sravan.korumilli@rug.nl}
\affiliation{Van Swinderen Institute, University of Groningen, 9747 AG, Groningen, The Netherlands}

\author{Shubham Maheshwari}
\email{s.maheshwari@rug.nl}
\affiliation{Van Swinderen Institute, University of Groningen, 9747 AG, Groningen, The Netherlands}

\author{Anupam Mazumdar}
\email{anupam.mazumdar@rug.nl}
\affiliation{Van Swinderen Institute, University of Groningen, 9747 AG, Groningen, The Netherlands}


\begin{abstract}
We study (covariant) scalar-vector-tensor (SVT) perturbations of infinite derivative gravity (IDG), at the quadratic level of the action, around conformally-flat, covariantly constant curvature backgrounds which are \textit{not} maximally symmetric spacetimes (MSS). This extends a previous analysis of perturbations done around MSS, which were shown to be ghost-free. We motivate our choice of backgrounds which arise as solutions of IDG in the UV, avoiding big bang and black hole singularities. Contrary to MSS, in this paper we show that, generically, all SVT modes are coupled to each other at the quadratic level of the action. We consider simple examples of the full IDG action, and illustrate this mixing and also a case where the action can be diagonalized and ghost-free solutions constructed. Our study is widely applicable for both non-singular cosmology and black hole physics where backgrounds depart from MSS. In appendices, we provide SVT perturbations around conformally-flat and \textit{arbitrary} backgrounds which can serve as a compendium of useful results when studying SVT perturbations of various higher derivative gravity models.
\end{abstract}

\maketitle


\tableofcontents

\section{Introduction}\label{intro}

Einstein's general relativity (GR) is an excellent theory in the infrared (IR) regime, far from the source, and at late times. Its success has been seen from solar system tests~\cite{Will:2014kxa} to the recent detection of gravitational waves from mergers of binary compact systems~\cite{Abbott:2016blz,TheLIGOScientific:2017qsa}. Despite this success, GR requires modifications at short distance and time scales, i.e., in the ultraviolet (UV) regime. At the classical level, there exist cosmological (big bang) and black hole singularities~\cite{Penrose:1964wq,Penrose:1969pc,Hawking:1969sw,Hawking:1973uf}. At the quantum level, GR is non-renormalizable~\cite{tHooft:1974toh,Goroff:1985sz}. In fact, we do not yet know, experimentally, if a quantum theory of gravity exists~\cite{Bose:2017nin}. There are various ways to address these problems, like string theory~\cite{Polchinski:1998rrsh}, loop quantum gravity~\cite{Ashtekar:2007tv}, causal set dynamics~\cite{Bombelli:1987aa} and asymptotic safety~\cite{Weinberg:1980gg}. There is another very insightful approach, perhaps not yet at the fundamental level, but rather from the bottom-up: adding higher derivative corrections to GR. In four spacetime dimensions, quadratic curvature gravity is renormalizable~\cite{Stelle:1976gc} but suffers from ghost instabilities~\footnote{Ghosts are degrees of freedom with `wrong' sign kinetic terms, compared to healthy sectors of the theory which have Hamiltonians bounded from below. When ghost and healthy systems interact, runaway production of ghost and normal particles from the vacuum is kinematically allowed, characterized by a divergent phase space integral \cite{Cline:2003gs}.}. In fact, any finite derivative theory of gravity suffers from classical and quantum instabilities in non-supersymmetric, time-dependent backgrounds~\cite{VanNieuwenhuizen:1973fi,Woodard:2015zca}. One can go beyond finite derivative modifications to GR in the spirit of string field theory (SFT)~\cite{Witten:1985cc,Freund:1987ck,Freund:1987kt,Frampton:1987sp,Vladimirov:1994wi} (for a review, see~\cite{deLacroix:2017lif}) and higher derivative ($\alpha'$) corrections that appear in the low energy string effective action~\cite{Zwiebach:1985uq,Gross:1986iv}. 

As mentioned above, one potential approach to tackle these problems is to invoke infinite covariant derivatives~\cite{Krasnikov:1987yj,Kuzmin:1989sp,Siegel:2003vt,Tseytlin:1995uq,Biswas:2005qr,Biswas:2011ar,Modesto:2011kw,Tomboulis:2015gfa,Buoninfante:2018mre}. It was shown that infinite derivative gravity (IDG) can potentially resolve cosmological (big bang) singularity~\cite{Biswas:2005qr,Biswas:2006bs} and black hole singularities in static~\cite{Biswas:2011ar,Frolov:2015bia,Frolov:2015usa,Edholm:2016hbt,Frolov:2015bta,Boos:2018bxf,Buoninfante:2018rlq,Koshelev:2018hpt,Kilicarslan:2018yxd} and dynamical setups~\cite{Frolov:2015bia,Frolov:2015usa, Kilicarslan:2019njc}. Furthermore, infinite derivative field theories~\cite{Tomboulis:2015gfa,Buoninfante:2018mre,Talaganis:2014ida} have UV behavior similar to that of SFT and p-adic strings~\cite{deLacroix:2017lif}. In infinite derivative theories, there exists a scale of non-locality, $M_s$, below the Planck scale in $4$ spacetime dimensions. The non-locality appears only at the level of interactions, which can potentially ameliorate the UV problems at higher loops for proper gravitational form factors~\cite{Tomboulis:1997gg,Modesto:2011kw,Talaganis:2014ida}. It has been shown that there exists a complementarity due to non-local interactions which suppresses scattering amplitudes with $2$~\cite{Biswas:2014yia,Talaganis:2016ovm}, or more interactions~\cite{Buoninfante:2018gce}, such that the fundamental non-local scale may shift in the IR. This allows the intriguing possibility of observing this non-local scale at low energies, allowing us to constrain or falsify the theory~\cite{Koshelev:2017bxd,Buoninfante:2019swn}. Furthermore, infinite derivative theories have very interesting quantum properties at finite temperatures~\cite{Biswas:2009nx,Biswas:2010xq,Biswas:2010yx,Biswas:2012ka,Boos:2019zml}, which mimics some properties of the Hagedorn behavior of strings at temperatures above the string scale. In addition to all these developments, it is important to study classical perturbations of IDG in four spacetime dimensions. In fact, earlier papers have studied cosmological perturbations around a bounce, ameliorating the cosmological (big bang) singularity~\cite{Biswas:2010zk,Biswas:2012bp}, around de Sitter(dS) and anti de Sitter (AdS) backgrounds~\cite{Biswas:2016etb,Biswas:2016egy}, and during cosmic inflation~\cite{Barnaby:2006hi,Koivisto:2008xfa,Koshelev:2016vhi,Koshelev:2017tvv,Craps:2014wga,Conroy:2016sac}. 

The aim of this paper is to study perturbations in IDG at the quadratic level of the action around backgrounds that go \textit{beyond} those usually studied in literature, viz. flat/dS/AdS. In particular, we study the SVT modes of metric perturbation around a background which is conformally-flat, and has covariantly constant curvatures motivated through physically relevant examples like cosmological bounces and non-singular, non-vacuum, spherically symmetric, static spacetimes. In appendices, we provide, for the first time, second order SVT perturbations around \textit{arbitrary} and conformally-flat (e.g. FLRW) backgrounds. Our treatment is general with many valuable results, which can be useful for future work in a variety of higher derivative gravity theories.

The paper is organized as follows: in sec.(\ref{sec.IDGac}), we introduce IDG, a general class of higher derivative gravity, and the action that we use in this paper. In sec.(\ref{pertgenbackgsec4}), we introduce the background field method, (covariant) SVT decomposition of metric perturbation and the general structure of the second order action. In sec.(\ref{sec-nonmss}), we will discuss typical backgrounds studied in literature, like dS/AdS, and then choose a specific \textit{non}-MSS background upon which we perform perturbations in later sections. This background is conformally-flat and satisfies covariantly constant curvature conditions. These conditions significantly reduce the convoluted perturbative expressions to be encountered in later sections, and also make contact with physically motivated examples from cosmological bounce and black hole-like scenarios. We also lay out the background equations of motion. In sec.(\ref{secondordervarsection}), we segregate, organize and present second order perturbations at the level of the action, in SVT form, for the action we started with in sec.(\ref{sec.IDGac}). They possess peculiar scalar-vector-tensor mixings, explicitly showing how perturbations around general backgrounds become difficult to analyze. In sec.(\ref{checkinglimitsmss}), we will verify our general analysis by deriving dS/AdS and flat space limits which were studied in previous papers \cite{Biswas:2016egy}, \cite{Biswas:2016etb}. In sec.(\ref{physicalmodes}), we analyze the physical spectrum for simple cases of the general action we started with. Finally, we conclude with some comments on possible directions for future work in sec.(\ref{conclusions}). Many relevant perturbative expressions and commutation relations are collected in Appendices (\ref{appendixperturbations}), (\ref{appendixsecondordervariations}) and (\ref{comrelsec}). They include perturbations and commutation relations around an \textit{arbitrary} background and are therefore useful for future work in gravity, and higher derivative gravity in particular.\\

\underline{Notation}: We use the metric signature $(-,+,+,+)$. $\hbar = c = 1$, and the reduced Planck mass is $M_{P}^{2} = (8 \pi G)^{-1}$. Relevant quantities like curvatures are split into background and perturbations as explained in Appendix (\ref{examplepertcurvatures}). Overbars on curvatures indicate their values on a fixed background.

\section{Infinite derivative gravity}
\label{sec.IDGac}

The most general action for gravity which is quadratic in Ricci scalar $R$, traceless Ricci tensor $S^{\mu}_{\ \nu}$ and Weyl tensor $C^{\rho \sigma}_{\ \ \ \mu \nu}$ each, parity invariant, torsion-free, and which can be made ghost-free around maximally symmetric backgrounds is given by \cite{Biswas:2011ar,Biswas:2016egy,Biswas:2016etb}~\footnote{It is worthwhile to mention that the most general quadratic curvature action which contains torsion and generalizes Poincar\'e gravity was constructed recently in \cite{delaCruz-Dombriz:2018aal}. See also \cite{Mazumdar:2018xjz} for a 3D version of the action in eq.(\ref{action}).}
\begin{equation} \label{action}
\begin{aligned}
	S &= \int d^4 x \frac{\sqrt{-g}}{2} \left[ M_P^2 \left( R -  2 \Lambda \right) + R \Fc_1\LF \square_{s} \RF R+ S^{\nu}_{\ \mu} \Fc_2\LF \square_{s} \RF S^{\mu}_{\ \nu} +   C^{\rho \sigma}_{\ \ \ \mu \nu}   \Fc_3\LF \square_{s} \RF   C^{\mu \nu}_{\ \ \ \rho \sigma} \right]\,\\
	&= \int d^4 x \  \Lc_{EH + \Lambda} + \Lc_{R^2} + \Lc_{S^2} + \Lc_{C^2}\,,
\end{aligned}
\end{equation}
where we have defined $\square_{s} \equiv \square / M_s^2 $, where $M_s (<M_P)$ is the scale of non-locality. The traceless Ricci tensor in $4$ dimensions is given by
\be
S_{\mu \nu} = R_{\mu \nu} - \frac{R}{4} g_{\mu \nu}.
\ee
$\Fc_i\LF  \square_{s} \RF$ are analytic gravitational form factors which are functions of d'Alembertian and have a power series expansion:
\begin{equation} \label{formfactor}
\Fc_i\LF  \square_{s} \RF =\sum_{n=0}^{\infty}f_{i,n}\LF\frac{ \square }{M_s^2}\RF^n\,.
\end{equation}
The coefficients go as $f_{i,n} \sim \Oc\LF M_{P}^{2}/M_{s}^{2} \RF$.
In IR, for low enough momenta $k \ll M_s$, $\Lc_{EH + \Lambda}$ dominates over the higher curvature terms. One can easily consider special cases of the above general action by considering arbitrary form factors, like ${\cal F}_{i}(\square_s)=1$, and truncating $\Fc_i\LF  \square_{s} \RF$ to a finite order in $\square_{s}$~\footnote{From eq.(\ref{action}) we can reach various limits, such as
pure Einstein-Hilbert term, or fourth order gravity: $R+R^2+R^{\mu\nu}R_{\mu\nu}+C^{\mu\nu\lambda\sigma}C_{\mu\nu\lambda\sigma}$, or 
sixth order gravity, by keeping $f_{i,0}$ and $f_{i,1}$ terms and so on and so forth around Minkowski background. Also higher curvature terms are also encapsulated in this action, because around arbitrary backgrounds (i.e., other than Minkowski), $\square_s$ also contributes to ${\cal O}(h)$. Therefore, in principle, expanding the action around an arbitrary background will give ${\cal O}(h^{n})$ contribution to the action. }. The full, non-linear equations of motion (EoM), around an arbitrary background, has been derived in \cite{Biswas:2013cha} (see Appendix (\ref{fullequations of motion})). In this paper, we take the above action in eq.(\ref{action}) and analyze second order perturbations around a specific (\textit{non}-MSS) background.

\section{Perturbation theory} \label{pertgenbackgsec4}

We will use the well-known background field method (see~\cite{Abbott:1981ke},\cite{Christensen:1979iy}) to study second order metric perturbations at the level of the action. We define the covariant scalar-vector-tensor (SVT) decomposition and provide the most general structure of the second order action ($\delta^{2} S$) one can expect in IDG (and a generic quadratic curvature gravity) around arbitrary backgrounds.
Restricting to $3+1$ dimensions, we split the full metric $g_{\mu \nu}$ into background $\bar{g}_{\mu \nu}$ and perturbation $h_{\mu \nu}$:
\be \nonumber
g_{\mu \nu} = \bar{g}_{\mu \nu} + h_{\mu \nu},
\ee
where $| h_{\mu \nu} | < | \bar{g}_{\mu \nu} |$, and expand the action $S$ up to $\mathcal{O}(h^2)$ to get $\delta^{2}S$. We are interested in perturbations around vacuum solutions so we put $T_{\mu \nu} = 0$. Extremizing $S$ to first order gives the background EoM, $\bar{E}_{\mu \nu} = 0$. We consider on-shell perturbations and therefore get rid of the linear in $h$ term in $\delta^{2}S$ by imposing $\bar{E}_{\mu \nu} = 0$. The background EoM are useful in simplifying the computation of $\delta^{2}S$ as we will see later.



\subsection{Scalar-vector-tensor decomposition}
We can covariantly decompose $h_{\mu\nu}$ into two scalar, one vector and one tensor (SVT) degrees of freedom \cite{VanNieuwenhuizen:1973fi,AIHPA_1967__7_2_149_0,York:1974psa,Gibbons:1978ji}:
\be \label{hdecomp}
h_{\mu \nu} = \widehat{h}_{\mu \nu} + \bar{\nabla}_{ \mu} A_{\nu} + \bar{\nabla}_{ \nu} A_{\mu} + \left( \bar{\nabla}_\mu \bar{\nabla}_\nu - \frac{1}{4} \bar{g}_{\mu \nu} \bar{\square}  \right) B + \frac{1}{4} \bar{g}_{\mu \nu}h,
\ee
where
\be \label{hdcond}
\bar{\nabla}^\mu \widehat{h}_{\mu \nu} = 0\,,\quad 
\bar{g}^{\mu \nu} \widehat{h}_{\mu \nu} = 0\,,\quad 
\bar{\nabla}^\mu A_{\mu} = 0\,.
\ee
The first two conditions make $\widehat{h}_{\mu \nu}$ transverse and traceless, respectively. The last condition ensures transversality of $A_{\mu}$, while $h$ is the trace $\bar{g}^{\mu \nu} h_{\mu \nu}$. For both GR and IDG, the above decomposition produces kinetic mixings between $B$ and $h$ modes in $\delta^2 S$ when the background is flat or dS/AdS. This mixing can be removed by the field redefinition: $\phi \equiv \bar{\square}  B - h$, so that the graviton quadratic form has one tensor $\widehat{h}_{\mu \nu}$ and one scalar $\phi$ mode, as expected \cite{Biswas:2016egy,Biswas:2016etb}. Following suit, we rewrite eq.(\ref{hdecomp}) as
\be \label{hdecomp2}
\ba
h_{\mu \nu} &= \widehat{h}_{\mu \nu} + \bar{\nabla}_{ \mu} A_{\nu} + \bar{\nabla}_{ \nu} A_{\mu} + \bar{\nabla}_\mu \bar{\nabla}_\nu B - \frac{1}{4} \bar{g}_{\mu \nu} \phi\,,
\ea
\ee
which is the decomposition we will use in this paper.
The theory has gauge redundancy expressed by the following gauge transformation:
\be
h_{\mu \nu} \to h_{\mu \nu} + \bar{\nabla}_{\nu} \xi_{\mu} + \bar{\nabla}_{\mu} \xi_{\nu}\,,
\ee
where the gauge transformation vector $\xi_{\mu}$ can be decomposed into its transverse and longitudinal parts as \cite{Alvarez-Gaume:2015rwa}:
\be
\xi_{\mu} = \widehat{\xi}_{\mu} + \partial_{\mu} \chi, \qquad \qquad \text{where } \ \bar{\nabla}^{\mu}  \widehat{\xi}_{\mu} = 0,
\ee
so that the SVT modes transform as following under the gauge transformations:
\be \label{gaugetransformationsofmodes}
\widehat{h}_{\mu \nu} \to \widehat{h}_{\mu \nu}\,, \qquad \qquad
\phi \to \phi\,, \qquad \qquad
A_{\mu} \to A_{\mu} + \widehat{\xi}_{\mu}\,, \qquad \qquad
B \to B + 2 \xi,
\ee
from which we see that $\widehat{h}_{\mu \nu}$ and $\phi$ are gauge invariant while $A_{\mu}$ and $B$ are not. Note that we have not yet canonically normalized $h_{\mu \nu}$, so $[h_{\mu \nu}] = 0, [\phi]=0, [B]=-2, [A_{\mu}]=-1$ and $[\widehat{h}_{\mu \nu}]=0$.


\subsection{General structure of $\delta^{2}S$}

Studying the propagating modes of the theory given by eq.(\ref{action}) around general backgrounds is quite involved. In order to find $\delta^{2}S$ for any theory, we need to compute perturbations of many quantities. We provide perturbations of curvatures around \textit{arbitrary} backgrounds in Appendix~\ref{appendixperturbations}. Decomposition of these perturbations in SVT form are given in Appendix~\ref{appendixsecondordervariations}. The second order action around \text{arbitrary} backgrounds has the following general form\footnote{We have suppressed Lorentz indices in each matrix element of $\cK$ which are needed to make the action a scalar.}:
\[
\delta^{2} S =
\int d^4x \sqrt{-\bar{g}} \ \begin{bmatrix}
B & \phi & A_{\rho} & \widehat{h}_{\mu \nu}
\end{bmatrix}
\begin{bmatrix}
\mathcal{K}_{00} & \mathcal{K}_{01} & \mathcal{K}_{02} & \mathcal{K}_{03}\\
\mathcal{K}_{10} & \mathcal{K}_{11} & \mathcal{K}_{12} & \mathcal{K}_{13}\\
\mathcal{K}_{20} & \mathcal{K}_{21} & \mathcal{K}_{22} & \mathcal{K}_{23}\\
\mathcal{K}_{30} & \mathcal{K}_{31} & \mathcal{K}_{32} & \mathcal{K}_{33}
\end{bmatrix}
\begin{bmatrix}
B\\
\phi \\
A_{\sigma} \\
\widehat{h}_{\alpha \beta}
\end{bmatrix}\,,
\]
where $\mathcal{K}$ is a complicated, generalized kinetic matrix given by the sum of kinetic and effective mass matrices of all SVT modes. Its entries can be read off from the general structure of perturbations given in Appendix (\ref{appendixsecondordervariations}). From there, one can see that $\cK$ is not diagonal in general and involves mixings between the different SVT modes. For GR and IDG around dS/AdS/flat backgrounds, $\cK$ is diagonal in $\phi$ and $\widehat{h}_{\mu \nu}$ \cite{Biswas:2016egy}:
\[ 
\begin{bmatrix}
B & \phi & A_{\rho} & \widehat{h}_{\mu \nu}
\end{bmatrix}
\begin{bmatrix} 
0 & 0 & 0 & 0 \\
0 & \mathcal{K}_{11} & 0 & 0 \\
0 & 0 & 0 & 0 \\
0 & 0 & 0 & \mathcal{K}_{33}
\end{bmatrix} 
\begin{bmatrix}
B\\
\phi \\
A_{\sigma} \\
\widehat{h}_{\alpha \beta}
\end{bmatrix}\,.\label{msskineticmatrix}
\]

Next, we discuss backgrounds typically studied in perturbative analyses, and then choose a specific \textit{non}-MSS background ($\bar{g}_{\mu \nu}$) for the purpose of this paper. We motivate our choice of background which occurs in the context of IDG and removes cosmological and black hole type singularities.

\section{Typical background spacetimes} \label{sec-nonmss}

\begin{figure}[!ht]
\includegraphics[scale=0.385]{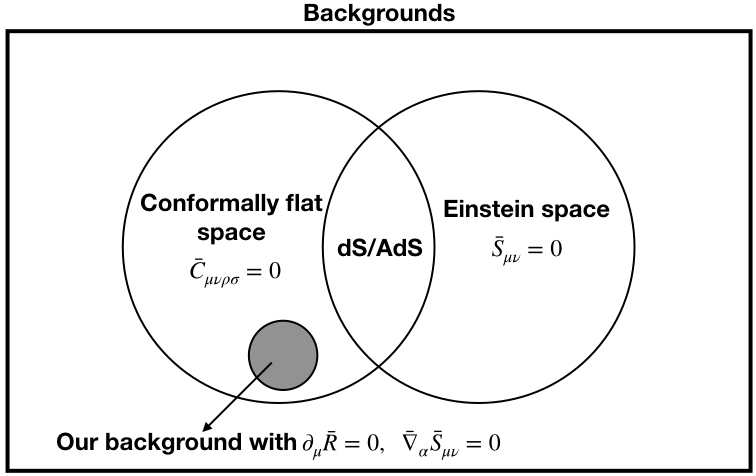}
\caption{Venn diagram with our specific conformally-flat, \textit{non}-maximally symmetric background. $\partial_{\mu} \bar{R}=0$ follows from $\bar{\nabla}_{\alpha} \bar{S}_{\mu \nu} = 0$. dS/AdS limit is achieved by putting $\bar{S}_{\mu \nu} = 0$.}
\label{fig:venn}
\end{figure}

We recall here aspects of pedagogical backgrounds one typically encounters when studying gravitational perturbations, and choose a specific background for the purpose of this paper. The lower the symmetry of the background, the harder it is to find $\delta^{2} S$, or equivalently, compute (equations of motion of) perturbations expressible in a form that is tractable to solving them, either analytically or numerically. It is useful to decompose all curvatures into their irreducible trace and traceless parts. Doing this for the Riemann tensor gives
\be \label{riemanndecomp}
\bar{R}_{\mu \nu \rho \sigma} = \underbrace{ \frac{\bar{R}}{12} \left( \bar{g}_{\mu \rho } \bar{g}_{\nu \sigma } - \bar{g}_{\mu \sigma } \bar{g}_{\nu \rho } \right) }_{\text{dS/AdS}} + \underbrace{ \frac{1}{2} \left( \bar{g}_{\nu \sigma } \bar{S}_{\mu \rho } +  \bar{g}_{\mu \rho } \bar{S}_{\nu \sigma }  -   \bar{g}_{\nu \rho } \bar{S}_{\mu \sigma } -   \bar{g}_{\mu \sigma } \bar{S}_{\nu \rho }  \right) + \bar{C}_{\mu \nu \rho \sigma } }_{\text{deviation from dS/AdS}}\,,
\ee
which manifestly separates the Riemann tensor for a maximally symmetric spacetime (MSS) and deviations from it. Two commonly studied spacetimes are conformally-flat (FLRW) and Einstein ($\bar{R}_{\mu \nu} \propto \bar{g}_{\mu \nu}$) manifolds, defined by $\bar{C}_{\mu \nu \rho \sigma} = 0$ and $\bar{S}_{\mu \nu} = 0$, respectively (see FIG. \ref{fig:venn}).

Given $\bar{S}_{\mu \nu} = \bar{C}_{\mu \nu \rho \sigma} = 0$, we get $\partial_{\mu} \bar{R} = 0$ from the (differential) Bianchi identity, implying that the manifold is MSS with the Riemann tensor assuming the dS/AdS form highlighted in eq.(\ref{riemanndecomp}). 
This means that the intersection of conformally-flat and Einstein manifolds is MSS as shown in FIG. \ref{fig:venn}. From FIG. \ref{fig:venn} and eq.(\ref{riemanndecomp}), it is clear that deviations from MSS can arise either when $\bar{S}_{\mu \nu}$ or $\bar{C}_{\mu \nu \rho \sigma} \neq 0$. In this paper, we will focus on conformally-flat, \textit{non}-MSS backgrounds. To be clear, the background we consider in this paper has the following properties:
\begin{subequations} \label{mss20}
\begin{align}
\bar{C}_{\mu \nu \rho \sigma} &= 0, \quad \bar{S}_{\mu \nu} \neq 0, \quad \bar{R} \neq 0 \label{mss19}\,, \\
\bar{\nabla}_{\alpha} \bar{S}_{\mu \nu} &= 0, \quad \partial_{\mu} \bar{R} = 0\,, \label{mss21}
\end{align}
\end{subequations}
where the `covariantly constant curvature' conditions\footnote{Given eq.(\ref{mss21}) and $\bar{C}_{\mu \nu \rho \sigma} = 0$ from eq.(\ref{mss19}), we can succinctly write eq.(\ref{mss21}) as $\bar{\nabla}_{\alpha}\bar{R}_{\mu \nu \rho \sigma} = 0$. This covariantly constant curvature condition has appeared elsewhere in literature, for example, to simplify calculations of asymptotic heat kernel expansions when developing a covariant technique of computing the one-loop effective action in the presence of arbitrary background fields in curved space \cite{Avramidi:1990je}.} in eq.(\ref{mss21}) greatly simplify SVT perturbations of the action to be discussed in sec.(\ref{secondordervarsection}). Actually, $\bar{\nabla}_{\alpha} \bar{S}_{\mu \nu} = 0$ implies $\partial_{\mu} \bar{R} = 0$ via the contracted Bianchi identity in eq.(\ref{contrbianchi}), so effectively we have only one condition, but we keep both of them explicit. Note that eq.(\ref{mss21}) does not imply that the background is MSS; a simple example is the Schwarzschild metric. It is easy to take the dS/AdS limit of $\delta^{2} S$ around this more general background in eq.(\ref{mss20}) by putting $\bar{S}_{\mu \nu} = 0$ (and $\bar{R} = 0$ for flat space limit) as we will see in sec.(\ref{checkinglimitsmss}). The  background in eq.(\ref{mss20}) is not as restrictive as MSS, and we will see from the analysis of perturbations later in sec.\ref{physicalmodes} that this is just enough to draw interesting qualitative and quantitative conclusions about the nature of gravity in less symmetric, non-trivial backgrounds, particularly scalar-vector-tensor mixing essentially because $\bar{S}_{\mu \nu} \neq 0$. In Table \ref{tablecurvatures}, we collect properties of curvatures of typical spacetimes and the background we consider in this paper. FIG. \ref{fig:venn} shows a Venn diagram of different backgrounds with our chosen background in the shaded region.

\begin{table} [!ht]
\begin{tabular}{|c|c|c|c|c|c|c|c|}
	\hline
	 &  &  &  &  &  & & \tabularnewline  & $\bar{R}$  & $\bar{S}_{\mu\nu}$ & $\bar{R}_{\mu\nu}$   & $\bar{C}_{\mu\nu\rho\sigma}$ & $\bar{R}_{\mu\nu\rho\sigma}$  & $\partial_{\mu}\bar{R}$  & $\bar{\nabla}_{\alpha}\bar{S}_{\mu\nu}$     \tabularnewline
	\hline 
	&  &  &  &  &  & & \tabularnewline Maximally symmetric spacetime (MSS) & constant  & $0$ & $\bar{R}\bar{g}_{\mu\nu}/4$  & $0$ & $\bar{R} \left( \bar{g}_{\mu \rho } \bar{g}_{\nu \sigma } - \bar{g}_{\mu \sigma } \bar{g}_{\nu \rho } \right) /12$ & $0$  & $0$    \tabularnewline
	\hline 
	&  &  &  &  &  & & \tabularnewline Einstein (Ricci flat) & $\neq 0$ $(0)$  & $0$ & $\bar{R}\bar{g}_{\mu\nu}/4$ $(0)$ & $\neq0$  & $\neq \bar{R} \left( \bar{g}_{\mu \rho } \bar{g}_{\nu \sigma } - \bar{g}_{\mu \sigma } \bar{g}_{\nu \rho } \right)/12$ & $\neq 0$ $(0)$  & $0$     \tabularnewline
	\hline 
	&  &  &  &  &  & & \tabularnewline Conformally-flat (e.g. FLRW) 
	 & $\neq 0$  & $\neq 0$ & $\neq 0$ &$0$ & $\neq \bar{R} \left( \bar{g}_{\mu \rho } \bar{g}_{\nu \sigma } - \bar{g}_{\mu \sigma } \bar{g}_{\nu \rho } \right)/12$ & $\neq0$  & $\neq0$       \tabularnewline
	\hline 
	&  &  &  &  &  & & \tabularnewline This paper (eq.(\ref{mss20}))
	 & constant  & $\neq0$ & $\neq 0$ &$0$ & $\neq \bar{R} \left( \bar{g}_{\mu \rho } \bar{g}_{\nu \sigma } - \bar{g}_{\mu \sigma } \bar{g}_{\nu \rho } \right)/12$ & $0$  & $0$       \tabularnewline
	\hline 
\end{tabular}
\caption{Properties of typical background spacetimes. `$\neq 0$' in particular implies an arbitrary function of $x^{\mu}$.}
\label{tablecurvatures}
\end{table}

\subsection{Background equations of motion}
\label{sec4Abeq}
The full, non-linear EoM for the action in eq.(\ref{action}) is given in Appendix (\ref{equations of motion}). On these, we impose our background conditions $\bar{C}_{\mu \nu \rho \sigma} = 0, \bar{\nabla}_{\alpha} \bar{S}_{\mu \nu} = 0$ and $\partial_{\mu} \bar{R} = 0$ from eq.(\ref{mss20}) to obtain:
\be \label{backgroundequations of motion}
\Omega \bar{S}^{\mu}_{\ \nu}  =  f_{2,0} \left( \frac{1}{2} \delta^{\mu}_{\ \nu} \bar{S}^{\alpha \beta}\bar{S}_{\alpha \beta}   - \frac{1}{2} \bar{R} \bar{S}^{\mu}_{\ \nu} - 2 \bar{S}^{\mu \alpha} \bar{S}_{\nu \alpha}  \right)\,,
\ee
where we have defined
\be
\Omega \equiv M_P^{2}  +  2 f_{1,0}  \bar{R} \,,
\ee
and used 
\be \label{lambdafrombgequations of motion}
\bar{R} = 4 \Lambda\,,
\ee
which comes from the trace of the equations of motion given in eq.(\ref{traceequations of motion}). We have a $3$ parameter family of solutions characterized by $(\Lambda, f_{1,0}, f_{2,0})$. Of course, one trivial solution is $\bar{S}_{\mu \nu} = 0$, but this is just MSS. For $\bar{S}_{\mu \nu} \neq 0$ for our background (see eq.(\ref{mss19})), we consider non-trivial solutions of eq.(\ref{backgroundequations of motion}), but the exact form of these solutions is irrelevant for our purposes. The background EoM in eq.(\ref{backgroundequations of motion}) will be useful in simplifying $\delta^{2}S$ as we will see later. Next, we consider examples where our background conditions in eq.(\ref{mss20}) may be realized physically.

\subsection{Physical examples}
\subsubsection{Near a cosmological bounce}

In bouncing cosmology, we usually deal with a flat Friedmann-Lema\^itre-Robertson-Walker (FLRW) metric given
by
\begin{equation}
ds^2= -dt^2+a^2\LF dr^2+r^2d\Omega^2 \RF\,,\qquad K= 12\LF H^2+\dot{H} \RF^2+12\LF \frac{k}{a^2} + H^2 \RF^2\,,
\end{equation}
where an overdot indicates differentiation with respect to cosmic time $t$ and $\displaystyle H = ({\dot{a}}/{a})$ is the Hubble factor. $K$ is the Kretschmann scalar which can be decomposed as
\begin{equation} \label{kret}
K = \bar{R}_{\mu\nu\rho\sigma}\bar{R}^{\mu\nu\rho\sigma} = \bar{C}_{\mu\nu\rho\sigma}\bar{C}^{\mu\nu\rho\sigma}+ 2\bar{S}_{\mu\nu}\bar{S}^{\mu\nu} + \frac{\bar{R}^{2}}{6}\,.
\end{equation}
Since FLRW metric is conformally-flat, the Weyl tensor in eq.(\ref{kret}) vanishes. So if $K$ is singular as $t \to 0$, it implies that $\bar{S}_{\mu\nu}\bar{S}^{\mu\nu}$ and/or $\bar{R}^{2}$ blow up. This is the case in big bang singularity where $ a\sim t^{\alpha},\,H\sim{\alpha}/{t}$, for $\alpha >0$. Now, let us consider a bouncing scenario.  General physical conditions at any symmetric bounce are \cite{Biswas:2010zk}
\begin{equation}
H=0,\quad \dot{H}\neq 0,\quad \ddot{H}=0,\quad \dddot{H}\neq 0,\quad \ddddot{H}=0, \cdots \implies H^{(2n)}=0.
\end{equation} 
If we consider a simple bouncing scenario, which happens very slowly, we then have
\begin{equation}\label{hidotH}
H^{(2n+1)}\approx 0,\qquad n\geq 1\,.
\end{equation}
To realize a slow-bouncing scenario, we can Taylor-expand the Hubble factor in time as
\begin{equation}
H=M_s\sum_{n=0}^{\infty} c_n(M_st)^{2n},
\end{equation}
where $M_s$ characterizes the time scale related to bounce. The Ricci scalar and tensor for flat FLRW are 
\begin{equation}
\bar{R}=12 H^2+6\dot{H},\quad \bar{R}^0_{\ 0}= 3H^2+3\dot{H},\quad \bar{R}^i_{\ i}= 3H^2+\dot{H}\,.
\end{equation}
At the point of slow bounce, we have $H=0$ and $\dot{H}\approx$ constant. In the limit $t\ll {1}/{M_s}$, we have
\begin{equation} \label{bounce1}
\bar{R}= \bar{R}_0 M_s^2+\bar{R}_1 M_s^2 t^2+ \cdots \implies \partial_\mu \bar{R} \approx 0,\qquad  \bar{\square}  \bar{R}\approx 0 \,~~  \text{if} \, ~~ \bar{R}_1\ll \bar{R}_0\,.
\end{equation}
The non-zero components of traceless Ricci tensor, in general, take the form 
\begin{equation}
\bar{S}^\mu_{\ \nu}=\frac{1}{2}\dot{H} \,\textit{diag}\LF 3,\,-1,\,-1,\,-1 \RF\,,
\end{equation}
during the bounce. Using the approximation of a slow bouncing scenario stated in eq.(\ref{hidotH}) which is nearly constant, we obtain  
\begin{equation}
\bar{\nabla}_{ \alpha}\bar{S}^\mu_{\ \nu} \approx 0\,,\qquad \bar{\square}\bar{S}^{\mu}_{\ \nu} \approx 0\,. 
\end{equation}
We see that we can satisfy our background conditions in eq.(\ref{mss20}) in a bouncing scenario that occurs sufficiently slow, or in the limit $t\to 0$. Therefore, studying perturbations around backgrounds satisfying the conditions in eq.(\ref{mss20}) enables us to address the stability of a bouncing universe.


\subsubsection{Non-vacuum, non-singular static compact objects}

In IDG, classical and quantum analyses suggest the existence of a non-singular, spherically symmetric metric, with an effective scale of non-locality $M_{\text{eff}}$ (which may be different from $M_s$, see \cite{Buoninfante:2018gce,Buoninfante:2019swn}), and is given by \cite{Buoninfante:2018rlq}:
\begin{equation} \label{idgmetric}
ds^2= -\LF 1+2\Phi \RF dt^2+\LF 1-2\Phi \RF \left[ dr^2+r^2d\Omega^2 \right] , \qquad \qquad \Phi = -\frac{G m}{r}\text{Erf}\LF \frac{rM_{\text{eff}}}{2} \RF.
\end{equation}
In IR, it asymptotes to the linearized Schwarzschild solution. Computing curvatures for the above metric in the limit $r\ll {1}/{M_{\text{eff}}}$, we get \cite{Buoninfante:2018xiw}
\begin{equation}
\begin{aligned}
\bar{R}_{00}&=\bar{R}_{11}\approx\frac{e^{-r^2M_{\text{eff}}^2}G m M_{\text{eff}}^3}{2\sqrt{\pi}}\,,\qquad \bar{R}_{22}=\bar{R}_{33}=0\,,\qquad \bar{R}=2\bar{R}_{00}\,, \\
\bar{S}_{00}&=\bar{S}_{11}=\frac{3 G m M_{\text{eff}}^{3} e^{-\frac{1}{4} M_{\text{eff}}^2 r^2}}{4 \sqrt{\pi }}\,,\qquad \bar{S}_{22}=\bar{S}_{33}=0, \qquad \bar{\nabla}_\alpha \bar{S}_{\mu\nu} \approx \frac{G m  M_{\text{eff}}^5 r e^{-\frac{1}{4} M_{\text{eff}}^2 r^2}}{4 \sqrt{\pi }}\,,
\end{aligned}
\end{equation}
so that, in the short distance regime $ \displaystyle r\ll {1}/{M_{\text{eff}}}$, we have
\begin{equation} \label{compactobjectconds}
\bar{\nabla}_\alpha \bar{S}_{\mu\nu} \approx 0\,, \qquad \qquad \partial_{\mu} \bar{R} \approx 0\,,
\end{equation}
as $r\rightarrow 0$. Also, the Kretschmann scalar is finite in this limit: $K \sim G^{2} m^{2} M_{\text{eff}}^{6}$. This solution is very different from a Schwarzschild black hole, which is a vacuum solution and has a  singularity as $r \to 0$. For differences between the two solutions, see \cite{Buoninfante:2018xiw}. Similar behavior persists if a charged system is studied~\cite{Buoninfante:2018stt}. The above conditions (eq.(\ref{compactobjectconds})) approximately satisfy the conditions we earlier imposed on our background choice  in eq.(\ref{mss20}). Furthermore, it was shown in \cite{Buoninfante:2018rlq} that the above metric solution becomes conformally-flat for length scales much smaller than the effective non-locality scale $1/M_{\text{eff}}$.


Also in \cite{Buoninfante:2018rlq}, it was shown that in the non-local region, the following metric is a solution of IDG EoM, with non-singular curvature scalar invariants:
\begin{equation}
ds^2=\LF\frac{2}{r M_{\text{eff}} }\RF^2\LF -dt^2+dr^2+r^2d\Omega^2 \RF\,,
\end{equation}
which gives
\begin{equation}
\bar{R}=0,\quad \bar{R}_{\mu\nu}=\frac{1}{r^2}\left(\begin{array}{cccc}
1 & 0 & 0 & 0\\
0 & 1 & 0 & 0\\
0 & 0 & r^{2} & 0\\
0 & 0 & 0 & r^{2}\sin^{2}\theta
\end{array}\right),\,\quad \bar{\nabla}_\alpha \bar{R}_{\mu\nu}=\bar{\square}  \bar{R}_{\mu\nu}=0\,.
\end{equation}
From the discussions above, there is enough physical motivation to study perturbations around our background in eq.(\ref{mss20}), satisfying $\bar{\nabla}_\alpha \bar{S}_{\mu\nu} \approx 0, \ \partial_\mu \bar{R} \approx 0$. Next, we study perturbations in SVT form around these backgrounds.


\section{Second order variations of the action} \label{secondordervarsection}

In this section, we will organize and present the second order perturbations for every term in the action given in eq.(\ref{action}) around the background in eq.(\ref{mss20}) and express them in SVT form (see eq.(\ref{hdecomp2})):
\be \label{delta2lagrangian}
\delta^2 S = \int d^4 x \ \delta^2 \mathcal{L} = \int d^4 x \ \delta^2 \mathcal{L}_{EH + \Lambda}  +  \delta^2 \mathcal{L}_{R^2}  +  \delta^2 \mathcal{L}_{S^2}  +  \delta^2 \mathcal{L}_{C^2}.
\ee
After obtaining perturbations in terms of $h_{\mu \nu}$ and background curvatures, a significant series of steps must be carried out in order to simplify them and express the final expressions in SVT form. We present only the final, most simplified expressions in this section, after applying our background curvature conditions in eq.(\ref{mss20}) and Bianchi identities. The step-by-step procedure we have followed is outlined below:

\begin{enumerate} [label=(\roman*)] \label{simplifyingsteps}

\item Decompose all background curvatures into their irreducible trace and traceless parts so that all expressions are specified in terms of Ricci scalar ($\bar{R}$), traceless Ricci tensor ($\bar{S}_{\mu \nu}$) and Weyl tensor ($\bar{C}_{\mu \nu \rho \sigma}$).

\item Commute covariant derivatives as much as possible so that they assume a canonical form, and end up acting on the divergenceless $\widehat{h}_{\mu \nu}$ and $A_{\mu}$, giving zero (see eq.(\ref{hdcond})).

\item Set $\bar{C}_{\mu \nu \rho \sigma} = 0$, since the background is conformally-flat (see eq.(\ref{mss19})).

\item The $nth$  order perturbation of an arbitrary tensor shares the same index symmetry and tracelessness properties as the original tensor. For example, the first order perturbation $c^{\mu \nu}_{\ \ \rho \sigma}$ inherits the properties of the full Weyl tensor $C^{\mu \nu}_{\ \ \rho \sigma}$, in particular, it is completely traceless.

\item Apply the contracted differential Bianchi identities:
\be \label{contrbianchi}
\bar{\nabla}^{\mu}\bar{R}_{\mu \nu} = \frac{1}{2} \partial_{\nu}\bar{R},  \qquad \qquad 
\bar{\nabla}^{\mu}\bar{S}_{\mu \nu} = \frac{1}{4} \partial_{\nu}\bar{R}\,.
\ee

\item Use the covariantly constant background curvature conditions in eq.(\ref{mss21}): $\bar{\nabla}_\alpha \bar{S}_{\mu\nu} = 0, \ \partial_\mu \bar{R} = 0$.
\end{enumerate}

The detailed derivation of perturbations can be found in Appendix (\ref{appendixsecondordervariations}) where the above steps have been followed. In particular, we have listed perturbations first around an \textit{arbitrary} background, then applied conformal-flatness ($\bar{C}_{\mu \nu \rho \sigma} = 0$) and finally applied the covariantly constant curvature conditions in eq.(\ref{mss21}). The general perturbations can be useful for future work when considering general or conformally-flat backgrounds (like FLRW) which do \textit{not} satisfy eq.(\ref{mss21}).

Perturbations at the quadratic level of the action around arbitrary, non-MSS backgrounds are quite convoluted (as can be seen from explicit expressions in Appendix (\ref{appendixsecondordervariations})) and although they tell us a great deal about the structure of possible higher order terms, some simplifying assumptions must be imposed if we are to make any headway in calculations. Our background conditions in eq.(\ref{mss20}) turn out to be very helpful in simplifying calculations to a large extent. $\delta^2 S$ is not too hard to find when the background is MSS (see~\cite{Biswas:2016egy}). The reason is the appearance in perturbative expressions of many terms proportional to $\partial_{\mu} \bar{R}, \bar{S}_{\mu \nu}$ and $\bar{C}_{\mu \nu \rho \sigma}$, which are zero for MSS. In this paper, we analyze perturbations beyond MSS because we choose to keep $\bar{S}_{\mu \nu} \neq 0$. Note that all perturbations have a common factor of $\sqrt{-\bar{g}}$, which we have omitted hereon for brevity.


\subsection{Perturbing Einstein-Hilbert plus cosmological constant term} \label{ehlambdapert}

The Einstein-Hilbert term with cosmological constant $\Lambda$ in our Lagrangian in eq.(\ref{action}) is:
\be
\ba
\mathcal{L}_{EH + \Lambda} & = \frac{M_P^2}{2} \sqrt{-g} \left(R - 2 \Lambda \right)\,.
\ea
\ee
Defining $\delta_0 \equiv \delta^2 \Big[ \sqrt{-g} (R - 2 \Lambda) \Big] $, we get  (see \cite{Biswas:2016egy}):
\be
\begin{aligned} \label{deltazero}
	\delta^2 \mathcal{L}_{EH + \Lambda} & =  \frac{M_P^2}{2} \delta_0\,,\\
	& =\frac{M_P^2}{2} \LT \delta^2 \left(\sqrt{-g} \right) \frac{\bar{R}}{2} + \delta^2 (R) \sqrt{-\bar{g}} + \delta (\sqrt{-g}) r \RT\,.
\end{aligned}
\ee
where we have used $\Lambda = \bar{R}/4$ which satisfies the background trace equations of motion (see eq.(\ref{lambdafrombgequations of motion})). We then first express eq.(\ref{deltazero}) in terms of $h_{\mu \nu}$ and then decompose the result in SVT form using eq.(\ref{hdecomp2}). $\delta_0$ contains terms purely quadratic in $\phi, B, A_{\mu}$ and $\widehat{h}_{\mu \nu}$, and mixed terms between them. Detailed calculation from scratch is given in Appendix (\ref{appendixehlambda}). Here we present the final expressions after following the simplifying steps given before in the beginning of sec.(\ref{simplifyingsteps}). The purely quadratic in scalar, vector and tensor modes each, are:
\be \label{eh33}
\ba
\delta_{0 (BB)} &= B \left[ \frac{1}{8} \bar{R} \bar{S}_{\alpha \beta } \bar{\nabla}^{\beta }\bar{\nabla}^{\alpha }
+ \frac{1}{2}  \bar{S}_{\alpha }{}^{\gamma } \bar{S}_{\beta \gamma } \bar{\nabla}^{\beta }\bar{\nabla}^{\alpha }
+ \frac{1}{4}  \bar{S}_{\beta \gamma } \bar{\nabla}^{\gamma }\bar{\nabla}^{\beta }\bar{\square}  \right] B\,,
\\ \\
\delta_{0 (AA)} &= A^\alpha \left[ \frac{1}{12}  \bar{R} \bar{S}_{\alpha \beta }
+  \bar{S}_{\alpha }{}^{\gamma } \bar{S}_{\beta \gamma }
-  \frac{1}{2} \bar{g}_{\alpha \beta} \bar{S}_{\mu \nu } \bar{S}^{\mu \nu }
-  \bar{S}_{\nu \mu } \bar{\nabla}^{\mu }\bar{\nabla}^{\nu } \bar{g}_{\alpha \beta} \right] A^\beta\,,
\ea
\ee
\be
\ba
\delta_{0 (\widehat{h}   \widehat{h} ) }= \widehat{h}_{\mu \nu} \left[ \frac{-\bar{R} + 6 \bar{\square}}{24}  \right] \widehat{h}^{\mu \nu}\,,
\qquad \qquad \delta_{0 (\phi \phi)} =\phi \left[ \frac{-\bar{R} - 3 \bar{\square}}{32} \right] \phi\,,
\ea
\ee
while the $5$ non-zero mix terms are:
\be \label{ehlambdamix1}
\ba
\delta_{0 (B \phi)} = B \left[ - \frac{1}{4} \bar{S}_{\alpha \beta } \bar{\nabla}^{\beta }\bar{\nabla}^{\alpha }  \right] \phi\,,
\qquad \qquad
 \delta_{0 (BA)} = B \left[ \frac{1}{2} \bar{R} \bar{S}_{\alpha \beta } \bar{\nabla}^{\beta } + 2 \bar{S}_{\alpha }{}^{\gamma } \bar{S}_{\beta \gamma } \bar{\nabla}^{\beta } \right] A^\alpha\,,
\ea
\ee
\be \label{ehlambdamix2}
\ba
\delta_{0 (B \widehat{h} ) }= B \left[ \frac{1}{3} \bar{R} \bar{S}^{\alpha \beta }
+ 2  \bar{S}^{\alpha \gamma } \bar{S}^{\beta}_{\ \gamma }
-  \frac{1}{2}  \bar{S}^{\alpha \beta } \bar{\square}  \right] \widehat{h}_{\alpha \beta }\,,
\qquad  
\delta_{0 (\phi A )} = \phi \left[  - \frac{1}{2} \bar{S}_{\alpha \beta } \bar{\nabla}^{\beta } \right] A^{\alpha } ,
\qquad   \delta_{0 (A  \widehat{h} )} = A^{\alpha } \left[  -2 \bar{S}^{\beta \gamma } \bar{\nabla}_{\gamma }  \right] \widehat{h}_{\alpha \beta}.
\ea
\ee
and the $\phi \widehat{h} $ scalar-tensor mixing is zero. Note from eq.(\ref{ehlambdamix1}) - eq.(\ref{ehlambdamix2}) that the mode mixing terms are proportional to $\bar{S}_{\mu \nu}$, which vanish around MSS and flat backgrounds, as we will see later in sec.(\ref{checkinglimitsmss}). But these mixings are non-zero for our background (see background EoM in eq.(\ref{backgroundequations of motion})).


\subsection{Perturbing quadratic in Ricci scalar term} \label{ricciscalarsquaredpert}

The Ricci scalar squared term in our Lagrangian in eq.(\ref{action}) is
\be \label{quadraticricciscalar}
\mathcal{L}_{R^{2}}  = \frac{1}{2} \sqrt{-g} \ R \Fc_1 (\square_{s}) R.
\ee
Let us lay out all the $10$ possible $\mathcal{O}(h^2)$ perturbations in $\delta^2 \left(\sqrt{-g} \ R \Fc_1 (\square_{s}) R \right)$. We label\footnote{For example, $(1100)$ denotes the expression which has $\mathcal{O}(h^{1})$ perturbation of $\sqrt{-g}$, $\mathcal{O}(h^{1})$ of $R$, $\mathcal{O}(h^{0})$ of $\Fc_1 (\square_{s})$ and $\mathcal{O}(h^{0})$ of $R$, all adding up to give $\mathcal{O}(h^2)$.} each term in $ \sqrt{-g} R \Fc_1 (\square_{s}) R $ by a number $n = 0,1$ or $2$ to indicate a contribution\footnote{
We follow the following notation: $R = \bar{R} + r + \delta^{2} R$,
$S_{\mu \nu} = \bar{S}_{\mu \nu} + s_{\mu \nu} + \delta^{2} S_{\mu \nu}$, and 
$C_{\mu \nu \rho \sigma} = \bar{C}_{\mu \nu \rho \sigma} + c_{\mu \nu \rho \sigma} + \delta^{2} C_{\mu \nu \rho \sigma}$. For details, see Appendix (\ref{appendixperturbations}).} of $\mathcal{O}(h^n)$:
\be \label{listofsecondvariationsricciscalar}
\ba
	(1100) &= \sqrt{-\bar{g}} \ \frac{h}{2} \ r  \mathcal{\bar{F}}_1(\bar{\square}_{s}) \bar{R}\,,\\
	(1010) &= \sqrt{-\bar{g}} \ \frac{h}{2} \ \bar{R} \ \delta(\Fc_1(\square_{s})) \bar{R}\,,\\
	(1001) &= \sqrt{-\bar{g}} \ \frac{h}{2} \ \bar{R}  \mathcal{\bar{F}}_1(\bar{\square}_{s}) r\,,\\
	(0110) &= \sqrt{-\bar{g}} \ r \ \delta(\Fc_1(\square_{s}))  \bar{R}\,,\\
	(0101) &= \sqrt{-\bar{g}} \ r \ \mathcal{\bar{F}}_1(\bar{\square}_{s}) r\,,\\
	(0011) &= \sqrt{-\bar{g}} \ \bar{R} \ \delta(\Fc_1(\square_{s}))  r\,,\\
	(2000) &= \sqrt{-\bar{g}} \left( - \frac{1}{4} h_{\mu \nu} h^{\mu \nu} + \frac{h^2}{8} \right) \bar{R} \mathcal{\bar{F}}_1(\bar{\square}_{s}) \bar{R}\,,\\
	(0200) &= \sqrt{-\bar{g}} \ \delta^{2}(R) \mathcal{\bar{F}}_1(\bar{\square}_{s}) \bar{R}\,,\\
	(0020) &= \sqrt{-\bar{g}} \  \bar{R} \delta^2(\Fc_1(\square_{s})) \bar{R}\,, \\
	(0002) &= \sqrt{-\bar{g}} \  \bar{R} \mathcal{\bar{F}}_1(\bar{\square}_{s}) \delta^2 (R)\,.
\ea
\ee
In total, there are 9 independent terms because $(0200)$ is equal to $(0002)$ after integration by parts.
Substituting the SVT decomposition eq.(\ref{hdecomp2}) in eq.(\ref{listofsecondvariationsricciscalar}) gives purely quadratic terms in $\phi, B, A $ and $\widehat{h}~$\footnote{Indices on $A_{\mu}$ and $\widehat{h}_{\mu \nu}$ suppressed for brevity.} each, and additionally, mixed terms between them. There are $6$ possible mix terms: $\phi B, \phi A , \phi \widehat{h} , BA, B \widehat{h} $ and $A \widehat{h} $, just as for $\delta^2 \mathcal{L}_{EH + \Lambda}$ in sec.(\ref{ehlambdapert}).

In $\delta^2 \mathcal{L}_{R^2}$, we collect some terms that actually turn out to be equal to $\delta^2 \mathcal{L}_{EH + \Lambda}$, as shown for maximally symmetric backgrounds in \cite{Biswas:2016egy}. Consider the part $\delta^2 \left( \sqrt{-g} R \right) \mathcal{\bar{F}}_1 (\bar{\square}_{s}) \bar{R}$ in $\delta^2 \mathcal{L}_{R^2}$. We see that it can be expanded as:
\be\label{rf1r}
\begin{aligned}
\delta^2 \left( \sqrt{-g} R \right) \mathcal{\bar{F}}_1(\bar{\square}_{s}) \bar{R} &= \delta^2 \left( \sqrt{-g} R \right) f_{1,0} \bar{R}
= \left[ \delta^2 \left( \sqrt{-g} \right) \bar{R} + 2 \sqrt{-\bar{g}} \delta^2 \left(  R \right) + \delta \left( \sqrt{-g} \right) r \right] f_{1,0} \bar{R}\,,
\end{aligned}
\ee
where in the first equality, the only non-zero term from $\mathcal{\bar{F}}_1 (\bar{\square}_{s}) \bar{R}$ was picked, the rest being zero from eq.(\ref{mss21}). In the second equality, the middle term has a coefficient of $2$ because it occurs twice in $\delta^2 \mathcal{L}_{R^2}$, as we just noted above: $(0200) = (0002)$. From eq.(\ref{rf1r}) and eq.(\ref{deltazero}), we notice that
\be \label{ehlambda3}
\delta^2 \left( \sqrt{-g} R \right) \mathcal{\bar{F}}_1 (\bar{\square}_{s}) \bar{R} = 2  \delta_0  f_{1,0} \bar{R} - \left[\delta \left( \sqrt{-g} \right) r  \right] f_{1,0} \bar{R}\,.
\ee
We recall here for convenience $\delta^2 \mathcal{L}_{R^2}$ from \cite{Biswas:2016egy}:
\be \label{mssricciscalarsquared}
\ba
\delta^2 \mathcal{L}_{R^2}&=\frac{1}{2}  \sqrt{-\bar g}\left[2\left(\frac h2r+\frac12\left(\frac{h^2}8-\frac{h_{\mu\nu}h^{\mu\nu}} 4\right)\bar R+\delta^2(R)\right){f_{1,0}}\bar R+r\Fc_1(\bar{\square}_s) r\right.\\
&\left.+\left(\frac h2\bar R+r\right)\delta(\Fc_1(\square_{s}))\bar R+\bar R\delta^2(\Fc_1(\square_{s}))\bar R+\frac h2\bar R( \Fc_1(\bar{\square}_s) - {f_{1,0}})r+\bar R\delta(\Fc_1(\square_{s}))r
\right]\,.
\ea
\ee
Comparing eq.(\ref{ehlambda3}) and eq.(\ref{mssricciscalarsquared}), the second term in eq.(\ref{ehlambda3}) forms a part of the second last term in the second line of eq.(\ref{mssricciscalarsquared}). In the second line of eq.(\ref{mssricciscalarsquared}), the first $2$ terms vanish upon using $\partial_{\mu}\bar{R} = 0$ from eq.(\ref{mss21}) and Appendix (\ref{deltaboxricciscalar}); the remaining $2$ terms are each non-zero, but sum to zero after using $\partial_{\mu}\bar{R} = 0$ and integrating by parts, exactly like the MSS case as shown in \cite{Biswas:2016egy}. Ultimately, this implies
\be \label{riccisquare}
\delta^2 \mathcal{L}_{R^2} = \sqrt{-\bar{g}} \left[   f_{1,0} \bar{R} \delta_0  + \frac{1}{2} r  \Fc_1 (\bar{\square}_{s})  r  \right]\,.
\ee
In deriving this result, we just used $\bar{R} =$ constant, which is true for both MSS and our background in eq.(\ref{mss20}). Therefore, for $\delta^2 \mathcal{L}_{R^2}$, the only non-trivial term we have to find is $(0101) = \sqrt{-\bar{g}} \ r \Fc_1 (\bar{\square}_{s})  r$. The detailed calculation from scratch is given in Appendix (\ref{appendixricciscalarsquare}) Finally, we have terms which are purely quadratic in scalar and tensor modes
\be \label{riccisquarefinal}
\ba
\left[ r  \Fc_1 (\bar{\square}_{s})  r \right]_{\phi \phi} = \phi \left[ \left(\bar{R} + 3 \bar{\square} \right)^{2} \frac{\Fc_1(\bar{\square}_{s})}{16} \right] \phi\,,
\qquad \qquad \left[ r  \Fc_1 (\bar{\square}_{s})  r \right]_{\widehat{h} \widehat{h} } = \widehat{h}^{\alpha \beta } \left[ \bar{S}_{\alpha \beta }   \Fc_1(\bar{\square}_{s})  \bar{S}_{\mu \nu } \right] \widehat{h}^{\mu \nu }\,,
\ea
\ee
and a scalar-tensor mix term
\be  \label{riccisquaremixterm}
\ba
\left[ r  \Fc_1 (\bar{\square}_{s})  r \right] _{\phi \widehat{h} } &= \phi \left[ -(\bar{R} + 3 \bar{\square} ) \bar{S}_{\alpha \beta} \frac{\Fc_1 (\bar{\square}_s)}{2}  \right]  \widehat{h}^{\alpha \beta }
\ea
\ee
with rest of the SVT perturbations zero for our chosen background in eq.(\ref{mss20}). Note that the scalar-tensor ($\phi \widehat{h} $) mixing in eq.(\ref{riccisquaremixterm}) is proportional to $\bar{S}_{\mu \nu}$ and so vanishes around flat and MSS backgrounds as we see later in sec.(\ref{checkinglimitsmss}). But it is non-zero for our background.


\subsection{Perturbing quadratic in traceless Ricci tensor term} \label{ssquarepert}

The traceless Ricci tensor squared term in our Lagrangian in eq.(\ref{action}) is
\be \label{quadratictracefreericciscalar}
\mathcal{L}_{S^{2}} = \frac{1}{2} \sqrt{-g} \ S^{\nu}_{\ \mu} \Fc_2 ( \square_{s} ) S^{\mu}_{\ \nu}.
\ee
$\delta^2 \left(\sqrt{-g} \ S^{\nu}_{\ \mu} \Fc_2 ( \square_{s} ) S^{\mu}_{\ \nu} \right)$ has the following contributions after imposing the background conditions, eq.(\ref{mss20}), and integration by parts:
\be \label{listofsecondvariationstfricci}
\ba
	(1100) &= \sqrt{-\bar{g}} \ \frac{h}{2} \ s^{\nu}_{\ \mu}  \cF_2(\bar{\square}_{s}) \bar{S}^{\mu}_{\ \nu}=\sqrt{-\bar{g}} \ \frac{h}{2} f_{2,0}\ s^{\nu}_{\ \mu} \bar{S}^{\mu}_{\ \nu}\,,\\
	(1010) &= \sqrt{-\bar{g}} \ \frac{h}{2} \ \bar{S}^{\nu}_{\ \mu}  \delta(\Fc_2(\square_{s})) \bar{S}^{\mu}_{\ \nu}\,,\\
	(1001) &= \sqrt{-\bar{g}} \ \frac{h}{2} \ \bar{S}^{\nu}_{\ \mu} \cF_2(\bar{\square}_{s}) s^{\mu}_{\ \nu} =  \sqrt{-\bar{g}} \frac{1}{2} \ \bar{S}^{\nu}_{\ \mu} s^{\mu}_{\ \nu} \cF_2(\bar{\square}_{s}) h\,, \\
	(0110) &= \sqrt{-\bar{g}} \ s^{\nu}_{\ \mu}  \delta(\Fc_2(\square_{s}))  \bar{S}^{\mu}_{\ \nu}\,,\\
	(0101) &= \sqrt{-\bar{g}} \ s^{\nu}_{\ \mu}  \cF_2(\bar{\square}_{s}) s^{\mu}_{\ \nu}\,,\\
	(0011) &= \sqrt{-\bar{g}} \ \bar{S}^{\nu}_{\ \mu}  \delta(\Fc_2(\square_{s}))  s^{\mu}_{\ \nu}\,,\\
	(2000) &= \sqrt{-\bar{g}} \left( - \frac{1}{4} h_{\alpha \beta} h^{\alpha \beta} + \frac{h^2}{8} \right) \bar{S}^{\nu}_{\ \mu} \cF_2(\bar{\square}_{s}) \bar{S}^{\mu}_{\ \nu} =  \sqrt{-\bar{g}} f_{2,0} \left( - \frac{1}{4} h_{\alpha \beta} h^{\alpha \beta} + \frac{h^2}{8} \right) \bar{S}^{\nu}_{\ \mu} \bar{S}^{\mu}_{\ \nu}\,,\\
	(0200) &= \sqrt{-\bar{g}} \ \delta^{2}(S^{\nu}_{\ \mu}) \cF_2(\bar{\square}_{s}) \bar{S}^{\mu}_{\ \nu} = \sqrt{-\bar{g}} \ f_{2,0} \bar{S}^{\mu}_{\ \nu} \delta^{2}(S^{\nu}_{\ \mu})\,,  \\
	(0020) &= \sqrt{-\bar{g}} \  \bar{S}^{\nu}_{\ \mu} \delta^2(\Fc_2(\square_{s})) \bar{S}^{\mu}_{\ \nu}\,, \\
	(0002) &= \sqrt{-\bar{g}} \  \bar{S}^{\nu}_{\ \mu} \cF_2(\bar{\square}_{s}) \delta^2 (S^{\mu}_{\ \nu}) = \sqrt{-\bar{g}} \ f_{2,0}  \bar{S}^{\nu}_{\ \mu} \delta^2 (S^{\mu}_{\ \nu})\,,
\ea
\ee
giving $9$ independent terms because $(0200) = (0002)$. For MSS, because $\bar{S}_{\mu \nu} = 0$, the only non-zero contribution in $\delta^2 \mathcal{L}_{S^2}$ is $(0101) = \sqrt{-\bar{g}} s^{\nu}_{\ \mu}  \cF_2 (\bar{\square}_{s})  s^{\mu}_{\ \nu}$. Expressing the above second order perturbation terms (eq.(\ref{listofsecondvariationstfricci})) in SVT decomposition is not straightforward. Although, most terms can be computed without much effort, the four terms $(1010), (0110), (0011)$ and $(0020)$ are non-trivial because they involve either $\delta(\Fc_2(\square_{s}))$ or $\delta^{2} (\Fc_2(\square_{s}))$ acting on a tensor of rank $2$, which have long expressions. We now simplify these terms as much as possible.


\begin{enumerate}
{
\item $\displaystyle (1010) = \sqrt{-\bar{g}} \ \frac{h}{2} \ \bar{S}^{\nu}_{\ \mu}  \delta(\Fc_2(\square_{s})) \bar{S}^{\mu}_{\ \nu}$. Recall that $\Fc_2(\square_{s})$ is a power series (see eq.(\ref{formfactor})):
\be
\ba
\Fc_2 \left(\square_{s} \right) &= \sum_{n=0}^{\infty} f_{2,n} \square_{s}^{n}\\
&= f_{2,0} + f_{2,1} \square_{s} + f_{2,2} \square_{s}^{2} + f_{2,2} \square_{s}^{3} + \cdots \nonumber\,,
\ea
\ee
This implies:
\be\label{deltaf2series}
\ba
\delta \Fc_2 \left(\square_{s} \right) &= f_{2,1} \delta \left( \square_{s} \right) + f_{2,2} \delta \left( \square_{s}^2 \right) + f_{2,3} \delta \left( \square_{s}^3 \right) + \cdots\\
&= f_{2,1} \delta \left( \square_{s} \right) + f_{2,2} \Big( \delta \left( \square_{s} \right) \square_{s} + \square_{s} \delta \left( \square_{s} \right) \Big) + f_{2,3} \Big( \delta(\square_{s}) \square_{s}^2 + \square_{s} \delta(\square_{s}) \square_{s} + \square_{s}^2 \delta(\square_{s}) \Big) + \cdots\,,
\ea
\ee
Now, using the covariantly constant curvature conditions in eq.(\ref{mss21}), we can put to zero each term in $\delta \Fc_2 \left(\bar{\square}_s \right)$ which has $\bar{\square}_s$ on its rightmost end. This enables us to effectively write $ \delta \Fc_2 \left(\square_{s} \right)$ in terms of $\delta (\square_{s})$ as:
\be
\ba
 \delta \Fc_2 \left(\square_{s} \right) &= \left( f_{2,1} + f_{2,2} \bar{\square}_s + f_{2,3} \bar{\square}_s^2 + \cdots \right) \delta (\square_{s})\\
 &= \left[ \frac{\Fc_2 \left(\bar{\square}_s \right) - f_{2,0}}{\bar{\square}_s} \right] \delta (\square_{s})\,.
\ea
\ee
We then finally obtain:
\be \label{final1010tfricci}
(1010) = \sqrt{-\bar{g}} \ \frac{h}{2} \ \bar{S}^{\nu}_{\ \mu} \left[ \frac{\Fc_2 \left(\bar{\square}_s \right) - f_{2,0}}{\bar{\square}_s} \right]  \delta (\square_{s}) \bar{S}^{\mu}_{\ \nu} = \sqrt{-\bar{g}} \ \frac{h}{2} \ \bar{S}^{\nu}_{\ \mu} \Fc_5(\bar{\square}_s) \delta (\square_{s}) \bar{S}^{\mu}_{\ \nu} \,,
\ee
where we define
\be \label{f4}
\Fc_4(\square_{s}) \equiv \left[ \frac{ \Fc_1 (\square_{s}) - f_{1,0} }{\square_{s}} \right]\,,
\ee
and
\be \label{deltaboxsmunu}
\ba
\delta (\square_{s}) \bar{S}^{\mu}_{\ \nu} &= \frac{1}{M_{s}^{2}} \Bigg[ - \frac{1}{6} h_{\nu }{}^{\alpha } \bar{R} \bar{S}^{\mu }{}_{\alpha } 
-  \frac{1}{4} h_{\nu }{}^{\alpha } \bar{S}_{\alpha \beta } \bar{S}^{\mu \beta } 
+ \frac{1}{12}  h \bar{R} \bar{S}^{\mu }{}_{\nu } 
+ \frac{1}{2} h^{\alpha \beta } \bar{S}_{\alpha \beta } \bar{S}^{\mu }{}_{\nu } 
-  \frac{1}{6} h^{\mu \alpha } \bar{R} \bar{S}_{\nu \alpha } 
-  \frac{3}{2} h^{\alpha \beta } \bar{S}^{\mu }{}_{\alpha } \bar{S}_{\nu \beta } \\
& \qquad + \frac{1}{2}  h \bar{S}^{\mu \beta } \bar{S}_{\nu \beta } 
-  \frac{1}{4} h^{\mu \alpha } \bar{S}_{\alpha \beta } \bar{S}_{\nu }{}^{\beta } 
+ \frac{1}{2} \bar{S}_{\nu }{}^{\alpha } \bar{\nabla}_{\beta }\bar{\nabla}_{\alpha }h^{\mu \beta } 
+ \frac{1}{2} \bar{S}^{\mu \alpha } \bar{\nabla}_{\beta }\bar{\nabla}_{\alpha }h_{\nu }{}^{\beta } 
+ \frac{1}{2} \bar{S}_{\nu }{}^{\alpha }  \bar{\square} h^{\mu }{}_{\alpha } \\
& \qquad -  \frac{1}{2} \bar{S}^{\mu \alpha }  \bar{\square} h_{\nu \alpha } 
-  \frac{1}{2} \bar{S}_{\nu }{}^{\alpha } \bar{\nabla}^{\mu }\bar{\nabla}_{\beta }h_{\alpha }{}^{\beta } 
-  \frac{1}{2} \bar{S}^{\mu \alpha } \bar{\nabla}_{\nu }\bar{\nabla}_{\beta }h_{\alpha }{}^{\beta } \Bigg]\,.
\ea
\ee

\item $\displaystyle (0110) = \sqrt{-\bar{g}} \ s^{\nu}_{\ \mu}  \delta(\Fc_2(\square_{s}))  \bar{S}^{\mu}_{\ \nu}$. Just like above, here we have
\be\label{final0110tfricci}
\ba
(0110) &= \sqrt{-\bar{g}} \ s^{\nu}_{\ \mu}  \delta(\Fc_2(\square_{s}))  \bar{S}^{\mu}_{\ \nu}\\
&= \sqrt{-\bar{g}} \ s^{\nu}_{\ \mu}  \left[ \frac{\Fc_2 \left(\bar{\square}_s \right) - f_{2,0}}{\bar{\square}_s} \right] \delta (\square_{s})   \bar{S}^{\mu}_{\ \nu} =   \sqrt{-\bar{g}} \ s^{\nu}_{\ \mu}  \Fc_5(\bar{\square}_s) \delta (\square_{s})   \bar{S}^{\mu}_{\ \nu}\,,
\ea
\ee
where we have defined
\be \label{f5}
\Fc_5(\square_{s}) \equiv \left[ \frac{ \Fc_2 (\square_{s}) - f_{2,0} }{\square_{s}} \right]\,,
\ee
and $s^{\nu}_{\ \mu}$ is given in Appendix (\ref{snumu}).

\item $\displaystyle (0011) = \sqrt{-\bar{g}} \ \bar{S}^{\nu}_{\ \mu}  \delta(\Fc_2(\square_{s}))  s^{\mu}_{\ \nu}$. Now the background $\bar{S}^{\nu}_{\ \mu}$ is to the left of $\delta(\Fc_2(\square_{s}))$. So we use $\bar{\square}_{s}$ operators in eq.(\ref{deltaf2series}) on the leftmost positions to integrate by parts and put these terms to zero. We finally get:
\be\label{final0011tfricci}
(0011) = \sqrt{-\bar{g}} \ \bar{S}^{\nu}_{\ \mu} \delta (\square_{s})  \left[ \frac{\Fc_2 \left(\bar{\square}_s \right) - f_{2,0}}{\bar{\square}_s} \right] s^{\mu}_{\ \nu} =   \sqrt{-\bar{g}} \ \bar{S}^{\nu}_{\ \mu} \delta (\square_{s})  \Fc_5(\bar{\square}_s) s^{\mu}_{\ \nu}
\ee
which has a long and complicated expression and there seems to be no straightforward way to obtain a readable form for $(0011)$.
\item $ \displaystyle (0020) = \sqrt{-\bar{g}} \  \bar{S}^{\nu}_{\ \mu} \delta^2(\Fc_2(\square_{s})) \bar{S}^{\mu}_{\ \nu}$ has the most involved expression of all:
\be\label{final0020tfricci}
\ba
(0020) =  \sqrt{-\bar{g}} \  \bar{S}^{\nu}_{\ \mu} \Bigg[ f_{2,1} \delta^{2}(\square_{s}) + 2 \delta(\square_{s}) \cF_{6} (\bar{\square}_{s}) \delta(\square_{s}) \Bigg]  \bar{S}^{\mu}_{\ \nu}\,,
\ea
\ee
where we have defined
\be \label{f6}
\cF_{6} (\bar{\square}_{s}) \equiv \left[ \frac{\Fc_2(\bar{\square}_{s}) - f_{2,0} - f_{2,1} \bar{\square}_{s} }{\bar{\square}_{s}^{2}} \right].
\ee
}
\end{enumerate}


\subsubsection*{Dominant terms in the UV} \label{dominanttermsf2}

The SVT decomposition of $\delta^2 \mathcal{L}_{S^2}$ in all its glory is quite complicated and we do not present it here. For a simplified analysis, in the UV limit $k\gtrsim M_s$, we consider the dominant contributions in eq.(\ref{listofsecondvariationstfricci}). Below we argue that $(1001)$ and $(0101)$ are the only contributions we need to consider in the UV limit. We assume $\bar{R},\,\bar{S}_{\mu\nu} < \Oc(M_s^2)$, which is reasonable, at least for the cosmological and black hole scenarios discussed in sec.(\ref{sec-nonmss}).

First of all, we neglect the local contributions in eq.(\ref{listofsecondvariationstfricci}), that is, which do not have the form factors $\cF_{i}(\bar{\square}_{s})$; in the UV limit, terms with $\cF_{i}(\bar{\square}_{s})$ dominate $\delta^{2} S$. These are $(1100), (2000), (0200)$ and $(0002)$. For example, $(1001)$ is dominant compared to $(1100)$ because $\cF_{2}(\bar{\square}_{s}) > f_{2,0}$. The remaining non-local contributions are $(1010), (1001), (0110), (0101), (0011)$ and $(0020)$. Let us compare all these terms with $(0101)$. Suppressing indices, $(0101)$ has the following terms schematically:
\begin{equation}\label{domiUVf2}
\ba
h \Bigg\{ \bar{\nabla} \bar{\nabla} \Fc_2(\bar{\square}_s) \bar{\nabla} \bar{\nabla} + \bar{S} \Fc_{2}(\bar{\square}_s)\bar{\nabla} \bar{\nabla} + \bar{R} \Fc_{2}(\bar{\square}_s) \bar{\nabla} \bar{\nabla} \Bigg\}h +
h\Bigg\{\bar{S} \Fc_{2}(\bar{\square}_s) \bar{S} + \bar{R} \Fc_{2}(\bar{\square}_s) \bar{R} + \bar{R} \Fc_{2}(\bar{\square}_s)\bar{S} \Bigg\}h\,,
\ea
\end{equation}
as can be seen from the expression for $s^{\nu}_{\ \mu}$ in Appendix (\ref{snumu}). Note that $h$ denotes the full perturbation before SVT decomposition. Assuming $k \gtrsim M_s$ in the UV limit and $\bar{R},\bar{S}_{\mu\nu} < \Oc(M_s^2)$, we can neglect terms in the second line of eq.(\ref{domiUVf2}) compared to terms in the first line. In this limit, we can also deduce that
\begin{equation}\label{apprperts}
\frac{1}{M_s^2} \delta(\square)\bar{S}_{\mu\nu} \ll s_{\mu\nu}, \qquad \qquad
\frac{1}{M_s^2} \delta(\square) s_{\mu \nu} \ll \frac{1}{M_s^2} \bar{\square} s_{\mu \nu}.
\end{equation}
To see this, we check contributions coming from the various dimensionful quantities. Consider the first inequality in eq.(\ref{apprperts}). Ignoring $h$, $\delta(\square)\bar{S}_{\mu\nu} / M_s^{2}$ given in eq.(\ref{deltaboxsmunu}) is schematically $\left( \bar{S} \bar{S} +  \bar{S}  \bar{R} +  \bar{S} \bar{\nabla} \bar{\nabla} \right) / M_s^{2}$, while $s^{\nu}_{\ \mu}$ given in eq.(\ref{snumu}) is schematically $ \left( \bar{R} + \bar{\nabla} \bar{\nabla} \right) $. Comparing these two and using the above-mentioned limit, $s_{\mu\nu}$ is dominant. The same argument can be used to see how the second inequality in eq.(\ref{apprperts}) comes about. Using the above reasoning, we see which terms in  eq.(\ref{listofsecondvariationstfricci}) can be neglected:
\begin{enumerate} [label=(\roman*)]
	\item $(1010)$:  given in eq.(\ref{final1010tfricci}), is schematically written as:
		\begin{equation}\label{1010les}
	\frac{1}{M_s^2} \bar{S} \bar{S} \ h \ \Fc_5(\bar{\square}_s) \Bigg\{ \bar{\nabla} \bar{\nabla} + \bar{S} + \bar{R} \Bigg\} h.
	\end{equation}
	Comparing eq.(\ref{1010les}) and $(0101)$ in eq.(\ref{domiUVf2}), we can neglect $(1010)$ in comparison with $(0101)$.
	\item $(1001)$: given in eq.(\ref{listofsecondvariationstfricci}), goes like $ \sqrt{-\bar{g}}\bar{S}^\nu_\mu s^\mu_\nu \Fc_s(\bar{\square}_s) h $, which we can compare with $(0101)$ given in eq.(\ref{domiUVf2}). Counting derivatives and curvatures, we can deduce that the contribution of $(1001)$ is of the same order as $(0101)$.
	\item $(0110)$: given in eq.(\ref{final0110tfricci}) can be neglected compared to $(0101)$ in eq.(\ref{domiUVf2}) by using the first approximation in eq.(\ref{apprperts}).
	\item $(0011)$: given in eq.(\ref{final0011tfricci}) can be similarly neglected compared to $(0101)$, using the second approximation in eq.(\ref{apprperts}). 
	\item $(0020)$: given in eq.(\ref{final0020tfricci}) contains two kinds of contributions. The first one is local as it has no $\cF_{2}(\bar{\square}_{s})$. This can be neglected compared to $(0101)$, which has $\cF_{2}(\bar{\square}_{s})$. The second part is non-local which can be neglected compared to $(0101)$ using the first approximation in eq.(\ref{apprperts}).
\end{enumerate}


\subsection{Perturbing quadratic in Weyl tensor term}
Lastly, the quadratic in Weyl tensor term in our Lagrangian in eq.(\ref{action}) is
\be \label{weylsquare}
\mathcal{L}_{C^2} =  \frac{1}{2} \sqrt{-g}  \  C^{\rho \sigma}_{\ \ \ \mu \nu}   \Fc_3 (\square_{s})  C^{\mu \nu}_{\ \ \ \rho \sigma}\,.
\ee
Since our background is conformally-flat, any term in $\delta^2 \left(  \sqrt{-g} \ C^{\rho \sigma}_{\ \ \ \mu \nu}   \Fc_3 (\square_{s})  C^{\mu \nu}_{\ \ \ \rho \sigma} \right)$ containing $\bar{C}_{\mu \nu \rho \sigma}$ vanishes. We are then left with only one non-zero term:
\be\label{weylpert}
(0101) = \sqrt{-\bar{g}}  \  c^{\rho \sigma}_{\ \ \mu \nu}  \mathcal{\bar{F}}_3 (\bar{\square}_{s})  c^{\mu \nu}_{\ \ \rho \sigma}\,,
\ee
which finally becomes:
\be \label{ocweyloperator}
\ba
c^{\rho \sigma}_{\ \ \mu \nu}  \mathcal{\bar{F}}_3 (\bar{\square}_{s})  c^{\mu \nu}_{\ \ \rho \sigma} &= \widehat{h}_{\alpha \beta} \Bigg\{ \bigg[ \delta^\beta_{\ \nu} \bar{g}^{\alpha \rho} \bar{S}_{\ \mu}^{\sigma} + 2 \delta^{\alpha}_{\ \mu} \bar{g}^{\beta \sigma}  \bar{\nabla}_{\nu} \bar{\nabla}^{\rho} \bigg] \mathcal{\bar{F}}_3 (\bar{\square}_{s}) \bigg[ -\frac{1}{12} \delta^{\nu}_{\ \sigma} \delta_{\ \rho}^{\lambda} \bar{g}^{\gamma \mu} \bar{R} + \frac{1}{12} \delta^{\nu}_{\ \rho} \delta^{\lambda}_{\ \sigma} \bar{g}^{\gamma \mu} \bar{R} - \frac{1}{12} \delta^{\mu}_{\ \rho} \delta^{\lambda}_{\ \sigma} \bar{g}^{\nu \gamma} \bar{R}\\
& \qquad + \frac{1}{3} \delta^{\mu}_{\ \rho} \delta^{\nu}_{\ \sigma} \bar{S}^{\gamma \lambda} - \frac{1}{2} \delta^{\nu}_{\ \sigma} \delta^{\gamma}_{\ \rho} \bar{S}^{\mu \lambda} + \frac{1}{2} \delta^{\nu}_{\ \rho} \delta^{\gamma}_{\ \sigma} \bar{S}^{\mu \lambda} - \frac{1}{4} \delta^{\lambda}_{\ \sigma} \bar{g}^{\gamma \nu} \bar{S}^{\mu}_{\ \rho} + \frac{1}{4} \delta^{\lambda}_{\ \rho} \bar{g}^{\gamma \nu} \bar{S}^{\mu}_{\ \sigma} - \frac{1}{2} \delta^{\mu}_{\ \rho} \delta^{\gamma}_{\ \sigma} \bar{S}^{\nu \lambda}\\
& \qquad + \frac{1}{4} \delta^{\lambda}_{\ \sigma} \bar{g}^{\gamma \mu} \bar{S}^{\nu}_{\ \rho} - \frac{1}{4} \delta^{\lambda}_{\ \rho} \bar{g}^{\gamma \mu} \bar{S}^{\nu}_{\ \sigma} - \frac{1}{4} \delta^{\nu}_{\ \sigma} \bar{g}^{\gamma \mu}  \bar{S}_{\rho}^{\ \lambda} + \frac{1}{4} \delta^{\nu}_{\ \rho} \bar{g}^{\gamma \mu} \bar{S}_{\sigma}^{\ \lambda} - \frac{1}{4} \delta^{\mu}_{\ \rho} \bar{g}^{\nu \gamma} \bar{S}_{\sigma}^{\ \lambda} + \frac{1}{4} \delta^{\nu}_{\ \sigma} \delta^{\lambda}_{\ \rho} \bar{g}^{\mu \gamma} \bar{\square}\\
& \qquad - \frac{1}{4} \delta^{\nu}_{\ \rho} \delta^{\lambda}_{\ \sigma} \bar{g}^{\mu \gamma} \bar{\square} + \frac{1}{4} \delta^{\mu}_{\ \rho} \delta^{\lambda}_{\ \sigma} \bar{g}^{\nu \gamma} \bar{\square} - \frac{1}{2} \delta^{\lambda}_{\ \sigma} \bar{g}^{\nu \gamma} \bar{\nabla}_{\rho} \bar{\nabla}^{\mu} + \frac{1}{2} \delta^{\lambda}_{\ \sigma}  \bar{g}^{\mu \gamma} \bar{\nabla}_{\rho} \bar{\nabla}^{\nu}
+ \frac{1}{2} \delta^{\lambda}_{\ \rho} \bar{g}^{\nu \gamma} \bar{\nabla}_{\sigma} \bar{\nabla}^{\mu} \\
& \qquad- \frac{1}{2} \delta^{\lambda}_{\ \rho} \bar{g}^{\mu \gamma} \bar{\nabla}_{\sigma} \bar{\nabla}^{\nu} \bigg] \Bigg\} \widehat{h}_{\gamma \lambda}\,,\\
& \equiv \widehat{h}_{\alpha \beta}  \cO_{(C)}^{\alpha \beta \gamma \lambda}  \widehat{h}_{\gamma \lambda}\,,
\ea
\ee
where the operator $\cO_{(C)}^{\alpha \beta \gamma \lambda}$ is introduced for brevity; $(C)$ indicates that it comes from $ \delta^{2} \mathcal{L}_{C^2}$. All other SVT contributions including mix terms are zero. Detailed calculation from scratch is given in Appendix (\ref{appendixweylsquare}). MSS and flat space limits are discussed in sec.(\ref{checkinglimitsmss}) which are greatly simplified 
after setting $\bar{S}_{\mu \nu} = 0$.


\section{\texorpdfstring{\MakeLowercase{d}}\texorpdfstring{\MakeUppercase{S}}/A\texorpdfstring{\MakeLowercase{d}} \texorpdfstring{\MakeUppercase{S}} and flat spacetime limits} \label{checkinglimitsmss}

In this section, we verify that $\delta^{2} \mathcal{L}$ in eq.(\ref{delta2lagrangian}) computed around our non-MSS, conformally-flat background given by eq.(\ref{mss20}) reduces~\footnote{The dS/AdS/flat space limit can also be verified from the most general, second order perturbations around \textit{arbitrary} backgrounds given in Appendix (\ref{appendixsecondordervariations}).} to the known expressions computed before in \cite{Biswas:2016egy}, \cite{Biswas:2016etb}. This section borrows many steps from there. Three important simplifications occur when computing $\delta^{2} \mathcal{L}$ around MSS. First, all perturbations that contain at least one $A_{\mu}$ or one $B$ mode go to zero since each term in these expressions is proportional to $\bar{S}_{\mu \nu}$ (see Appendix (\ref{appendixsecondordervariations}) or sec.(\ref{secondordervarsection})). This kills all terms that are purely quadratic in each $A_{\mu}$ or $B$, and also modes that are mixed with them. Second, all mode mixings between $\widehat{h}_{\mu \nu}$ and $\phi$ vanish upon using the Riemann tensor for MSS in eq.(\ref{riemanndecomp}), transversality constraints on SVT modes in eq.(\ref{hdcond}) and integration by parts. Third, the complicated piece $\delta^2 \mathcal{L}_{S^2}$ in eq.(\ref{quadratictracefreericciscalar}) has only one non-zero contribution coming from $(0101)$, with the rest vanishing because $\bar{S}_{\mu \nu} = 0$.

The non-zero SVT contributions coming from $\delta^2 \mathcal{L}_{EH + \Lambda}$ are (see eq.(\ref{deltazero})):
\be
\ba
\delta^2 \mathcal{L}_{EH + \Lambda} = \sqrt{- \bar{g}} \frac{M_{P}^{2}}{2} \delta_{0} &= \sqrt{- \bar{g}} \frac{M_{P}^{2}}{2} \left[ \delta_{0 (\widehat{h}   \widehat{h} )} +  \delta_{0 (\phi \phi)} \right]\\
&=  \sqrt{- \bar{g}} \frac{M_{P}^{2}}{2} \Bigg[ \frac{1}{4} \widehat{h}_{\mu \nu} \left( \bar{\square} - \frac{\bar{R}}{6}  \right) \widehat{h}^{\mu \nu} - \frac{1}{32} \phi \left( 3 \bar{\square} + \bar{R}  \right) \phi  \Bigg]\,.
\ea
\ee
The non-zero SVT contributions coming from $\delta^2 \mathcal{L}_{R^2}$ are (see eq.(\ref{riccisquare})):
\be
\ba
\delta^2 \mathcal{L}_{R^2} &=  \sqrt{- \bar{g}} f_{1,0} \bar{R} \Bigg[ \frac{1}{4} \widehat{h}_{\mu \nu} \left( \bar{\square} - \frac{\bar{R}}{6}  \right) \widehat{h}^{\mu \nu}  - \frac{1}{32} \phi \left( 3 \bar{\square} + \bar{R}  \right) \phi   \Bigg]   + \sqrt{- \bar{g}} \frac{1}{32}  \phi  \left( 3 \bar{\square} +  \bar{R} \right)^2 \Fc_1(\bar{\square}_{s})  \phi \,.
\ea
\ee
The non-zero SVT contributions coming from $\delta^2 \mathcal{L}_{S^2}$ are (see (eq.(\ref{listofsecondvariationstfricci})):
\be
\ba
\delta^2 \mathcal{L}_{S^2} &= (0101) = \frac{1}{2} \sqrt{-\bar{g}} \ s^{\nu}_{\ \mu}  \cF_2(\bar{\square}_{s}) s^{\mu}_{\ \nu}\\
&= \sqrt{- \bar{g}} \frac{1}{2} \left[ \left( \frac{1}{4}  \bar{D}_{\mu}^{\ \nu} \phi \right) \Fc_{2} (\bar{\square}_{s}) \left( \frac{1}{4} \bar{D}^{\mu}_{\ \nu} \phi \right) \right] +  \sqrt{- \bar{g}} \frac{1}{2} \left[ \frac{1}{2} \left(  \bar{\square} - \frac{\bar{R}}{6}  \right) \widehat{h}_{\mu}^{\ \nu}  \cF_2(\bar{\square}_{s})  \frac{1}{2} \left(  \bar{\square} - \frac{\bar{R}}{6}  \right) \widehat{h}^{\mu}_{\ \nu}  \right]   \\
& = \sqrt{- \bar{g}} \frac{1}{32} \left[ \phi \Fc_{2} \left( \bar{\square}_{s} + \frac{2 \bar{R}_{s}}{3} \right) \left(  \frac{3 \bar{\square} + \bar{R}}{4}  \right)\bar{\square} \phi \right] + \sqrt{- \bar{g}} \frac{1}{8} \left[ \left(  \bar{\square} - \frac{\bar{R}}{6}  \right) \widehat{h}_{\mu}^{\ \nu}  \cF_2(\bar{\square}_{s})   \left(  \bar{\square} - \frac{\bar{R}}{6}  \right) \widehat{h}^{\mu}_{\ \nu}  \right]\,,
\ea
\ee
where we have defined
\be
\ba
\bar{D}_{\mu}^{\ \nu} \equiv \bar{\nabla}_{\mu} \bar{\nabla}^{\nu} - \delta_{\mu}^{\ \nu} \frac{\bar{\square}}{4}\,, \qquad \qquad \bar{D}_{\mu}^{\ \mu} = 0\,,~~~~~~~~~~~~~~~
\bar{R}_{s} &\equiv \frac{\bar{R}}{M_{s}^{2}}\,,
\ea
\ee
and used the following recursion relations applicable \textit{only} for MSS\footnote{It is difficult and perhaps impossible to derive recursion relations for an arbitrary background. Nevertheless, many useful commutation relations and some recursion relations for an arbitrary background are presented in Appendix (\ref{comrelsec}).}:
\be \label{mssrecursive1}
\ba
\bar{\nabla}_{\mu} (\bar{\square}_{s})^{n} H^{\mu \nu} &= \left( \bar{\square}_{s} + \frac{5 \bar{R}_{s}}{12} \right)^{n} \bar{\nabla}_{\mu} H^{\mu \nu}\,, \qquad \qquad \text{where }~~~~~~~~ H^{\mu}_{\ \mu} = 0 \,,\\
\bar{\nabla}_{\mu} (\bar{\square}_{s})^{n} t^{\mu} &= \left( \bar{\square}_{s} + \frac{\bar{R}_{s}}{4} \right)^{n} \bar{\nabla}_{\mu} t^{\mu}\,.
\ea
\ee
Finally, the non-zero SVT contributions coming from $\delta^2 \mathcal{L}_{C^2}$ are (see (eq.(\ref{weylpert})):
\be
\ba
(0101) &= \frac{1}{2} \sqrt{-\bar{g}}  \  c^{\rho \sigma}_{\ \ \mu \nu}  \mathcal{\bar{F}}_3 (\bar{\square}_{s})  c^{\mu \nu}_{\ \ \rho \sigma} \\
&= \sqrt{-\bar g} \frac{1}{4} \left[ \widehat{h}_\mu^{\ \nu}
\Fc_3\left(\bar{\square}_s + \frac{\bar R_s}3\right)\left(\bar{\square} -\frac
{\bar R}6\right)\left(\bar{\square}-\frac {\bar R}3\right) \widehat{h}^{\mu}_{\ \nu}
\right]\,,
\ea
\ee
where we have used the following recursion relations applicable \textit{only} for MSS:
\be \label{mssrecursive2}
\ba
\bar{\nabla}_\nu \bar{\square}^n c^{\nu  \beta \mu \alpha } &= \left( \bar{\square} + \frac{\bar{R}}{4} \right)^n \bar{\nabla}_\nu c^{\nu  \beta \mu \alpha } \\
\bar{\nabla}_\alpha \bar{\square}^n V^{\beta \mu \alpha} &= \left( \bar{\square} + \frac{\bar{R}}{12} \right)^n \bar{\nabla}_\alpha V^{\beta \mu \alpha} \qquad \qquad \text{where } V^{\beta \mu \alpha}  \equiv  \bar{\nabla}_\nu c^{\nu  \beta \mu \alpha }\,.
\ea
\ee
Full $\delta^{2}S$ can be split into its tensor ($\delta^{2} S_{\widetilde{h}}$) and scalar ($\delta^{2} S_{\widetilde{\phi}}$) parts, and we get
\be \label{mssfinalsvt}
\ba
\delta^{2} S_{\widetilde{h}} &= \frac{1}{2} \int d^{4}x \sqrt{-\bar g} \ \widetilde{h}^{\mu\nu} \LF \bar{\square} -\frac {\bar R}6
\RF \Bigg\{1+{ \frac{2}{M_P^2}} {f_{1,0}}\bar R+
\frac{1}{M_P^{2}}
\left[\LF \bar{\square} -\frac {\bar R}6
\RF{\Fc}_2(\bar{\square}_{s} )+2\left(\bar{\square} -\frac
{\bar R}3\right)\Fc_3\left(\bar{\square}_{s} +\frac{\bar{R}_{s}}3\right)
\right] \Bigg\} \widetilde{h}_{\mu\nu}\,,  \\
\delta^{2} S_{\widetilde{\phi}} &=- \frac{1}{2} \int d^{4}x \sqrt{-\bar g} \ \widetilde{\phi}  \LF \bar{\square} +{\frac{\bar R}{3}}\RF
\Bigg\{1+{\frac{2}{M_P^2}  }{f_{1,0}}\bar R-{\frac{1} {M_P^2}} \left[2(3\bar{\square} +\bar R)\Fc_1(\bar{\square}_{s} )+
\frac1{2}\bar{\square} {\Fc}
_2\left(\bar{\square}_{s} +\frac{2 \bar {R}_{s}}{3} \right)\right]
\Bigg\} \widetilde{\phi}\,,
\ea
\ee
where we have introduced canonically normalized fields
\be
\widetilde{h}_{\mu \nu} =\frac{1}{2} M_P \widehat{h}_{\mu\nu},  \qquad \qquad
\widetilde{\phi} = \sqrt{\frac{3}{32}}M_P \phi, \qquad \qquad \qquad [\widetilde{h}_{\mu \nu}] = [\widetilde{\phi}] = 1.
\ee
This matches with the results obtained in \cite{Biswas:2016egy} and verifies the MSS limit. Effectively, eq.(\ref{mssfinalsvt}) gives spin-2 and spin-0 components of the graviton propagator around dS/AdS backgrounds.

Now, going to the flat space limit is easy by substituting $\bar{R} = 0$ above. We get the spin-2 and spin-0 components of the propagator: $\displaystyle \Pi_{2(0)} = \frac{i}{ p^2 \zeta_{2(0)}}$, where
\be
\ba
\zeta_2 = 1 - \frac{2 p^2}{M_P^2} \Big[ \Fc_2 (-p_s^2) + 2 \Fc_3 (-p_s^2)  \Big] \qquad \qquad
\zeta_0 = - \Bigg\{ 1 + \frac{2 p^2}{M_P^2} \Big[ 6\Fc_1 (-p_s^2) + \frac{1}{2} \Fc_2 (-p_s^2)  \Big] \Bigg\}
\ea
\ee

and we have defined $p_s^2 \equiv p^2/M_s^2$. We finally get
\be
\Pi_{\text{flat}} = \Pi_{2} + \Pi_{0} = \frac{i}{p^{2}} \left( \frac{1}{\zeta_{2}} - \frac{1}{\zeta_{0}} \right)
\ee

This matches with the result in \cite{Biswas:2016egy} and verifies the flat space limit.

Before moving on to the next section, recall that the background curvature dependence in expressions comprising $\delta^{2} S$ appears in terms of $\bar{R}$  and $\bar{S}_{\mu \nu}$. These curvatures are generated in two ways: first, when we compute the perturbation of a quantity (like Ricci scalar) to some order, and second, when covariant derivatives ($\bar{\nabla}_{ \mu}$) are commuted across the infinite tower of $\bar{\square}_{s}$ operators (in $\Fc_i (\bar{\square}_s)$) to the other side in order to contract with corresponding covariant derivatives ($\bar{\nabla}^{ \mu}$) to form scalar $\bar{\square}$ operators. Perturbations around our background satisfying eq.(\ref{mss20}) reduce to those around MSS when $\bar{S}_{\mu \nu}$ is set to zero. Naively, we might be able to separate the non-MSS and MSS parts of perturbations by separating expressions with and without $\bar{S}_{\mu \nu}$, respectively. This is not possible, however, because simple recursive relations when commuting $\bar{\nabla}^{ \mu}$, like in eqs.(\ref{mssrecursive1}) and (\ref{mssrecursive2}) for MSS, do not seem to exist for non-MSS backgrounds (except for the simplest cases when $\bar{\nabla}^{ \mu}$ operates on a low-rank tensor (see for e.g. Appendix (\ref{recursiongeneral1}) and (\ref{recursiongeneral2})).


\section{Illustrative examples} \label{physicalmodes}

To facilitate studying equations of motion of SVT modes in IDG around our background (eq.(\ref{mss20})), we consider $3$ examples, each of which has only one $\Fc_{i} (\square_{s})$ non-zero. $\Lc_{EH + \Lambda} $ is never zero since we want the higher derivative theory to reduce to GR in IR. In two cases, we will show that the different SVT degrees of freedom can be decoupled and solutions can be obtained in the UV regime. The third case shows non-trivial mode mixing.


\subsection{$ \cL = \cL_{EH + \Lambda} + \Lc_{C^2} \quad (\Fc_1 = \Fc_{2} = 0, \Fc_{3} \neq 0)$} \label{illustrative1}

In this case, the action is composed of Einstein-Hilbert plus cosmological constant and quadratic in Weyl tensor parts of the full action in eq.(\ref{action}). The background EoM in eq.(\ref{backgroundequations of motion}) in this case reduces to $M_P^{2} \bar{S}^{\mu}_{\ \nu}  = 0$ whose solution is $\bar{S}^{\mu}_{\ \nu} = 0$, which is just MSS, and has been studied before in \cite{Biswas:2016egy}. We briefly recall the results from there. The second order action can be read off from eq.(\ref{mssfinalsvt}):
\be
\delta^2 S = \int d^{4} x \sqrt{- \bar{g}} \ \widetilde{h}_{\mu \nu} \underbrace{\Bigg[ \frac{1}{2} \left( \bar{\square} - \frac{\bar{R}}{6} \right) \left[ 1 + \frac{2}{M_P^2} \left( \bar{\square} - \frac{\bar{R}}{3}  \right) \Fc_3 \left( \bar{\square}_s + \frac{\bar{R}_{s} }{3}  \right)  \right] \bar{g}^{\alpha \mu} \bar{g}^{\beta \nu} \Bigg]}_{\cK_{33}} \widetilde{h}_{\alpha \beta}
-   \widetilde{\phi} \underbrace{\Bigg[ \frac{3 \bar{\square} + \bar{R}}{6}   \Bigg]}_{\cK_{11}} \widetilde{\phi}\,,
\ee

with a kinetic matrix $\cK$  diagonal in $\widetilde{\phi}$ and $\widetilde{h}_{\mu \nu}$ modes (see eq.(\ref{msskineticmatrix})). We see in the quadratic form for $\widetilde{h}_{\mu \nu}$ above that in addition to the usual pole for GR at $\bar{R}/6$ expressed by $( \bar{\square} - \bar{R}/6 )$, there is a multiplicative factor which contains the higher derivative, non-local form factor $\cF_{3}(\square_{s})$. By demanding this multiplicative factor to be exponential of an entire function\footnote{According to the Weierstrass product theorem, any entire function $f(z)$ with no zeros can be written as $f(z) = e^{g(z)}$, where $g(z)$ is an entire function. We can use this to construct $\cF_{i} (\square_{s})$ without poles.}, it becomes possible to modify the graviton propagator such that there are no extra degrees of freedom, and also no ghosts. For example, for an entire function $\alpha (\bar{\square}_{s})$, and
\be \label{f3formfactorfirstcase}
\Fc_3 (\bar{\square}_s) = \frac{M_P^2}{2} \left[ \frac{e^{\alpha \left( \bar{\square}_{s} - \frac{2\bar{R}_{s}}{3} \right)} -1 }{\left( \bar{\square} - \frac{2 \bar{R}}{3} \right)} \right]\,,
\ee
which is holomorphic, we get the following tensor quadratic form:
\be
S_{\widetilde{h}} = \int d^{4} x  \frac{\sqrt{- \bar{g}}}{2} \ \widetilde{h}_{\mu \nu} \left(  \bar{\square} - \frac{\bar{R}}{6} \right)  e^{ \alpha \left( \bar{\square}_s - \frac{2\bar{R}_s}{3} \right) }  \widetilde{h}^{ \mu \nu} \,,
\ee
which has the usual graviton pole corresponding to dS/AdS radius, and, additionally, exponential suppression at large momenta $k \gtrsim M_s$.


\subsection{$ \cL = \cL_{EH + \Lambda} + \Lc_{S^2} \quad (\Fc_1 = \Fc_{3} = 0, \Fc_{2} \neq 0)$}

In this case, the Lagrangian consists of Einstein-Hilbert plus cosmological constant and quadratic in traceless Ricci tensor parts of the full Lagrangian in eq.(\ref{action}), and we have:
\be \label{tracelessriccilagrangian}
\delta^{2} \cL = \frac{M_{P}^{2}}{2} \delta_{0} + \delta^2 \mathcal{L}_{S^2}.
\ee
In this case, the background EoM from eq.(\ref{backgroundequations of motion}) is $M_{P}^{2} \bar{S}^{\mu}_{\ \nu}  =  f_{2,0} \left( \frac{1}{2} \delta^{\mu}_{\ \nu} \bar{S}^{\alpha \beta}\bar{S}_{\alpha \beta}   - \frac{1}{2} \bar{R} \bar{S}^{\mu}_{\ \nu} - 2 \bar{S}^{\mu \alpha} \bar{S}_{\nu \alpha}  \right)$ which is non-trivial to solve. Furthermore, from explicit perturbative expressions given in Appendices (\ref{appendixperturbations}) and (\ref{appendixsecondordervariations}) (see also sec.(\ref{ssquarepert})), we see that $\delta^{2} \cL$ is highly non-trivial and possesses all the $6$ possible mixings between scalar, vector and tensor modes. Therefore, in this case the generalized kinetic matrix $\Kc_{\mu \nu} \neq 0$ for all $\mu,\nu$. The full $\delta^{2} \cL$ in eq.(\ref{tracelessriccilagrangian}) for all SVT modes is quite long and not very enlightening, and so we will not present them here.

However, following the discussion in sec.(\ref{dominanttermsf2}), we can greatly simplify $\delta^2 \mathcal{L}_{S^2}$ and extract the two dominant contributions in the UV, where infinite covariant derivatives dominate, i.e. $(0101)$ and $(1001)$. In particular, in this limit, we neglect contributions coming from $M_{P}^{2} \delta_{0}$ and terms in $\delta^2 \mathcal{L}_{S^2}$ that are quadratic in background curvatures because, in this UV limit, the momenta scale as $k\gtrsim M_s$ with $\bar{R},\,\bar{S}_{\mu\nu}< \Oc(M_s^2)$. In Appendix (\ref{appendixtfriccisquare}), we have presented the $(0101)$ and $(1001)$ contributions in $\delta^2 \mathcal{L}_{S^2}$ for all SVT modes. For now, let us just consider $\phi$ and $\widehat{h}_{\mu\nu}$. There exist non-trivial SVT mode mixings in $\delta^{2} \cL$ in eq.(\ref{tracelessriccilagrangian}) like $ B (\cK_{01}) \phi, \ A^{\mu} (\cK_{21})_{\mu} \phi, \ \widehat{h}_{\mu \nu} (\cK_{31})^{\mu \nu} \phi, \ \widehat{h}_{\mu \nu} (\cK_{30})^{\mu \nu} B $  and  $ \widehat{h}_{\mu \nu} (\cK_{32})^{\mu \nu}_{\ \ \lambda} A^{\lambda} $, along with the usual purely quadratic in one SVT mode terms $\phi (\cK_{11}) \phi$ and $\widehat{h}_{\mu \nu}  (\cK_{33})^{\mu \nu \rho \sigma}  \widehat{h}_{\rho \sigma}$. The respective kinetic matrix elements are:
\begin{equation} \label{tfriccimixingsapprox}
\begin{aligned}
\Kc_{01} &= \Bigg[
\frac{1}{8} \bar{S}^{\nu}_{\ \mu}  \cF_2 (\bar{\square}_{s} ) \bar{\square} \bar{\nabla}_{\nu } \bar{\nabla}^{\mu } 
\Bigg], \qquad \qquad
\Kc_{21}  = \Bigg[  
\frac{1}{2}   \bar{\nabla}^{\alpha } \cF_2(\bar{\square}_{s}) \bar{S}_{\nu \alpha } \bar{\nabla}^{\nu }\bar{\nabla}_{\mu } 
-  \bar{\nabla}_{\nu } \cF_2(\bar{\square}_{s})  \bar{S}^{\alpha }{}_{\mu } \bar{\nabla}^{\nu }\bar{\nabla}_{\alpha }
\Bigg], \\ 
\Kc_{31} &= \Bigg[ 
\frac{1}{24}  \bar{R}  \cF_2(\bar{\square}_{s})  \bar{\nabla}^{\nu }\bar{\nabla}^{\mu }
+ \frac{1}{2}  \bar{S}^{ \sigma \mu }  \cF_2(\bar{\square}_{s})  \bar{\nabla}^{\nu }\bar{\nabla}_{\sigma }
-  \frac{1}{4} \bar{S}^{\mu \nu } \bar{\square}  \cF_2(\bar{\square}_{s})    
-  \frac{1}{4} \bar{\square}  \cF_2(\bar{\square}_{s})  \bar{\nabla}^{\nu} \bar{\nabla}^{\mu}
- \frac{1}{8}  \bar{S}^{\mu \nu}  \cF_2(\bar{\square}_{s}) \bar{\square}\\
&\qquad \quad + \frac{1}{4} \bar{S}^{\mu \nu}  \cF_2 (\bar{\square}_{s} ) \bar{\square}
\Bigg], \qquad \qquad
\Kc_{30}  = \Bigg[ 
-  \frac{1}{4} \bar{S}^{\mu  \nu}   \cF_2 (\bar{\square}_{s} ) \bar{\square}^{2}
\Bigg],\\
\Kc_{32} &=  \Bigg[ 
\cF_2(\bar{\square}_{s})   \bar{\square} \bar{S}^{\nu}_{\ \rho } \bar{\nabla}^{\rho } \delta^{\mu}_{\ \lambda}
- \cF_2(\bar{\square}_{s})   \bar{\square} \bar{S}^{\mu }{}_{\rho } \bar{\nabla}^{\nu } \delta^{\rho}_{\ \lambda}
\Bigg], \qquad \qquad
\Kc_{11} = \Bigg[
\frac{1}{16}  \bar{\nabla}_{\mu } \bar{\nabla}^{\nu } \cF_2(\bar{\square}_{s}) \bar{\nabla}_{\nu }\bar{\nabla}^{\mu } 
- \frac{1}{64}  \cF_2(\bar{\square}_{s}) \bar{\square}^{2}
\Bigg], \\
\cK_{33} &=
\Bigg[ 
-  \frac{1}{12}  
\cF_2(\bar{\square}_{s}) \bar{R} \bar{\square} \bar{g}^{\rho \mu} \bar{g}^{\sigma \nu} 
-   \cF_2(\bar{\square}_{s})  \bar{S}^{ \sigma \nu }  \bar{\square} \bar{g}^{\mu \rho} 
+ \frac{1}{4} 
\cF_2(\bar{\square}_{s}) \bar{\square}^{2}  \bar{g}^{\rho \mu} \bar{g}^{\sigma \nu}
\Bigg]
\end{aligned}
\end{equation}

Unlike the previous case in sec.(\ref{illustrative1}), here we see that decoupling the different SVT degrees of freedom at the  quadratic level of the action is not an easy task, and perhaps impossible. This indicates having quadratic in (traceless) Ricci tensor terms in the action will, in general, give rise to all SVT modes being coupled to each other when the background is not MSS\footnote{Notice that when taking the MSS limit ($\bar{S}_{\mu \nu} = 0$), all mode mixing operators in eq.(\ref{tfriccimixingsapprox})  manifestly vanish, except $ \frac{1}{24}  \bar{R}  \cF_2(\bar{\square}_{s})  \bar{\nabla}^{\nu }\bar{\nabla}^{\mu } -  \frac{1}{4} \bar{\square}  \cF_2(\bar{\square}_{s})  \bar{\nabla}^{\nu }\bar{\nabla}^{\mu }$ in $(\cK_{31})^{\mu \nu}$. This goes to zero too, after integrating by parts and using the transversality condition $\bar{\nabla}^\mu \widehat{h}_{\mu \nu} = 0$.}. We defer the analysis of eq.(\ref{tracelessriccilagrangian}) to future.


\subsection{$ \cL = \cL_{EH + \Lambda} + \Lc_{R^2} \quad (\Fc_2 = \Fc_{3} = 0, \Fc_{1} \neq 0)$}

In this case, the Lagrangian consists of Einstein-Hilbert plus cosmological constant and quadratic in Ricci scalar terms of the full Lagrangian in eq.(\ref{action}):
\be \label{psiaction0}
\cL=  \frac{1}{2} \left[ M_P^2 \left( R -  2 \Lambda \right) + R \Fc_1\LF \square_{s} \RF R \right].
\ee
The second order Lagrangian in this case is (see eqs.(\ref{deltazero}) and (\ref{riccisquare})):
 \be \label{delta2l}
 \ba
 \delta^2 \mathcal{L} &= \frac{1}{2} \left[ \Omega \delta_0 + r \Fc_1 (\bar{\square}_s) r \right],
 \ea
 \ee
where $\Omega=M_p^2+2f_{1,0}\bar{R}$ 
and the background EoM from eq.(\ref{backgroundequations of motion}) is:
 \begin{equation} \label{bgkeomf1nonzero}
 \Omega \bar{S}^{\mu}_{\ \nu}  = 0.
 \end{equation}
 If we trace back to how we obtained the background EoM in eq.(\ref{bgkeomf1nonzero}) from Appendix (\ref{equations of motion}), we notice that only the condition $\partial_{\mu} \bar{R} = 0 $ was used, not $\bar{\nabla}_{\alpha} \bar{S}_{\mu \nu} = 0$. Now, the background EoM in eq.(\ref{bgkeomf1nonzero}) has $3$ possible solutions. Let us consider them one by one:
\begin{enumerate}
\item \underline{$\Omega \neq 0, \bar{S}_{\mu\nu} = 0$}: Since $\bar{S}_{\mu\nu} = 0$, this leads us trivially to the  MSS case presented in sec.(\ref{checkinglimitsmss}) (also see \cite{Biswas:2016egy}) and we get the scalar and tensor parts of the second order action as:
\be \label{omeganonzeromss}
\ba
\delta^{2} S_{\widetilde{h}} &= \frac{\Omega}{2 M_{P}^{2}} \int d^{4}x \sqrt{-\bar g} \  \widetilde{h}^{\mu\nu} \LF \bar{\square} -\frac {\bar R}6 \RF  \widetilde{h}_{\mu\nu}\,,  \\
\delta^{2} S_{\widetilde{\phi}} &=- \frac{1}{2 M_{P}^{2}} \int d^{4}x \sqrt{-\bar g} \ \widetilde{\phi}  \LF \bar{\square} +{\frac{\bar R}{3}}\RF
\left[ \Omega - 6 \left( \bar{\square} +  \frac{\bar R}{3} \right) \Fc_1(\bar{\square}_{s} )
\right] \widetilde{\phi}.
\ea
\ee
In order to have no extra degrees of freedom in the $\widetilde{\phi}$ sector, in addition to a possible Brans-Dicke scalar of mass $m$, we can choose
\be \label{mssforf1part}
\left[ \Omega - 6 \left( \bar{\square} +  \frac{\bar R}{3} \right) \Fc_1(\bar{\square}_{s} ) \right] =  \left( 1 - \frac{\bar{\square}}{m^{2}} \right)^{\epsilon} e^{\alpha(\bar{\square}_{s})},
\ee
where $\alpha(\square_{s})$ is an entire function (like in eq.(\ref{f3formfactorfirstcase})) and $\epsilon  = 0 $ or $1$. $\epsilon  = 0$ corresponds to no extra scalar mode, while $\epsilon  = 1$  corresponds to one extra Brans-Dicke scalar. Imposing $m^{2} > 0$ ensures that the scalar is not tachyonic.

\item \underline{$\Omega = 0, \bar{S}_{\mu\nu} \neq 0$}: Since $\bar{S}_{\mu\nu} \neq 0$, we have a \textit{non}-MSS background, as required by our background conditions in eq.(\ref{mss20}). $\Omega = 0$ then determines the background Ricci scalar to be:
 \begin{equation}\label{backgrf1}
 \bar{R} = - \frac{M_{P}^{2}}{2 f_{1,0} }\,. 
 \end{equation}

The first term in eq.(\ref{delta2l}) vanishes because $\Omega = 0$. From eq.(\ref{riccisquarefinal}) and eq.(\ref{riccisquaremixterm}), we see that $\delta^2 \mathcal{L}$ contains only $\phi$ and $\widehat{h}_{\mu\nu}$ modes:
\be \label{lagphi}
\ba
\delta^2 \mathcal{L} &= 
\frac{1}{2} \phi \left[  \left( \bar{R} + 3 \bar{\square}  \right)^2 \frac{{\Fc}_1(\bar{\square}_{s})}{16} \right]  \phi +
\frac{1}{2} \widehat{h}^{\alpha \beta } \bar{S}_{\alpha \beta }  \Fc_1(\bar{\square}_{s})  \bar{S}_{\mu \nu }\widehat{h}^{\mu \nu } - \frac{1}{2} \widehat{h}^{\mu \nu } \left[  (\bar{R} + 3 \bar{\square} )\bar{S}_{\mu \nu} \frac{{\Fc}_1 (\bar{\square}_s)}{2}  \right]   \phi \,.
\ea
\ee
where, now, we have used the condition $\bar{\nabla}_{\alpha} \bar{S}_{\mu \nu} = 0$ (see eq.(\ref{mss21})). We can diagonalize $\delta^2 \mathcal{L}$ using the field redefinition
\be \label{psidefinition}
\psi \equiv \LF \bar{R} + 3\bar{\square} \RF\frac{\phi}{4}-\bar{S}_{\mu\nu}\widehat{h}^{\mu\nu},
\ee
so that the quadratic action becomes
\begin{equation} \label{psiaction}
\delta^2 S = \int d^{4} x \sqrt{- \bar{g}} \ \frac{1}{2} \psi \Fc_{1}\LF \bar{\square}_s \RF \psi
\end{equation}
and varying $\delta^2 S$ with respect to $\psi$ gives us the EoM for $\psi$:
\be \label{eomforpsi}
\Fc_{1}\LF \bar{\square}_s \RF \psi = 0.
\ee
This new scalar $\psi$ is gauge-invariant as it is composed of gauge-invariant fields $\widehat{h}_{\mu \nu}$ and $\phi$ (see eq.(\ref{gaugetransformationsofmodes})). Now, the propagating degrees of freedom in the theory described by the action in eq.(\ref{psiaction}) depends on the form of $\cF_{1} (\bar{\square}_{s})$ we choose in our Lagrangian in eq.(\ref{psiaction0}). For example, if we choose $\cF_{1} (\bar{\square}_{s}) =1$, then $\psi$ has no kinetic term. Also, from the EoM in eq.(\ref{eomforpsi}), we get $\psi=0$, which from eq.(\ref{psidefinition}) gives the constraint equation
\be
\LF \bar{R} + 3\bar{\square} \RF\frac{\phi}{4}  =  \bar{S}_{\mu\nu}\widehat{h}^{\mu\nu}.
\ee
If we choose $\cF_{1} (\bar{\square}_{s}) =\bar{\square}_{s} = \bar{\square} / M_{s}^{2}$, the quadratic action  becomes
\be
\delta^2 S = \int d^{4} x \sqrt{- \bar{g}} \ \frac{1}{2} \widetilde{\psi}  \bar{\square} \widetilde{\psi} 
\ee
where $\widetilde{\psi} \equiv \psi / M_{s}$ now has canonical dimensions $[\widetilde{\psi}]=1$. This theory describes a propagating massless scalar in a \textit{non}-MSS ($\bar{S}_{\mu \nu} \neq0$) background. It is also ghost-free because the sign of the kinetic term is correct. If we choose $\cF_{1} (\bar{\square}_{s})=  \left( 1 - \frac{\bar{\square}}{m^{2}} \right)^{\epsilon} e^{\alpha(\bar{\square}_{s})}$, for some entire function $\alpha(\bar{\square}_{s})$, we have 
\be
\delta^2 S = \int d^{4} x \sqrt{- \bar{g}} \ \frac{1}{2} \psi  \left( 1 - \frac{\bar{\square}}{m^{2}} \right)^{\epsilon} e^{\alpha(\bar{\square}_{s})} \psi,
\ee
where $\epsilon=0$ or $1$. The case $\epsilon=0$ furnishes no poles for $\psi$, while $\epsilon=1$ and furnishes a massive scalar of mass $m$ without any additional poles.

\item \underline{$\Omega = 0, \bar{S}_{\mu\nu} = 0$}: Since $\bar{S}_{\mu\nu} = 0$, we again have an MSS background, but this dS/AdS vacuum is different from that in eq.(\ref{omeganonzeromss}) where $\Omega$ was non-zero. Here, $\Omega = 0$, and the coefficient $f_{1,0}$ in $\cF_{1} (\square_{s})$ is related to the cosmological constant via
\be
\bar{R} = - \frac{M_{P}^{2}}{2 f_{1,0} } = 4 \Lambda.
\ee
The second order action in this case can be obtained by putting $\Omega = 0$ in eq.(\ref{omeganonzeromss}), where its tensor part $\delta^{2} S_{\widetilde{h}}$ vanishes, leaving us with just the scalar part:
\be \label{omegazeromss}
\ba
\delta^{2} S_{\widetilde{\phi}} = \frac{3}{ M_{P}^{2}} \int d^{4}x \sqrt{-\bar g} \  \widetilde{\phi}  \left( \bar{\square} +{\frac{\bar R}{3}} \right)^{2}
 \Fc_1(\bar{\square}_{s} )
 \widetilde{\phi}
\ea
\ee
which has the EoM for $\widetilde{\phi}$ given by
\be \label{eomlogmode}
\left( \bar{\square} +{\frac{\bar R}{3}} \right)^{2} \Fc_1(\bar{\square}_{s} )\widetilde{\phi} = 0.
\ee
As before, we can choose $\Fc_1(\bar{\square}_{s} ) = e^{\alpha(\bar{\square}_{s})}$ for some entire function $\alpha(\bar{\square}_{s})$ to avoid any additional poles for $\widetilde{\phi}$. In a local theory, with say $\Fc_1(\bar{\square}_{s} ) = 1$, the EoM in eq.(\ref{eomlogmode}) reduces to that of a 4th order wave operator acting on $\widetilde{\phi}$:
\be \label{eomlogmodelocal}
\left( \bar{\square} +{\frac{\bar R}{3}} \right)^{2} \widetilde{\phi} = 0.
\ee
In addition to modes $\widetilde{\phi}_{0}$ which satisfy both $\left( \bar{\square} +{\frac{\bar R}{3}} \right) \widetilde{\phi} = 0$ and eq.(\ref{eomlogmodelocal}), there also exist log modes $\widetilde{\phi}_{\text{log}} $ which satisfy eq.(\ref{eomlogmodelocal}) but not $\left( \bar{\square} +{\frac{\bar R}{3}} \right) \widetilde{\phi} = 0$ \cite{Bergshoeff:2011ri}. We see that this dS/AdS vacuum has a different spectrum compared to the ($\Omega \neq 0, \bar{S}_{\mu\nu} = 0$) case discussed before. In that case, $\delta^{2} S$ had both tensor $\widetilde{h}$ and scalar $\widetilde{\phi}$ modes (see eq.(\ref{omeganonzeromss})). In this case, we have only the scalar $\widetilde{\phi}$ mode, that too which satisfies a different EoM given in eq.(\ref{eomlogmode}).

\end{enumerate}




\section{Conclusions} \label{conclusions}

In this paper, we studied perturbations at the quadratic level of the action for Infinite Derivative Gravity, involving quadratic in curvature terms, proposed in \cite{Biswas:2011ar}. We computed (covariant) scalar-vector-tensor perturbations ($\phi, B, A_{\mu}, \widehat{h}_{\mu \nu}$) around conformally-flat, covariantly constant curvature backgrounds which are more general than, and go beyond maximally symmetric spacetimes. We have illustrated that conformally-flat, covariantly constant curvature backgrounds generically arise in IDG theories when resolving big bang and black hole singularities. Testing the stability of these backgrounds requires studying perturbations, which was the focus of this paper.  We found that around our chosen background, the different SVT modes do not decouple at the quadratic level of the action. This means that, for instance, the scalar mode $\phi$ is sourced by the remaining vector $A_{\mu}$, tensor $\widehat{h}_{\mu \nu}$ and scalar $B$ modes. To show the consistency of computations around our \textit{non}-maximally symmetric background, we derived dS/AdS and flat space limits, previously computed in \cite{Biswas:2016egy,Biswas:2011ar}.

Studying quadratic order SVT perturbations at the level of the action, around our conformally-flat, covariantly constant curvature background in various limits of IDG, we learn the following:
\begin{itemize}
	\item In the case where $\Fc_1=\Fc_2=0$ and $\Fc_3\neq 0$ i.e., when the action contains only the non-local, quadratic Weyl term in addition to the usual Einstein-Hilbert plus cosmological constant term, our background equations of motion imply that the only possible background solutions are maximally symmetric dS/AdS ($\bar{S}_{\mu\nu}=0$). So in this case, there is only one propagating degree of freedom, $\widehat{h}_{\mu\nu}$, and we can construct a ghost-free propagator for it, as was shown previously in \cite{Biswas:2016egy}.
		\item The case with $\Fc_1= 0,\,\Fc_2\neq 0$ and $\Fc_3= 0$ i.e., when the action contains non-local,  quadratic in traceless Ricci tensor term in addition to the usual GR term, is quite involved. The second order action has a complicated form which was expected because of our non-trivial background that goes beyond dS/AdS. We present the framework for perturbations in this case, but for economical reasons, present the second order action with only the terms that are highly dominant in the UV. Here, we see that all the SVT modes have non-local (containing $\cF_{2}(\square_{s})$) kinetic terms and also non-trivial mixings with all the other SVT modes.
		\item In the case with $\Fc_1\neq 0,\,\Fc_2=0$ and $\Fc_3= 0$ i.e., when the action contains non-local,  quadratic in Ricci scalar term along with the usual GR term, we found that the scalar $B$ and vector $A_\mu$  modes do not appear at all. The remaining $\phi$ and $\widehat{h}_{\mu\nu}$ modes are mixed but the second order action can be diagonalized. We have analysed all the vacuum solutions of this case and studied the spectrum of propagating modes. 
\end{itemize}

In addition to the fact that SVT modes are generically coupled, we notice that their kinetic terms are not only functions of $\square$, but also the background traceless Ricci tensor and Ricci scalar. This is similar to the case when background was dS/AdS (see \cite{Biswas:2016egy}) where we had form factors (obtained from ghost-free conditions) depending not only on covariant derivatives, but also the background Ricci scalar (for e.g., see eq.(\ref{mssfinalsvt})). In our case, form factors do depend on the background traceless Ricci tensor since our background is not MSS. In order to clearly see this, however, one must use the commutation relations  presented in Appendix (\ref{comrelsec}) to commute $\bar{\nabla}_{\mu}$ through $\cF_{i}(\square_{s})$ and contract with $\bar{\nabla}^{\mu}$ on the other side of the form factors. However, these commutation relations are not recursive in most cases and it is not possible to write a closed form expression for form factors where explicit background curvature dependence is manifest. Our study indicates that in order to formulate a background independent theory of IDG, we must start with form factors depending on curvatures i.e., $\Fc_i\LF \square_{s}, \frac{\bar{R}}{M_s^2}, \frac{\bar{S}_{\mu\nu}}{M_s^2} \RF$ in such a way that we do not have to fix the form factors depending on any particular background. One might also consider studying perturbations using a decomposition different from the one in eq.(\ref{hdecomp2}). We leave these interesting endeavors for future investigations.

The perturbations we have computed in this paper around backgrounds beyond MSS can have wide applications for probing UV physics for any higher derivative theory of gravity. We illustrated, with explicit computation, that SVT modes in these backgrounds couple. This indicates gravitational waves around these backgrounds can have non-trivial (\textit{extra}) polarizations, which in future we might be able to test in the light of LIGO/VIRGO detections \cite{Abbott:2018utx}.

\acknowledgements  

We thank Alexey S. Koshelev and Luca Buoninfante for useful discussions. KSK and AM are supported by  Netherlands Organization for Scientific Research (NWO) grant number 680-91-119. We acknowledge the use of \textit{xAct} package for Mathematica for carrying out various computations (\url{http://xact.es/}, Jos\'e M. Mart\'in-Garc\'ia, 2002-2018) \cite{Brizuela:2008ra,Nutma:2013zea}. SM particularly thanks Zhan-Feng Mai, Leo C. Stein, Alfonso Garc\'ia-Parrado, Jos\'e M. Mart\'in-Garc\'ia, Teake Nutma and Thomas B\"ackdahl for their useful advice on working with \textit{xAct}.

\appendix

\section{Full, non-linear equations of motion} \label{fullequations of motion}

Varying the action in eq.(\ref{action}) with respect to the metric gives the eom derived in \cite{Biswas:2013cha,Conroy:2014eja}:
\begin{equation}
\ba \label{equations of motion}
M_P^2 G^\mu_{\ \nu} &= T^\mu_{\ \nu} - M_P^2 \Lambda \delta^\mu_{\ \nu} - 2 S^\mu_{\ \nu} \Fc_1\LF \square_{s} \RF R + 2 (\nabla^\mu \nabla_\nu - \delta^\mu_{\ \nu} \square) \Fc_1\LF  \square_{s} \RF R - \frac{1}{2} R \Fc_2\LF  \square_{s} \RF S^\mu_{\ \nu}
- 2  S^\mu_{\ \beta} \Fc_2\LF  \square_{s} \RF S^\beta_{\ \nu} \\
& \qquad  + \frac{1}{2} \delta^\mu_{\ \nu} S^\alpha_{\ \beta} \Fc_2\LF  \square_{s} \RF S^\beta_{\ \alpha}
 + 2  \left( \nabla_\rho \nabla_\nu \Fc_2\LF  \square_{s} \RF S^{\mu \rho} 
 - \frac{1}{2} \square \Fc_2\LF  \square_{s} \RF S^\mu_{\ \nu} 
 - \frac{1}{2} \delta^\mu_{\ \nu} \nabla_\sigma \nabla_\rho \Fc_2\LF  \square_{s} \RF S^{\sigma \rho} \right)\\
& \qquad + L_{1 \nu}^{\ \mu} - \frac{1}{2} \delta^\mu_{\ \nu} \left( L_{1 \sigma}^{\ \sigma} + \bar{L}_1 \right) + L_{2 \nu}^{\ \mu} - \frac{1}{2} \delta^\mu_{\ \nu} \left( L_{2 \sigma}^{\ \sigma} + \bar{L}_2 \right)
+ 2 \Delta^\mu_{\ \nu}
 + 2 \left( S_{\alpha \beta}
  + 2 \nabla_\alpha \nabla_\beta \right) \Fc_3\LF  \square_{s} \RF C_\nu^{\ \alpha \beta \mu}
   + \Omega_{3\nu}^{\mu} \\
& \qquad    - \frac{1}{2}\delta_{\nu}^{\mu}(\Omega_{3\sigma}^{\;\sigma} 
   + \bar{\Omega}_{3})+4\Delta_{3\nu}^{\mu} \,,
\ea
\end{equation}
where $G_{\mu\nu} = R_{\mu\nu}-\frac{1}{2}g_{\mu\nu}R$ is the Einstein tensor and
\begin{align}
L_{1\nu}^{\mu}= \frac{1}{M_{s}^2} &\sum_{n=1}^{\infty}f_{1n}\sum_{l=0}^{n-1}\partial^{\mu} R^{(l)}\partial_{\nu}R^{(n-l-1)}\,,~~~\quad
\bar{L}_1=  \sum_{n=1}^{\infty}f_{1n}\sum_{l=0}^{n-1}R^{(l)}R^{(n-l)}\,,\\
{L_2}^\mu_\nu=\frac{1}{M_{s}^2} &\sum_{n=1}
^\infty
{f_2}_n\sum_{l=0}^{n-1}\nabla^\mu {{S}^{(l)} }{}^\alpha_\beta 
\nabla_\nu  {{S}^{(n-l-1)}}{}^\beta_\alpha
,~~~\bar{L}_2=\sum_{n=1}
^\infty
{f_2}_n\sum_{l=0}^{n-1}{S^{(l)}}{}^\alpha_\beta
{S^{(n-l)}}{}^\beta_\alpha\\
\Delta^\mu_{\nu}=
\frac{1}{M_{s}^2} & \sum_{n=1}
^\infty
{f_2}_n\sum_{l=0}^{n-1}\nabla_\beta[ {S^{(l)}}{}^\beta_\gamma
\nabla^\mu{S^{(n-l-1)}}{}^\gamma_\nu-\nabla^\mu
{S^{(l)}}{}^\beta_\gamma {S^{(n-l-1)}}{}^\gamma_\nu]\\
\Omega_{3\nu}^{\mu}=\frac{1}{M_{s}^{2}} & \sum_{n=1}^{\infty}f_{3{n}}\sum_{l=0}^{n-1}C_{\:\:\beta\lambda\sigma;\nu}^{\alpha(l)}C_{\alpha}^{\;\beta\lambda\sigma;\mu(n-l-1)},~~~\quad\bar{\Omega}_{3}=\sum_{n=1}^{\infty}f_{3_{n}}\sum_{l=0}^{n-1}C_{\:\:\beta\lambda\sigma}^{\alpha(l)}C_{\alpha}^{\;\beta\lambda\sigma(n-l)}\,,\label{details-1}\\
\Delta_{3\nu}^{\mu}=\frac{1}{M_{s}^{2}} & \sum_{n=1}^{\infty}f_{3{n}}\sum_{l=0}^{n-1}[C_{\quad\sigma\alpha}^{\lambda\beta(l)}C_{\lambda;\nu}^{\;\mu\sigma\alpha(n-l-1)}-C_{\quad\sigma\alpha;\nu}^{\lambda\beta\;\;(l)}C_{\lambda}^{\:\mu\sigma\alpha(n-l-1)}]_{;\beta}\,.\label{details-2}
\end{align}
where 
\begin{equation}
X^{(m)}= \square_{s}^m X
\end{equation}
Taking the trace of eq.(\ref{equations of motion}), we get
\be \label{traceequations of motion}
-M_P^2R=-4 M_P^2 \Lambda-6 \square \Fc_1(\square_{s}) R-({L_1} +2\bar{L}_{1})- 2\nabla_\mu\nabla_\nu\Fc_2(\square_{s}){S}^{\mu\nu}-({L_2} +2\bar{L}_{2})+2\Delta\,.
\ee


\section{Perturbations around an arbitrary background}\label{appendixperturbations}

The metric is perturbed as
\be
g_{\mu \nu} = \bar{g}_{\mu \nu} + h_{\mu \nu}\,.
\ee
In general, an expression $Q$ (arbitrary rank tensor, indices suppressed) which depends on $g_{\mu \nu}$, has the following expansion:
\be \label{generalpertexp}
Q = \bar{Q} + Q^{(1)} + Q^{(2)} + \cdots + Q^{(n)} + \cdots\,,
\ee
where the superscript in parentheses $(n)$ denotes that the perturbation is $\cO (h^n)$. $\bar{Q} = Q^{(0)}$ is the background value of $Q$. We use lowercase letters to indicate $\mathcal{O} (h)$ perturbation of the relevant quantity, while higher order perturbations are indicated explicitly by their order $n$. So, $q = Q^{(1)} = \delta{Q}$, $Q^{(2)} = \delta^2 Q$, etc.. Specifically, $Q$ could be the inverse metric, Ricci scalar ($R$), Ricci tensor ($R_{\mu \nu}$), traceless Ricci tensor ($S_{\mu \nu}$), Weyl tensor ($C_{\mu \nu \rho \sigma}$) or the Levi-Civita connection.  In order to set notation, we expand the curvatures up to $\mathcal{O} (h^{2})$:
\be \label{examplepertcurvatures}
\ba
R &= \bar{R} + r + \delta^{2} R\,,\\ 
R_{\mu \nu} &= \bar{R}_{\mu \nu} + r_{\mu \nu} + \delta^{2} R_{\mu \nu}\,,\\
S_{\mu \nu} &= \bar{S}_{\mu \nu} + s_{\mu \nu} + \delta^{2} \bar{S}_{\mu \nu}\,,\\
C_{\mu \nu \rho \sigma} &= \bar{C}_{\mu \nu \rho \sigma} + c_{\mu \nu \rho \sigma} + \delta^{2} C_{\mu \nu \rho \sigma}\,,
\ea
\ee
The perturbations of all other quantities involve perturbations to all orders in $h$. We will henceforth expand all other quantities up to $O(h^2)$ at most. As an example, suppose we want to find the $\cO(h^{2})$ contribution in $(QP)$ where $Q$ and $P$ are some quantities that depend on $g_{\mu \nu}$. To find $(QP)^{(2)} $, we expand each term and collect all $\cO(h^{2})$ contributions:
\be
\ba
(QP)^{(2)} &= \left[ Q^{(0)} + Q^{(1)} + Q^{(2)} + \cdots \right] \left[ P^{(0)} + P^{(1)} + P^{(2)} + \cdots \right]  \\
&=  \underbrace{Q^{(0)} P^{(2)} + Q^{(2)} P^{(0)} + Q^{(1)} P^{(1)}}_{\text{collect}} + \underbrace{\cdots \text{other terms} \cdots}_{\text{discard}}.
\ea
\ee
The inverse metric is
\be
g^{\mu \nu } = \bar{g}^{\mu \nu} - h^{\mu \nu } + h^{\mu \alpha } h^{\nu }{}_{\alpha }\,,
\ee
while square root of the metric determinant is
\be \label{sqrtdetg}
\sqrt{-g} = \sqrt{-\bar{g}} \left( 1 + \frac{h}{2} + \frac{h^2}{8} - \frac{1}{4} h_{\mu \nu} h^{\mu \nu} \right)\,.
\ee
Note that $\delta g_{\mu \nu} = h_{\mu \nu}$ and $\delta g^{\mu \nu} = -h^{\mu \nu}$. All traces are taken with respect to the background metric, like $h = \bar{g}^{\mu \nu} h_{\mu \nu}$. Covariant derivatives ($\bar{\nabla}_{\mu}$) are defined with respect to the background $\bar{g}_{\mu \nu}$. The Levi-Civita connection is
\be
\Gamma_{\mu \nu}^{\lambda} = \bar{\Gamma}_{\mu \nu}^{\lambda} + \frac{1}{2} \left( \bar{\nabla}_\mu h^\lambda_{\ \nu} + \bar{\nabla}_\nu h^\lambda_{\ \mu} - \bar{\nabla}^\lambda h_{\mu \nu} \right)\,.
\ee
Expansions for curvatures  are then obtained by substituting the above expansions in eq.(\ref{examplepertcurvatures}). We use the notation in eq.(\ref{generalpertexp}) for the remaining part of Appendix (\ref{appendixperturbations}). The following expansions are valid for an \textit{arbitrary} background. We get
\be \label{ricciscalarpert}
\ba
R^{(1)}={}& - h^{\alpha \beta } \bar{R}_{\alpha \beta } + \bar{\nabla}_{\beta }\bar{\nabla}_{\alpha }h^{\alpha \beta } -  \bar{\square}  h\,,\\
R^{(2)} ={}& - h^{\alpha \beta } h^{\gamma \lambda } \bar{R}_{\alpha \gamma \beta \lambda } + 2 h^{\alpha \beta } h_{\lambda }{}^{\gamma } \bar{R}_{\beta \gamma \alpha }{}^{\lambda } + h^{\alpha \beta } \bar{\nabla}_{\beta }\bar{\nabla}_{\alpha } h - 2 h^{\alpha \beta } \bar{\nabla}_{\beta }\bar{\nabla}_{\gamma }h_{\alpha }{}^{\gamma } -  \frac{1}{4} \bar{\nabla}_{\beta } h \bar{\nabla}^{\beta } h\,, \\
&-  \bar{\nabla}_{\alpha }h^{\alpha \beta } \bar{\nabla}_{\gamma }h_{\beta }{}^{\gamma } + \bar{\nabla}^{\beta } h \bar{\nabla}_{\gamma }h_{\beta }{}^{\gamma } + h^{\alpha \beta }  \bar{\square} h_{\alpha \beta } -  \frac{1}{2} \bar{\nabla}_{\beta }h_{\alpha \gamma } \bar{\nabla}^{\gamma }h^{\alpha \beta } + \frac{3}{4} \bar{\nabla}_{\gamma }h_{\alpha \beta } \bar{\nabla}^{\gamma }h^{\alpha \beta }\,,
\ea
\ee
\be
\ba
R_{\mu \nu}^{(1)} ={}& \frac{1}{2} h_{\nu }{}^{\alpha } \bar{R}_{\mu \alpha } + \frac{1}{2} h_{\mu }{}^{\alpha } \bar{R}_{\nu \alpha } -  h^{\alpha \beta } \bar{R}_{\mu \alpha \nu \beta } -  \frac{1}{2}  \bar{\square} h_{\mu \nu } + \frac{1}{2} \bar{\nabla}_{\mu }\bar{\nabla}_{\alpha }h_{\nu }{}^{\alpha } + \frac{1}{2} \bar{\nabla}_{\nu }\bar{\nabla}_{\alpha }h_{\mu }{}^{\alpha } -  \frac{1}{2} \bar{\nabla}_{\nu }\bar{\nabla}_{\mu } h\,, \\
R_{\mu \nu}^{(2)} ={}& - \frac{1}{2} h^{\beta \gamma } h_{\nu }{}^{\alpha } \bar{R}_{\mu \beta \alpha \gamma } + h_{\alpha }{}^{\gamma } h^{\alpha \beta } \bar{R}_{\mu \beta \nu \gamma } -  \frac{1}{2} h^{\beta \gamma } h_{\mu }{}^{\alpha } \bar{R}_{\nu \beta \alpha \gamma } -  \frac{1}{4} \bar{\nabla}_{\alpha } h \bar{\nabla}^{\alpha }h_{\mu \nu } + \frac{1}{2} \bar{\nabla}^{\alpha }h_{\mu \nu } \bar{\nabla}_{\beta }h_{\alpha }{}^{\beta }\\
& + \frac{1}{2} h^{\alpha \beta } \bar{\nabla}_{\beta }\bar{\nabla}_{\alpha }h_{\mu \nu } -  \frac{1}{2} \bar{\nabla}_{\alpha }h_{\nu \beta } \bar{\nabla}^{\beta }h_{\mu }{}^{\alpha } + \frac{1}{2} \bar{\nabla}_{\beta }h_{\nu \alpha } \bar{\nabla}^{\beta }h_{\mu }{}^{\alpha } + \frac{1}{4} \bar{\nabla}_{\alpha } h \bar{\nabla}_{\mu }h_{\nu }{}^{\alpha }-  \frac{1}{2} \bar{\nabla}_{\beta }h_{\alpha }{}^{\beta } \bar{\nabla}_{\mu }h_{\nu }{}^{\alpha }\\
&-  \frac{1}{2} h^{\alpha \beta } \bar{\nabla}_{\mu }\bar{\nabla}_{\beta }h_{\nu \alpha } + \frac{1}{4} \bar{\nabla}_{\mu }h^{\alpha \beta } \bar{\nabla}_{\nu }h_{\alpha \beta } + \frac{1}{4} \bar{\nabla}_{\alpha } h \bar{\nabla}_{\nu }h_{\mu }{}^{\alpha }-  \frac{1}{2} \bar{\nabla}_{\beta }h_{\alpha }{}^{\beta } \bar{\nabla}_{\nu }h_{\mu }{}^{\alpha } -  \frac{1}{2} h^{\alpha \beta } \bar{\nabla}_{\nu }\bar{\nabla}_{\beta }h_{\mu \alpha }\\
&+ \frac{1}{2} h^{\alpha \beta } \bar{\nabla}_{\nu }\bar{\nabla}_{\mu }h_{\alpha \beta } ,
\\
S_{\mu \nu}^{(1)} ={}& \frac{1}{4} \bar{g}_{\mu \nu } h^{\alpha \beta } \bar{R}_{\alpha \beta } + \frac{1}{2} h_{\nu }{}^{\alpha } \bar{R}_{\mu \alpha } + \frac{1}{2} h_{\mu }{}^{\alpha } \bar{R}_{\nu \alpha } -  \frac{1}{4} h_{\mu \nu } \bar{R} -  h^{\alpha \beta } \bar{R}_{\mu \alpha \nu \beta } -  \frac{1}{2}  \bar{\square} h_{\mu \nu } -  \frac{1}{4} \bar{g}_{\mu \nu } \bar{\nabla}_{\beta }\bar{\nabla}_{\alpha }h^{\alpha \beta }\\
& + \frac{1}{4} \bar{g}_{\mu \nu }  \bar{\square}  h + \frac{1}{2} \bar{\nabla}_{\mu }\bar{\nabla}_{\alpha }h_{\nu }{}^{\alpha } + \frac{1}{2} \bar{\nabla}_{\nu }\bar{\nabla}_{\alpha }h_{\mu }{}^{\alpha } -  \frac{1}{2} \bar{\nabla}_{\nu }\bar{\nabla}_{\mu } h
\ea
\ee
\be
\ba
S_{\mu \nu}^{(2)} ={}&\frac{1}{4} h^{\alpha \beta } h_{\mu \nu } \bar{R}_{\alpha \beta } 
+ \frac{1}{4} \bar{g}_{\mu \nu } h^{\alpha \beta } h^{\gamma \lambda } \bar{R}_{\alpha \gamma \beta \lambda } 
-  \frac{1}{2} \bar{g}_{\mu \nu } h^{\alpha \beta } h_{\lambda }{}^{\gamma } \bar{R}_{\beta \gamma \alpha }{}^{\lambda } 
-  \frac{1}{2} h^{\beta \gamma } h_{\nu }{}^{\alpha } \bar{R}_{\mu \beta \alpha \gamma } 
+ h_{\alpha }{}^{\gamma } h^{\alpha \beta } \bar{R}_{\mu \beta \nu \gamma } \\&
-  \frac{1}{2} h^{\beta \gamma } h_{\mu }{}^{\alpha } \bar{R}_{\nu \beta \alpha \gamma } 
-  \frac{1}{4} \bar{\nabla}_{\alpha } h \bar{\nabla}^{\alpha }h_{\mu \nu } 
+ \frac{1}{2} \bar{\nabla}^{\alpha }h_{\mu \nu } \bar{\nabla}_{\beta }h_{\alpha }{}^{\beta } 
-  \frac{1}{4} h_{\mu \nu } \bar{\nabla}_{\beta }\bar{\nabla}_{\alpha }h^{\alpha \beta } 
-  \frac{1}{4} \bar{g}_{\mu \nu } h^{\alpha \beta } \bar{\nabla}_{\beta }\bar{\nabla}_{\alpha } h \\&
+ \frac{1}{2} h^{\alpha \beta } \bar{\nabla}_{\beta }\bar{\nabla}_{\alpha }h_{\mu \nu } 
+ \frac{1}{4} h_{\mu \nu }  \bar{\square}  h 
+ \frac{1}{2} \bar{g}_{\mu \nu } h^{\alpha \beta } \bar{\nabla}_{\beta }\bar{\nabla}_{\gamma }h_{\alpha }{}^{\gamma } 
+ \frac{1}{16} \bar{g}_{\mu \nu } \bar{\nabla}_{\beta } h \bar{\nabla}^{\beta } h 
-  \frac{1}{2} \bar{\nabla}_{\alpha }h_{\nu \beta } \bar{\nabla}^{\beta }h_{\mu }{}^{\alpha } \\&
+ \frac{1}{2} \bar{\nabla}_{\beta }h_{\nu \alpha } \bar{\nabla}^{\beta }h_{\mu }{}^{\alpha } 
+ \frac{1}{4} \bar{g}_{\mu \nu } \bar{\nabla}_{\alpha }h^{\alpha \beta } \bar{\nabla}_{\gamma }h_{\beta }{}^{\gamma } 
-  \frac{1}{4} \bar{g}_{\mu \nu } \bar{\nabla}^{\beta } h \bar{\nabla}_{\gamma }h_{\beta }{}^{\gamma } 
-  \frac{1}{4} \bar{g}_{\mu \nu } h^{\alpha \beta }  \bar{\square} h_{\alpha \beta } \\&
+ \frac{1}{8} \bar{g}_{\mu \nu } \bar{\nabla}_{\beta }h_{\alpha \gamma } \bar{\nabla}^{\gamma }h^{\alpha \beta } 
-  \frac{3}{16} \bar{g}_{\mu \nu } \bar{\nabla}_{\gamma }h_{\alpha \beta } \bar{\nabla}^{\gamma }h^{\alpha \beta } 
+ \frac{1}{4} \bar{\nabla}_{\alpha } h \bar{\nabla}_{\mu }h_{\nu }{}^{\alpha } 
-  \frac{1}{2} \bar{\nabla}_{\beta }h_{\alpha }{}^{\beta } \bar{\nabla}_{\mu }h_{\nu }{}^{\alpha } 
-  \frac{1}{2} h^{\alpha \beta } \bar{\nabla}_{\mu }\bar{\nabla}_{\beta }h_{\nu \alpha } \\&
+ \frac{1}{4} \bar{\nabla}_{\mu }h^{\alpha \beta } \bar{\nabla}_{\nu }h_{\alpha \beta } 
+ \frac{1}{4} \bar{\nabla}_{\alpha } h \bar{\nabla}_{\nu }h_{\mu }{}^{\alpha } 
-  \frac{1}{2} \bar{\nabla}_{\beta }h_{\alpha }{}^{\beta } \bar{\nabla}_{\nu }h_{\mu }{}^{\alpha } 
-  \frac{1}{2} h^{\alpha \beta } \bar{\nabla}_{\nu }\bar{\nabla}_{\beta }h_{\mu \alpha } 
+ \frac{1}{2} h^{\alpha \beta } \bar{\nabla}_{\nu }\bar{\nabla}_{\mu }h_{\alpha \beta }\,,
\ea
\ee
\be
\ba
R_{\mu \nu \rho \sigma}^{(1)} ={}& \frac{1}{2} h_{\sigma }{}^{\alpha } \bar{R}_{\mu \alpha \nu \rho } -  \frac{1}{2} h_{\rho }{}^{\alpha } \bar{R}_{\mu \alpha \nu \sigma } + \frac{1}{2} h_{\sigma }{}^{\alpha } \bar{R}_{\mu \nu \rho \alpha } -  \frac{1}{2} h_{\rho }{}^{\alpha } \bar{R}_{\mu \nu \sigma \alpha } -  \frac{1}{2} h_{\sigma }{}^{\alpha } \bar{R}_{\mu \rho \nu \alpha } -  \frac{1}{2} h_{\nu }{}^{\alpha } \bar{R}_{\mu \rho \sigma \alpha } + \frac{1}{2} h_{\rho }{}^{\alpha } \bar{R}_{\mu \sigma \nu \alpha } \\
&+ \frac{1}{2} h_{\nu }{}^{\alpha } \bar{R}_{\mu \sigma \rho \alpha } + \frac{1}{2} h_{\mu }{}^{\alpha } \bar{R}_{\nu \rho \sigma \alpha } -  \frac{1}{2} h_{\mu }{}^{\alpha } \bar{R}_{\nu \sigma \rho \alpha } -  \frac{1}{2} \bar{\nabla}_{\rho }\bar{\nabla}_{\mu }h_{\nu \sigma } + \frac{1}{2} \bar{\nabla}_{\rho }\bar{\nabla}_{\nu }h_{\mu \sigma } + \frac{1}{2} \bar{\nabla}_{\sigma }\bar{\nabla}_{\mu }h_{\nu \rho } -  \frac{1}{2} \bar{\nabla}_{\sigma }\bar{\nabla}_{\nu }h_{\mu \rho }\,,
\\
R_{\mu \nu \rho \sigma}^{(2)} ={}& - \frac{1}{4} \bar{\nabla}_{\alpha }h_{\nu \sigma } \bar{\nabla}^{\alpha }h_{\mu \rho } 
+ \frac{1}{4} \bar{\nabla}_{\alpha }h_{\nu \rho } \bar{\nabla}^{\alpha }h_{\mu \sigma } 
+ \frac{1}{4} \bar{\nabla}^{\alpha }h_{\nu \sigma } \bar{\nabla}_{\mu }h_{\rho \alpha } 
-  \frac{1}{4} \bar{\nabla}^{\alpha }h_{\nu \rho } \bar{\nabla}_{\mu }h_{\sigma \alpha } 
-  \frac{1}{4} \bar{\nabla}^{\alpha }h_{\mu \sigma } \bar{\nabla}_{\nu }h_{\rho \alpha } \\
&\quad+ \frac{1}{4} \bar{\nabla}_{\mu }h_{\sigma \alpha } \bar{\nabla}_{\nu }h_{\rho }{}^{\alpha } 
+ \frac{1}{4} \bar{\nabla}^{\alpha }h_{\mu \rho } \bar{\nabla}_{\nu }h_{\sigma \alpha } 
-  \frac{1}{4} \bar{\nabla}_{\mu }h_{\rho }{}^{\alpha } \bar{\nabla}_{\nu }h_{\sigma \alpha } 
+ \frac{1}{4} \bar{\nabla}_{\alpha }h_{\nu \sigma } \bar{\nabla}_{\rho }h_{\mu }{}^{\alpha } 
-  \frac{1}{4} \bar{\nabla}_{\nu }h_{\sigma \alpha } \bar{\nabla}_{\rho }h_{\mu }{}^{\alpha } \\
&\quad-  \frac{1}{4} \bar{\nabla}^{\alpha }h_{\mu \sigma } \bar{\nabla}_{\rho }h_{\nu \alpha } 
+ \frac{1}{4} \bar{\nabla}_{\mu }h_{\sigma \alpha } \bar{\nabla}_{\rho }h_{\nu }{}^{\alpha } 
-  \frac{1}{4} \bar{\nabla}_{\alpha }h_{\nu \rho } \bar{\nabla}_{\sigma }h_{\mu }{}^{\alpha } 
+ \frac{1}{4} \bar{\nabla}_{\nu }h_{\rho \alpha } \bar{\nabla}_{\sigma }h_{\mu }{}^{\alpha } 
+ \frac{1}{4} \bar{\nabla}_{\rho }h_{\nu \alpha } \bar{\nabla}_{\sigma }h_{\mu }{}^{\alpha } \\
&\quad+ \frac{1}{4} \bar{\nabla}^{\alpha }h_{\mu \rho } \bar{\nabla}_{\sigma }h_{\nu \alpha } 
-  \frac{1}{4} \bar{\nabla}_{\rho }h_{\mu }{}^{\alpha } \bar{\nabla}_{\sigma }h_{\nu \alpha } 
-  \frac{1}{4} \bar{\nabla}_{\mu }h_{\rho \alpha } \bar{\nabla}_{\sigma }h_{\nu }{}^{\alpha }\,,
\ea
\ee
\be
\ba
C_{\mu \nu \rho \sigma}^{(1)} ={}&\frac{1}{6} \bar{g}_{\mu \sigma } \bar{g}_{\nu \rho } h^{\alpha \beta } \bar{R}_{\alpha \beta } 
-  \frac{1}{6} \bar{g}_{\mu \rho } \bar{g}_{\nu \sigma } h^{\alpha \beta } \bar{R}_{\alpha \beta } 
-  \frac{1}{4} \bar{g}_{\nu \sigma } h_{\rho }{}^{\alpha } \bar{R}_{\mu \alpha } 
+ \frac{1}{4} \bar{g}_{\nu \rho } h_{\sigma }{}^{\alpha } \bar{R}_{\mu \alpha } 
-  \frac{1}{2} h_{\nu \sigma } \bar{R}_{\mu \rho } 
+ \frac{1}{2} h_{\nu \rho } \bar{R}_{\mu \sigma } \\&
+ \frac{1}{4} \bar{g}_{\mu \sigma } h_{\rho }{}^{\alpha } \bar{R}_{\nu \alpha } 
-  \frac{1}{4} \bar{g}_{\mu \rho } h_{\sigma }{}^{\alpha } \bar{R}_{\nu \alpha } 
+ \frac{1}{2} h_{\mu \sigma } \bar{R}_{\nu \rho } 
-  \frac{1}{2} h_{\mu \rho } \bar{R}_{\nu \sigma } 
-  \frac{1}{4} \bar{g}_{\nu \sigma } h_{\mu }{}^{\alpha } \bar{R}_{\rho \alpha } 
+ \frac{1}{4} \bar{g}_{\mu \sigma } h_{\nu }{}^{\alpha } \bar{R}_{\rho \alpha } \\&
+ \frac{1}{4} \bar{g}_{\nu \rho } h_{\mu }{}^{\alpha } \bar{R}_{\sigma \alpha } 
-  \frac{1}{4} \bar{g}_{\mu \rho } h_{\nu }{}^{\alpha } \bar{R}_{\sigma \alpha } 
+ \frac{1}{6} \bar{g}_{\nu \sigma } h_{\mu \rho } \bar{R} 
-  \frac{1}{6} \bar{g}_{\nu \rho } h_{\mu \sigma } \bar{R} 
-  \frac{1}{6} \bar{g}_{\mu \sigma } h_{\nu \rho } \bar{R}
+ \frac{1}{6} \bar{g}_{\mu \rho } h_{\nu \sigma } \bar{R}\\&
+ \frac{1}{2} h_{\sigma }{}^{\alpha } \bar{R}_{\mu \alpha \nu \rho } 
-  \frac{1}{2} h_{\rho }{}^{\alpha } \bar{R}_{\mu \alpha \nu \sigma } 
+ \frac{1}{2} \bar{g}_{\nu \sigma } h^{\alpha \beta } \bar{R}_{\mu \alpha \rho \beta } 
-  \frac{1}{2} \bar{g}_{\nu \rho } h^{\alpha \beta } \bar{R}_{\mu \alpha \sigma \beta } 
+ \frac{1}{2} h_{\sigma }{}^{\alpha } \bar{R}_{\mu \nu \rho \alpha } 
-  \frac{1}{2} h_{\rho }{}^{\alpha } \bar{R}_{\mu \nu \sigma \alpha } \\&
-  \frac{1}{2} h_{\sigma }{}^{\alpha } \bar{R}_{\mu \rho \nu \alpha } 
-  \frac{1}{2} h_{\nu }{}^{\alpha } \bar{R}_{\mu \rho \sigma \alpha } 
+ \frac{1}{2} h_{\rho }{}^{\alpha } \bar{R}_{\mu \sigma \nu \alpha } 
+ \frac{1}{2} h_{\nu }{}^{\alpha } \bar{R}_{\mu \sigma \rho \alpha } 
-  \frac{1}{2} \bar{g}_{\mu \sigma } h^{\alpha \beta } \bar{R}_{\nu \alpha \rho \beta } 
+ \frac{1}{2} \bar{g}_{\mu \rho } h^{\alpha \beta } \bar{R}_{\nu \alpha \sigma \beta } \\&
+ \frac{1}{2} h_{\mu }{}^{\alpha } \bar{R}_{\nu \rho \sigma \alpha } 
-  \frac{1}{2} h_{\mu }{}^{\alpha } \bar{R}_{\nu \sigma \rho \alpha } 
+ \frac{1}{4} \bar{g}_{\nu \sigma }  \bar{\square} h_{\mu \rho } 
-  \frac{1}{4} \bar{g}_{\nu \rho }  \bar{\square} h_{\mu \sigma } 
-  \frac{1}{4} \bar{g}_{\mu \sigma }  \bar{\square} h_{\nu \rho } 
+ \frac{1}{4} \bar{g}_{\mu \rho }  \bar{\square} h_{\nu \sigma } \\&
-  \frac{1}{6} \bar{g}_{\mu \sigma } \bar{g}_{\nu \rho } \bar{\nabla}_{\beta }\bar{\nabla}_{\alpha }h^{\alpha \beta } 
+ \frac{1}{6} \bar{g}_{\mu \rho } \bar{g}_{\nu \sigma } \bar{\nabla}_{\beta }\bar{\nabla}_{\alpha }h^{\alpha \beta } 
+ \frac{1}{6} \bar{g}_{\mu \sigma } \bar{g}_{\nu \rho }  \bar{\square}  h 
-  \frac{1}{6} \bar{g}_{\mu \rho } \bar{g}_{\nu \sigma }  \bar{\square}  h 
-  \frac{1}{4} \bar{g}_{\nu \sigma } \bar{\nabla}_{\mu }\bar{\nabla}_{\alpha }h_{\rho }{}^{\alpha } \\&
+ \frac{1}{4} \bar{g}_{\nu \rho } \bar{\nabla}_{\mu }\bar{\nabla}_{\alpha }h_{\sigma }{}^{\alpha } 
+ \frac{1}{4} \bar{g}_{\mu \sigma } \bar{\nabla}_{\nu }\bar{\nabla}_{\alpha }h_{\rho }{}^{\alpha } 
-  \frac{1}{4} \bar{g}_{\mu \rho } \bar{\nabla}_{\nu }\bar{\nabla}_{\alpha }h_{\sigma }{}^{\alpha } 
-  \frac{1}{4} \bar{g}_{\nu \sigma } \bar{\nabla}_{\rho }\bar{\nabla}_{\alpha }h_{\mu }{}^{\alpha } 
+ \frac{1}{4} \bar{g}_{\mu \sigma } \bar{\nabla}_{\rho }\bar{\nabla}_{\alpha }h_{\nu }{}^{\alpha } \\&
+ \frac{1}{4} \bar{g}_{\nu \sigma } \bar{\nabla}_{\rho }\bar{\nabla}_{\mu } h 
-  \frac{1}{2} \bar{\nabla}_{\rho }\bar{\nabla}_{\mu }h_{\nu \sigma } 
-  \frac{1}{4} \bar{g}_{\mu \sigma } \bar{\nabla}_{\rho }\bar{\nabla}_{\nu } h 
+ \frac{1}{2} \bar{\nabla}_{\rho }\bar{\nabla}_{\nu }h_{\mu \sigma } 
+ \frac{1}{4} \bar{g}_{\nu \rho } \bar{\nabla}_{\sigma }\bar{\nabla}_{\alpha }h_{\mu }{}^{\alpha }
-  \frac{1}{4} \bar{g}_{\mu \rho } \bar{\nabla}_{\sigma }\bar{\nabla}_{\alpha }h_{\nu }{}^{\alpha }\\& 
-  \frac{1}{4} \bar{g}_{\nu \rho } \bar{\nabla}_{\sigma }\bar{\nabla}_{\mu } h 
+ \frac{1}{2} \bar{\nabla}_{\sigma }\bar{\nabla}_{\mu }h_{\nu \rho } 
+ \frac{1}{4} \bar{g}_{\mu \rho } \bar{\nabla}_{\sigma }\bar{\nabla}_{\nu } h 
-  \frac{1}{2} \bar{\nabla}_{\sigma }\bar{\nabla}_{\nu }h_{\mu \rho }\,,
\ea
\ee
Note that $C_{\mu \nu \rho \sigma}^{(2)}$ has $144$ terms and is not very enlightening and so we do not present it here.

Perturbations of mixed rank tensors can be written down once the perturbation of its fully covariant (or contravariant) counterpart is known, for instance
\be
\delta R^\mu_{\ \nu} = \delta ( R_{\alpha \nu} g^{\alpha \mu} ) = \delta R_{\alpha \nu} \bar{g}^{\alpha \mu} + \bar{R}_{\alpha \nu} \delta{g}^{\alpha \mu} =  \delta R_{\alpha \nu} \bar{g}^{\alpha \mu} - \bar{R}_{\alpha \nu} h^{\alpha \mu}\,,
\ee
We also need perturbations of covariant derivatives. For $\nabla_\mu Q$, this means that the Levi-Civita connection in the definition of $\nabla_\mu$ is expanded in $h$, keeping $Q$ unperturbed. Then $\delta(\nabla_\mu)$ means that after $\nabla_\mu$ operates on $\bar{Q}$, it is expanded in $h$, and the first order is picked out. We have
\be
\begin{aligned}
	\delta(\nabla_\mu) Q = {\nabla_\mu}^{(1)} Q &=  {\nabla_\mu}^{(1)} Q^{(0)}\,,\\
	&= (\nabla_\mu Q)^{(1)} - {\nabla_\mu}^{(0)}  Q^{(1)}\,,\\
	&= \delta(\nabla_\mu Q) - {\nabla_\mu} \delta Q\,.
\end{aligned}
\ee
For example, for a scalar $\phi$ which does not depend on $g_{\mu \nu}$, we have
\be
\ba
\delta(\square ) \phi &= \delta(g^{\mu \nu} \nabla_\mu \nabla_\nu) \phi = -h^{\mu \nu} \bar{\nabla}_\mu \partial_\nu \phi - \bar{g}^{\mu \nu} \delta \Gamma_{\mu \nu}^{\lambda} \partial_\lambda \phi\,,\\
&= \delta(\square \phi) - \bar{\square} \delta \phi\,.
\ea
\ee
Similarly, for Ricci scalar $R$, which depends on $g_{\mu \nu}$, we have
\be \label{deltaboxricciscalar}
\ba
\delta (\square) R &=  \delta(\square R) - \bar{\square}  \delta R\,,\\
&= \frac{1}{2} \bar{\nabla}_{\alpha }h \bar{\nabla}^{\alpha }\bar{R} -  \bar{\nabla}^{\alpha }\bar{R} \bar{\nabla}_{\beta }h_{\alpha }{}^{\beta } -  h_{\alpha \beta } \bar{\nabla}^{\beta }\bar{\nabla}^{\alpha } \bar{R}\,,\\
&=0 \qquad \text{for $\bar{R}= $ constant}\,.
\ea
\ee
We can now expand covariant derivative $\nabla_{\mu}$, box operator $\square$ and the form factors $\Fc_{i} (\square_{s} )$ in eq.(\ref{formfactor}) in a series in $h$
\be
\begin{aligned}
	\nabla_\mu &= {\nabla_\mu}^{(0)} + {\nabla_\mu}^{(1)} + {\nabla_\mu}^{(2)} + \cdots\,,\\
	\square  &= \square^{(0)} +  \square^{(1)} +  \square^{(2)} + \cdots\,,\\
	\Fc_{i} (\square_{s}) &= \Fc_{i} (\square_{s})^{(0)} + \Fc_{i} (\square_{s})^{(1)} + \Fc_{i} (\square_{s})^{(2)}  + \cdots\,.
\end{aligned}
\ee

\section{Second order variation of the action}\label{appendixsecondordervariations}

This appendix is self-contained. In Appendix (\ref{appendixperturbations}) earlier, we had listed perturbations of all curvatures in terms of $h_{\mu \nu}$. In this appendix, we perform SVT decomposition of $\delta^{2} S$ eq.(\ref{delta2lagrangian}) and apply our background conditions eq.(\ref{mss20}).  We closely follow the steps enumerated in eq.(\ref{simplifyingsteps}) in order to get the most simplified expressions. In particular,
\begin{enumerate} [label=(\roman*)]
\item We first present the most general perturbations in terms of the SVT modes $\phi, B, A_{\mu}$ and $\widehat{h}_{\mu \nu}$, and background curvatures $\bar{R}, \bar{R}_{\mu \nu}$ and $\bar{R}_{\mu \nu \rho \sigma}$ which are kept \textit{arbitrary}. We also use the second Bianchi identity eq.(\ref{simplifyingsteps}).

\item All background curvatures are then decomposed into their irreducible trace and traceless components $\bar{R}, \bar{S}_{\mu \nu}$ and $\bar{C}_{\mu \nu \rho \sigma}$. We also impose background conformal flatness $\bar{C}_{\mu \nu \rho \sigma} = 0$ eq.(\ref{mss19}). And use the Bianchi identity again.

\item Covariantly constant background curvature conditions $\bar{\nabla}_{\alpha} \bar{S}_{\mu \nu} = 0$ and $\partial_{\mu} \bar{R} = 0$ in eq.(\ref{mss21}) are applied.
\end{enumerate}

\subsection{$ \displaystyle \delta^2 \mathcal{L}_{EH + \Lambda} =  \frac{M_P^2}{2} \delta^2 \Big[ \sqrt{-g} (R - 2 \Lambda) \Big]  =  \frac{M_P^2}{2} \delta_{0}$} \label{appendixehlambda}

For reference, around an \textit{arbitrary} background, we have \cite{Biswas:2016egy}:
\be \nonumber
\ba
\delta^2 \Big[ \sqrt{-g} (R - 2 \Lambda) \Big] = \sqrt{-\bar g} \ & \Bigg[\left(\frac{1}{4} h_{\mu\nu}\bar{\square}
h^{\mu\nu} - \frac{1}{4} h\bar{\square} h+\frac{1}{2} h\bar{\nabla}_\mu\bar{\nabla}_\rho
h^{\mu\rho}+\frac{1}{2}\bar{\nabla}_\mu h^{\mu\rho}\bar{\nabla}_\nu h^\nu_{\ \rho}\right)\\
&\quad +(hh^{\mu\nu}-2h^\mu_{\ \sigma} h^{\sigma\nu})\left(\frac{1}{8}
\bar R \bar g_{\mu\nu}-\frac{1}{4} \Lambda \bar g_{\mu\nu} - \frac{1}{2} \bar R_{\mu\nu}\right) 
- \left(\frac{1}{2}\bar R_{\sigma\nu}h^\sigma_{\ \rho}
h^{\nu\rho}+\frac{1}{2}\bar R^\sigma_{\ \rho\nu\mu}h^\mu_{\ \sigma} h^{\nu\rho}\right)\Bigg]\,.
\ea
\ee
(i) Performing SVT decomposition eq.(\ref{hdecomp2}) of $\delta_0$ eq.(\ref{deltazero}) around an \textit{arbitrary} background, we get the terms which are purely quadratic in one SVT mode as
\be
\ba
\delta_{0(BB)} ={}&\frac{1}{8} B \bar{R}^{\beta \gamma } \bar{\nabla}_{\alpha }\bar{R}_{\beta \gamma } \bar{\nabla}^{\alpha }B
+ \frac{1}{4} \Lambda B \bar{\nabla}_{\alpha }\bar{R} \bar{\nabla}^{\alpha }B
-  \frac{1}{8} B \bar{R} \bar{\nabla}_{\alpha }\bar{R} \bar{\nabla}^{\alpha }B 
-  \frac{1}{8} B \bar{\nabla}^{\alpha }\bar{R} \bar{\square} \bar{\nabla}_{\alpha }B
+ \frac{1}{4} \Lambda B \bar{\square}^2 B
-  \frac{1}{8} B \bar{R} \bar{\square}^2 B  \\
& -  \frac{1}{8} B \bar{R}_{\alpha \beta } \bar{\nabla}^{\alpha }B \bar{\nabla}^{\beta }\bar{R}
+ \frac{1}{2} \Lambda B \bar{R}_{\alpha \beta } \bar{\nabla}^{\beta }\bar{\nabla}^{\alpha }B
-  \frac{7}{8} B \bar{R}_{\alpha }{}^{\gamma } \bar{R}_{\beta \gamma } \bar{\nabla}^{\beta }\bar{\nabla}^{\alpha }B
-  \frac{1}{4} B \bar{R}_{\alpha \beta } \bar{R} \bar{\nabla}^{\beta }\bar{\nabla}^{\alpha }B
-  \frac{1}{4} B \bar{R}^{\gamma \lambda } \bar{R}_{\alpha \gamma \beta \lambda } \bar{\nabla}^{\beta }\bar{\nabla}^{\alpha }B  \\
& -  \frac{1}{2} B \bar{R}_{\alpha }{}^{\gamma \lambda \mu} \bar{R}_{\beta \gamma \lambda \mu} \bar{\nabla}^{\beta }\bar{\nabla}^{\alpha }B
-  \frac{11}{8} B \bar{R}^{\beta \gamma } \bar{\nabla}^{\alpha }B \bar{\nabla}_{\gamma }\bar{R}_{\alpha \beta } 
+ \frac{1}{2} B \bar{R}^{\beta \gamma \lambda \mu} \bar{\nabla}^{\alpha }B \bar{\nabla}_{\gamma }\bar{R}_{\alpha \beta \lambda \mu}
-  \frac{1}{8} B \bar{R}_{\alpha \beta } \bar{\square} \bar{\nabla}^{\beta }\bar{\nabla}^{\alpha }B  \\
& + \frac{13}{8} B \bar{R}_{\alpha \gamma } \bar{\nabla}^{\gamma }\bar{\square} \bar{\nabla}^{\alpha }B
-  \frac{5}{4} B \bar{R}_{\beta \gamma } \bar{\nabla}^{\gamma }\bar{\nabla}^{\beta }\bar{\square} B 
-  \frac{3}{4} B \bar{\nabla}_{\alpha }\bar{R}_{\beta \gamma } \bar{\nabla}^{\gamma }\bar{\nabla}^{\beta }\bar{\nabla}^{\alpha }B
+ \frac{11}{4} B \bar{\nabla}_{\beta }\bar{R}_{\alpha \gamma } \bar{\nabla}^{\gamma }\bar{\nabla}^{\beta }\bar{\nabla}^{\alpha }B  \\
& -  \frac{7}{4} B \bar{\nabla}_{\gamma }\bar{R}_{\alpha \beta } \bar{\nabla}^{\gamma }\bar{\nabla}^{\beta }\bar{\nabla}^{\alpha }B
+ \frac{3}{2} B \bar{R}_{\alpha \beta \gamma \lambda } \bar{\nabla}^{\alpha }B \bar{\nabla}^{\lambda }\bar{R}^{\beta \gamma } 
+ \frac{5}{2} B \bar{R}_{\alpha \beta \gamma \lambda } \bar{\nabla}^{\lambda }\bar{\nabla}^{\gamma }\bar{\nabla}^{\beta }\bar{\nabla}^{\alpha }B
-  \frac{1}{2} B \bar{R}_{\alpha \gamma \beta \lambda } \bar{\nabla}^{\lambda }\bar{\nabla}^{\gamma }\bar{\nabla}^{\beta }\bar{\nabla}^{\alpha }B  \\
& -  \frac{1}{2} B \bar{R}_{\alpha \lambda \beta \gamma } \bar{\nabla}^{\lambda }\bar{\nabla}^{\gamma }\bar{\nabla}^{\beta }\bar{\nabla}^{\alpha }B\,,
\\
\delta_{0 (\phi \phi)} ={}&- \frac{1}{8} \Lambda \phi^2
-  \frac{3}{32} \phi \bar{\square}\phi\,,
\\
\delta_{0 (AA )} ={}&- \Lambda A^{\alpha } A^{\beta } \bar{R}_{\alpha \beta }
+ \frac{1}{2} A^{\alpha } A^{\beta } \bar{R}_{\alpha \beta } \bar{R}
-  A^{\alpha } A^{\beta } \bar{R}^{\gamma \lambda } \bar{R}_{\alpha \gamma \beta \lambda }
+ \frac{1}{2} A^{\alpha } \bar{\nabla}_{\alpha }A^{\beta } \bar{\nabla}_{\beta }\bar{R}
+ \frac{1}{2} A^{\alpha } A^{\beta } \bar{\nabla}_{\beta }\bar{\nabla}_{\alpha }\bar{R}
-  \Lambda A^{\alpha } \bar{\square} A_{\alpha } \\
& + \frac{1}{2} A^{\alpha } \bar{R} \bar{\square} A_{\alpha }
-  A^{\alpha } \bar{R}_{\beta \gamma } \bar{\nabla}^{\gamma }\bar{\nabla}^{\beta }A_{\alpha }
-  \frac{3}{2} A^{\alpha } \bar{R}_{\alpha \beta \gamma \lambda } \bar{\nabla}^{\lambda }\bar{\nabla}^{\gamma }A^{\beta }
+ \frac{3}{2} A^{\alpha } \bar{R}_{\alpha \gamma \beta \lambda } \bar{\nabla}^{\lambda }\bar{\nabla}^{\gamma }A^{\beta }
-  \frac{3}{2} A^{\alpha } \bar{R}_{\alpha \lambda \beta \gamma } \bar{\nabla}^{\lambda }\bar{\nabla}^{\gamma }A^{\beta }\,,
\\
\delta_{0 (\widehat{h} \widehat{h} )} ={}&\frac{1}{2} \Lambda \widehat{h}_{\alpha \beta } \widehat{h}^{\alpha \beta }
+ \frac{1}{2} \widehat{h}_{\alpha }{}^{\gamma } \widehat{h}^{\alpha \beta } \bar{R}_{\beta \gamma }
-  \frac{1}{4} \widehat{h}_{\alpha \beta } \widehat{h}^{\alpha \beta } \bar{R}
+ \frac{1}{2} \widehat{h}^{\alpha \beta } \widehat{h}^{\gamma \lambda } \bar{R}_{\alpha \gamma \beta \lambda }
+ \frac{1}{4} \widehat{h}^{\alpha \beta } \bar{\square} \widehat{h}_{\alpha \beta }\,,
\ea
\ee
and six mixed terms given by
\be
\ba
\delta_{0 (B \phi)}={}&\frac{1}{4} \Lambda B \bar{\square}\phi
-  \frac{1}{8} B \bar{\nabla}_{\alpha }\phi \bar{\nabla}^{\alpha }\bar{R}
-  \frac{1}{4} B \bar{R}_{\alpha \beta } \bar{\nabla}^{\beta }\bar{\nabla}^{\alpha }\phi\,,
\ea
\ee
\be
\ba
\delta_{0 (BA)}={}&- A^{\alpha } B \bar{R}^{\beta \gamma } \bar{\nabla}_{\alpha }\bar{R}_{\beta \gamma }
+ \Lambda A^{\alpha } B \bar{\nabla}_{\alpha }\bar{R}
-  \frac{1}{2} A^{\alpha } B \bar{R} \bar{\nabla}_{\alpha }\bar{R}
-  \frac{1}{2} B \bar{\nabla}^{\alpha }\bar{R} \bar{\square} A_{\alpha }
+ 2 \Lambda B \bar{R}_{\alpha \beta } \bar{\nabla}^{\beta }A^{\alpha }
+ 4 B \bar{R}_{\alpha }{}^{\gamma } \bar{R}_{\beta \gamma } \bar{\nabla}^{\beta }A^{\alpha }\\
& -  B \bar{R}_{\alpha \beta } \bar{R} \bar{\nabla}^{\beta }A^{\alpha }
- 4 B \bar{R}^{\gamma \lambda } \bar{R}_{\alpha \gamma \beta \lambda } \bar{\nabla}^{\beta }A^{\alpha }
- 2 B \bar{R}_{\alpha }{}^{\gamma \lambda \mu} \bar{R}_{\beta \lambda \gamma \mu} \bar{\nabla}^{\beta }A^{\alpha }
-  \frac{1}{2} A^{\alpha } B \bar{R}_{\alpha \beta } \bar{\nabla}^{\beta }\bar{R}
+ 2 A^{\alpha } B \bar{R}^{\beta \gamma } \bar{\nabla}_{\gamma }\bar{R}_{\alpha \beta }\\
&- 2 B \bar{R}_{\alpha \beta } \bar{\square} \bar{\nabla}^{\beta }A^{\alpha }
+ 2 B \bar{R}_{\alpha \gamma } \bar{\nabla}^{\gamma }\bar{\square} A^{\alpha }
+ B \bar{\nabla}_{\alpha }\bar{R}_{\beta \gamma } \bar{\nabla}^{\gamma }\bar{\nabla}^{\beta }A^{\alpha }
+ 6 B \bar{\nabla}_{\beta }\bar{R}_{\alpha \gamma } \bar{\nabla}^{\gamma }\bar{\nabla}^{\beta }A^{\alpha }
- 6 B \bar{\nabla}_{\gamma }\bar{R}_{\alpha \beta } \bar{\nabla}^{\gamma }\bar{\nabla}^{\beta }A^{\alpha }\\
&+ 5 A^{\alpha } B \bar{R}_{\alpha \beta \gamma \lambda } \bar{\nabla}^{\lambda }\bar{R}^{\beta \gamma }
+ 3 B \bar{R}_{\alpha \beta \gamma \lambda } \bar{\nabla}^{\lambda }\bar{\nabla}^{\gamma }\bar{\nabla}^{\beta }A^{\alpha }
- 2 B \bar{R}_{\alpha \gamma \beta \lambda } \bar{\nabla}^{\lambda }\bar{\nabla}^{\gamma }\bar{\nabla}^{\beta }A^{\alpha }
- 2 A^{\alpha } B \bar{R}^{\beta \gamma \lambda \mu} \bar{\nabla}_{\mu}\bar{R}_{\alpha \beta \gamma \lambda }\,,
\ea
\ee
\be
\ba
\delta_{0 (B \widehat{h})}={}&B \widehat{h}^{\alpha \beta } \bar{R}_{\alpha }{}^{\gamma } \bar{R}_{\beta \gamma }
-  B \widehat{h}^{\alpha \beta } \bar{R}^{\gamma \lambda } \bar{R}_{\alpha \gamma \beta \lambda }
-  \frac{1}{2} B \bar{R}^{\alpha \beta } \bar{\square} \widehat{h}_{\alpha \beta } 
+ B \bar{\nabla}_{\beta }\bar{R}_{\alpha \gamma } \bar{\nabla}^{\gamma }\widehat{h}^{\alpha \beta }
-  \frac{1}{2} B \bar{\nabla}_{\gamma }\bar{R}_{\alpha \beta } \bar{\nabla}^{\gamma }\widehat{h}^{\alpha \beta }\,,
\\ \\
\delta_{0 (\phi A )}={}&- \frac{1}{4} A^{\alpha } \phi \bar{\nabla}_{\alpha }\bar{R}
-  \frac{1}{2} \bar{R}_{\alpha \beta } \phi \bar{\nabla}^{\beta }A^{\alpha }\,,
\\ \\
\delta_{0 (\phi \widehat{h})}={}&0\,,
\qquad \qquad
\delta_{0 (A  \widehat{h})}={}- A^{\alpha } \widehat{h}^{\beta \gamma } \bar{\nabla}_{\alpha }\bar{R}_{\beta \gamma }
- 2 A^{\alpha } \bar{R}^{\beta \gamma } \bar{\nabla}_{\gamma }\widehat{h}_{\alpha \beta }\,.
\ea
\ee

(ii) Rewriting the above in terms of irreducible $\bar{R}, \bar{S}_{\mu \nu}, \bar{C}_{\mu \nu \rho \sigma}$ and imposing $\bar{C}_{\mu \nu \rho \sigma} = 0$, the terms purely quadratic in one SVT mode are then
\be
\ba
\delta_{0(BB)} ={}&\frac{1}{8} \Lambda B \bar{R} \bar{\square} B
-  \frac{1}{32} B \bar{R}^2 \bar{\square} B
-  \frac{5}{8} B \bar{S}_{\beta \gamma } \bar{S}^{\beta \gamma } \bar{\square} B
+ \frac{1}{4} \Lambda B \bar{\nabla}_{\alpha }\bar{R} \bar{\nabla}^{\alpha }B
-  \frac{3}{64} B \bar{R} \bar{\nabla}_{\alpha }\bar{R} \bar{\nabla}^{\alpha }B
-  \frac{7}{8} B \bar{S}^{\beta \gamma } \bar{\nabla}_{\alpha }\bar{S}_{\beta \gamma } \bar{\nabla}^{\alpha }B \\
& -  \frac{1}{16} B \bar{\nabla}^{\alpha }\bar{R} \bar{\square} \bar{\nabla}_{\alpha }B
+ \frac{1}{4} \Lambda B \bar{\square}^2 B 
-  \frac{1}{16} B \bar{R} \bar{\square}^2 B
-  \frac{11}{96} B \bar{S}_{\alpha \beta } \bar{\nabla}^{\alpha }B \bar{\nabla}^{\beta }\bar{R}
+ \frac{1}{2} \Lambda B \bar{S}_{\alpha \beta } \bar{\nabla}^{\beta }\bar{\nabla}^{\alpha }B \\
& -  \frac{11}{96} B \bar{R} \bar{S}_{\alpha \beta } \bar{\nabla}^{\beta }\bar{\nabla}^{\alpha }B
+ \frac{7}{8} B \bar{S}_{\alpha }{}^{\gamma } \bar{S}_{\beta \gamma } \bar{\nabla}^{\beta }\bar{\nabla}^{\alpha }B
-  \frac{3}{8} B \bar{S}^{\beta \gamma } \bar{\nabla}^{\alpha }B \bar{\nabla}_{\gamma }\bar{S}_{\alpha \beta } 
-  \frac{5}{8} B \bar{S}_{\alpha \beta } \bar{\square} \bar{\nabla}^{\beta }\bar{\nabla}^{\alpha }B
+ \frac{17}{8} B \bar{S}_{\alpha \gamma } \bar{\nabla}^{\gamma }\bar{\square} \bar{\nabla}^{\alpha }B \\
& -  \frac{5}{4} B \bar{S}_{\beta \gamma } \bar{\nabla}^{\gamma }\bar{\nabla}^{\beta }\bar{\square} B
-  \frac{3}{4} B \bar{\nabla}_{\alpha }\bar{S}_{\beta \gamma } \bar{\nabla}^{\gamma }\bar{\nabla}^{\beta }\bar{\nabla}^{\alpha }B 
+ \frac{11}{4} B \bar{\nabla}_{\beta }\bar{S}_{\alpha \gamma } \bar{\nabla}^{\gamma }\bar{\nabla}^{\beta }\bar{\nabla}^{\alpha }B
-  \frac{7}{4} B \bar{\nabla}_{\gamma }\bar{S}_{\alpha \beta } \bar{\nabla}^{\gamma }\bar{\nabla}^{\beta }\bar{\nabla}^{\alpha }B ,
\ea
\ee
\be
\ba
\delta_{0 (\phi \phi)} ={}&- \frac{1}{8} \Lambda \phi^2
-  \frac{3}{32} \phi \bar{\square}\phi \,,
\\ \\
\delta_{0 (AA )} ={}&- \frac{1}{4} \Lambda A_{\alpha } A^{\alpha } \bar{R}
+ \frac{1}{16} A_{\alpha } A^{\alpha } \bar{R}^2
-  \Lambda A^{\alpha } A^{\beta } \bar{S}_{\alpha \beta }
+ \frac{1}{3} A^{\alpha } A^{\beta } \bar{R} \bar{S}_{\alpha \beta } 
+ A^{\alpha } A^{\beta } \bar{S}_{\alpha }{}^{\gamma } \bar{S}_{\beta \gamma }
-  \frac{1}{2} A_{\alpha } A^{\alpha } \bar{S}_{\beta \gamma } \bar{S}^{\beta \gamma }\\
& + \frac{1}{2} A^{\alpha } \bar{\nabla}_{\alpha }A^{\beta } \bar{\nabla}_{\beta }\bar{R} 
+ \frac{1}{2} A^{\alpha } A^{\beta } \bar{\nabla}_{\beta }\bar{\nabla}_{\alpha }\bar{R}
-  \Lambda A^{\alpha } \bar{\square} A_{\alpha }
+ \frac{1}{4} A^{\alpha } \bar{R} \bar{\square} A_{\alpha } 
-  A^{\alpha } \bar{S}_{\beta \gamma } \bar{\nabla}^{\gamma }\bar{\nabla}^{\beta }A_{\alpha }\,,
\\ \\
\delta_{0 (\widehat{h} \widehat{h} )} ={}&\frac{1}{2} \Lambda \widehat{h}_{\alpha \beta } \widehat{h}^{\alpha \beta }
-  \frac{1}{6} \widehat{h}_{\alpha \beta } \widehat{h}^{\alpha \beta } \bar{R}
+ \frac{1}{4} \widehat{h}^{\alpha \beta } \bar{\square} \widehat{h}_{\alpha \beta }\,,
\ea
\ee
with the mixed terms
\be
\ba
\delta_{0 (B \phi)}={}&\frac{1}{4} \Lambda B \bar{\square}\phi
-  \frac{1}{16} B \bar{R} \bar{\square}\phi
-  \frac{1}{8} B \bar{\nabla}_{\alpha }\phi \bar{\nabla}^{\alpha }\bar{R}
-  \frac{1}{4} B \bar{S}_{\alpha \beta } \bar{\nabla}^{\beta }\bar{\nabla}^{\alpha }\phi \,,
\\
\delta_{0 (BA)}={}&\Lambda A^{\alpha } B \bar{\nabla}_{\alpha }\bar{R}
-  \frac{5}{24} A^{\alpha } B \bar{R} \bar{\nabla}_{\alpha }\bar{R}
- 4 A^{\alpha } B \bar{S}^{\beta \gamma } \bar{\nabla}_{\alpha }\bar{S}_{\beta \gamma } 
-  \frac{1}{4} B \bar{\nabla}^{\alpha }\bar{R} \bar{\square} A_{\alpha }
+ 2 \Lambda B \bar{S}_{\alpha \beta } \bar{\nabla}^{\beta }A^{\alpha }
+ \frac{5}{4} B \bar{R} \bar{S}_{\alpha \beta } \bar{\nabla}^{\beta }A^{\alpha } \\
& + 11 B \bar{S}_{\alpha }{}^{\gamma } \bar{S}_{\beta \gamma } \bar{\nabla}^{\beta }A^{\alpha }
+ \frac{13}{12} A^{\alpha } B \bar{S}_{\alpha \beta } \bar{\nabla}^{\beta }\bar{R}
+ 5 A^{\alpha } B \bar{S}^{\beta \gamma } \bar{\nabla}_{\gamma }\bar{S}_{\alpha \beta } 
- 3 B \bar{S}_{\alpha \beta } \bar{\square} \bar{\nabla}^{\beta }A^{\alpha }
+ 3 B \bar{S}_{\alpha \gamma } \bar{\nabla}^{\gamma }\bar{\square} A^{\alpha }\\
& + B \bar{\nabla}_{\alpha }\bar{S}_{\beta \gamma } \bar{\nabla}^{\gamma }\bar{\nabla}^{\beta }A^{\alpha } 
+ 6 B \bar{\nabla}_{\beta }\bar{S}_{\alpha \gamma } \bar{\nabla}^{\gamma }\bar{\nabla}^{\beta }A^{\alpha }
- 6 B \bar{\nabla}_{\gamma }\bar{S}_{\alpha \beta } \bar{\nabla}^{\gamma }\bar{\nabla}^{\beta }A^{\alpha }\,,
\\
\delta_{0 (B \widehat{h})}={}&\frac{1}{3} B h^{\alpha \beta } \bar{R} \bar{S}_{\alpha \beta }
+ 2 B \widehat{h}^{\alpha \beta } \bar{S}_{\alpha }{}^{\gamma } \bar{S}_{\beta \gamma }
-  \frac{1}{2} B \bar{S}^{\alpha \beta } \bar{\square} \widehat{h}_{\alpha \beta } 
+ B \bar{\nabla}_{\beta }\bar{S}_{\alpha \gamma } \bar{\nabla}^{\gamma }\widehat{h}^{\alpha \beta }
-  \frac{1}{2} B \bar{\nabla}_{\gamma }\bar{S}_{\alpha \beta } \bar{\nabla}^{\gamma }\widehat{h}^{\alpha \beta } \,,
\\
\delta_{0 (\phi A )}={}&- \frac{1}{4} A^{\alpha } \phi \bar{\nabla}_{\alpha }\bar{R}
-  \frac{1}{2} \bar{S}_{\alpha \beta } \phi \bar{\nabla}^{\beta }A^{\alpha }\,,
\qquad \quad
\delta_{0 (\phi \widehat{h})}={}0\,,
\qquad \quad 
\delta_{0 (A  \widehat{h})}={}- A^{\alpha } \widehat{h}^{\beta \gamma } \bar{\nabla}_{\alpha }\bar{S}_{\beta \gamma }
- 2 A^{\alpha } \bar{S}^{\beta \gamma } \bar{\nabla}_{\gamma }\widehat{h}_{\alpha \beta }\,.
\ea
\ee

(iii) Imposing $\bar{\nabla}_{\alpha} \bar{S}_{\mu \nu} = 0, \partial_{\mu} \bar{R} = 0$ and substituting $\Lambda = \bar{R}/4$ which solves the background equations of motion (see eq.(\ref{lambdafrombgequations of motion})), the terms purely quadratic in one SVT mode are
\be
\ba
\delta_{0(BB)} &= {}\frac{1}{8} B \bar{R} \bar{S}_{\alpha \beta } \bar{\nabla}^{\beta }\bar{\nabla}^{\alpha }B
 + \frac{1}{2} B \bar{S}_{\alpha }{}^{\gamma } \bar{S}_{\beta \gamma } \bar{\nabla}^{\beta }\bar{\nabla}^{\alpha }B
 + \frac{1}{4} B \bar{S}_{\beta \gamma } \bar{\nabla}^{\gamma }\bar{\nabla}^{\beta }\bar{\square} B\\
\delta_{0 (\phi \phi)} &={}- \frac{1}{32} \bar{R} \phi^2
 -  \frac{3}{32} \phi \bar{\square} \phi\,,\\
\delta_{0 (AA )} &={}\frac{1}{12} A^{\alpha } A^{\beta } \bar{R} \bar{S}_{\alpha \beta }
 + A^{\alpha } A^{\beta } \bar{S}_{\alpha }{}^{\gamma } \bar{S}_{\beta \gamma }
 -  \frac{1}{2} A_{\alpha } A^{\alpha } \bar{S}_{\beta \gamma } \bar{S}^{\beta \gamma }
 -  A^{\alpha } \bar{S}_{\beta \gamma } \bar{\nabla}^{\gamma }\bar{\nabla}^{\beta }A_{\alpha }\\
\delta_{0 (\widehat{h} \widehat{h} )} &=- \frac{1}{24} \widehat{h}_{\alpha \beta } \widehat{h}^{\alpha \beta } \bar{R}
 + \frac{1}{4} \widehat{h}^{\alpha \beta } \bar{\square} \widehat{h}_{\alpha \beta }\,,
 \ea
 \ee
and the mixings are
\be
\ba
\delta_{0 (B \phi)}&={}- \frac{1}{4} B \bar{S}_{\alpha \beta } \bar{\nabla}^{\beta }\bar{\nabla}^{\alpha }\phi,  \qquad \qquad
\delta_{0 (BA)}={}\frac{1}{2} B \bar{R} \bar{S}_{\alpha \beta } \bar{\nabla}^{\beta }A^{\alpha }
 + 2 B \bar{S}_{\alpha }{}^{\gamma } \bar{S}_{\beta \gamma } \bar{\nabla}^{\beta }A^{\alpha }\,,\\
\delta_{0 (B \widehat{h})}&={}\frac{1}{3} B \widehat{h}^{\alpha \beta } \bar{R} \bar{S}_{\alpha \beta }
 + 2 B \widehat{h}^{\alpha \beta } \bar{S}_{\alpha }{}^{\gamma } \bar{S}_{\beta \gamma }
 -  \frac{1}{2} B \bar{S}^{\alpha \beta } \bar{\square} \widehat{h}_{\alpha \beta }, \qquad \qquad
\delta_{0 (\phi A )}={}- \frac{1}{2} \bar{S}_{\alpha \beta } \phi \bar{\nabla}^{\beta }A^{\alpha }\,,\\
\delta_{0 (\phi \widehat{h})}&={}0 , \qquad \qquad
\delta_{0 (A  \widehat{h})}={}-2 A^{\alpha } \bar{S}^{\beta \gamma } \bar{\nabla}_{\gamma }\widehat{h}_{\alpha \beta}\,.
\ea
\ee


\subsection{$\displaystyle \delta^2 \mathcal{L}_{R^2}  = \delta^2 \left[ \frac{1}{2} \sqrt{-g} \ R \Fc_1(\square_{s}) R \right]$} \label{appendixricciscalarsquare}

We need to compute $\delta^2 \Big[ \sqrt{-g} \ R \Fc_1(\square_{s}) R \Big]$ given in eq.(\ref{riccisquare}).

(i) We just need to find perturbed Ricci scalars which, for an \textit{arbitrary} background, are
\be
\ba
r_{(\phi)} ={}\frac{1}{4} \bar{R} \phi
+ \frac{3}{4} \bar{\square}\phi \,,
\qquad \qquad 
r_{(B)}  ={}\frac{1}{2} \bar{\nabla}_{\alpha }\bar{R} \bar{\nabla}^{\alpha }B\,,
\qquad \qquad 
r_{(A)}  ={}A^{\alpha } \bar{\nabla}_{\alpha }\bar{R}\,,
\qquad \qquad 
r_{(\widehat{h})}  = -\widehat{h}^{\alpha \beta } \bar{R}_{\alpha \beta }\,.
\ea
\ee

(ii) Rewriting the above in terms of irreducible $\bar{R}, \bar{S}_{\mu \nu}, \bar{C}_{\mu \nu \rho \sigma}$ and imposing $\bar{C}_{\mu \nu \rho \sigma} = 0$ gives
\be
\ba
r_{(\phi)} ={}\frac{1}{4} \bar{R} \phi
+ \frac{3}{4} \bar{\square}\phi \,,
\qquad \qquad 
r_{(B)}  ={}\frac{1}{2} \bar{\nabla}_{\alpha }\bar{R} \bar{\nabla}^{\alpha }B\,,
\qquad \qquad 
r_{(A)}  ={}A^{\alpha } \bar{\nabla}_{\alpha }\bar{R}\,,
\qquad \qquad 
r_{(\widehat{h})} ={}- \widehat{h}^{\alpha \beta } \bar{S}_{\alpha \beta }\,.
\ea
\ee

(iii) Imposing $\bar{\nabla}_{\alpha} \bar{S}_{\mu \nu} = 0, \partial_{\mu} \bar{R} = 0$ to above gives
\be
\ba
r_{(\phi)} ={}\frac{1}{4} \bar{R} \phi + \frac{3}{4} \bar{\square}\phi \,,\qquad \qquad
r_{(B)}  ={}0 \,,\qquad \qquad
r_{(A)}  ={}0 \,,\qquad \qquad
r_{(\widehat{h})}   ={}-\widehat{h}^{\alpha \beta } \bar{S}_{\alpha \beta }\,.
\ea
\ee
Finally, the non-zero terms are
\be\label{C14}
\ba
(0101)_{\phi \phi} &= \left( \frac{1}{4} \bar{R} \phi + \frac{3}{4} \bar{\square}\phi \right)  \Fc_1(\bar{\square}_{s}) \left( \frac{1}{4} \bar{R} \phi + \frac{3}{4} \bar{\square}\phi \right)\,,\\
(0101)_{\widehat{h} \widehat{h}} &= \left(\widehat{h}^{\alpha \beta } \bar{S}_{\alpha \beta } \right) \Fc_1(\bar{\square}_{s}) \left( \widehat{h}_{\mu \nu } \bar{S}^{\mu \nu } \right)\,,\\
(0101)_{\phi \widehat{h}} &= \left( - \frac{1}{2} \bar{R} \phi - \frac{3}{2} \bar{\square}\phi \right)  \Fc_1(\bar{\square}_{s}) \left( \widehat{h}^{\alpha \beta } \bar{S}_{\alpha \beta } \right)\,.
\ea
\ee
All other SVT perturbations are zero. The final expressions for $\delta^{2} \mathcal{L}_{R^{2}}$ which use the above expressions are given in eq.(\ref{riccisquarefinal}), eq.(\ref{riccisquaremixterm}).


\subsection{ $ \displaystyle \delta^2 \mathcal{L}_{S^2} =  \delta^2 \left[ \frac{1}{2} \sqrt{-g} \ S^{\nu}_{\ \mu} \Fc_2 ( \square_{s} ) S^{\mu}_{\ \nu} \right]$} \label{appendixtfriccisquare}

$\delta^2 \left[ \sqrt{-g} \ S^{\nu}_{\ \mu} \Fc_2 ( \square_{s} ) S^{\mu}_{\ \nu} \right]$ has perhaps the most complicated terms eq.(\ref{listofsecondvariationstfricci}). We present some of them here. We need $s^{\nu}_{\ \mu}$, which in terms of $h_{\mu \nu}$ around an \textit{arbitrary} background is
\be \label{snumu}
\ba
s^{\nu}_{\ \mu} &= \frac{1}{4} \delta_{\mu }{}^{\nu } h^{\alpha \beta } \bar{R}_{\alpha \beta } 
+ \frac{1}{2} h^{\nu \alpha } \bar{R}_{\mu \alpha } 
+ \frac{1}{2} h_{\mu }{}^{\alpha } \bar{R}^{\nu }{}_{\alpha } 
-  \frac{1}{4} h_{\mu }{}^{\nu } \bar{R} 
-  h^{\alpha \beta } \bar{R}_{\mu \alpha }{}^{\nu }{}_{\beta } 
-  h^{\nu \alpha } \bar{S}_{\mu \alpha } 
-  \frac{1}{2} \bar{\square} h_{\mu }{}^{\nu }\\
&\qquad -  \frac{1}{4} \delta_{\mu }{}^{\nu } \bar{\nabla}_{\beta }\bar{\nabla}_{\alpha }h^{\alpha \beta }
+ \frac{1}{4} \delta_{\mu }{}^{\nu } \bar{\square} h 
+ \frac{1}{2} \bar{\nabla}_{\mu }\bar{\nabla}_{\alpha }h^{\nu \alpha } 
+ \frac{1}{2} \bar{\nabla}^{\nu }\bar{\nabla}_{\alpha }h_{\mu }{}^{\alpha } 
-  \frac{1}{2} \bar{\nabla}^{\nu }\bar{\nabla}_{\mu }h\,,
\ea
\ee
(i) Its SVT decomposition is
\be
\ba
s^{\nu}_{\ \mu} &= \frac{1}{4} \delta_{\mu }{}^{\nu } \widehat{h}^{\alpha \beta } \bar{R}_{\alpha \beta } 
+ \frac{1}{2} \widehat{h}^{\nu \alpha } \bar{R}_{\mu \alpha } 
+ \frac{1}{2} \widehat{h}_{\mu }{}^{\alpha } \bar{R}^{\nu }{}_{\alpha } 
-  \frac{1}{4} \widehat{h}_{\mu }{}^{\nu } \bar{R} 
-  \widehat{h}^{\alpha \beta } \bar{R}_{\mu \alpha }{}^{\nu }{}_{\beta } 
-  \widehat{h}^{\nu \alpha } \bar{S}_{\mu \alpha } 
+ \frac{1}{4} \bar{S}_{\mu }{}^{\nu } \phi 
+ A^{\alpha } \bar{\nabla}_{\alpha }\bar{R}_{\mu }{}^{\nu } \\
&\qquad -  \frac{1}{4} A^{\alpha } \delta_{\mu }{}^{\nu } \bar{\nabla}_{\alpha }\bar{R} 
-  \frac{1}{2} \bar{\square} \widehat{h}_{\mu }{}^{\nu } 
-  \frac{1}{16} \delta_{\mu }{}^{\nu } \bar{\square} \phi 
-  \bar{S}_{\mu \alpha } \bar{\nabla}^{\alpha }A^{\nu } 
+ \frac{1}{2} \bar{\nabla}_{\alpha }\bar{R}_{\mu }{}^{\nu } \bar{\nabla}^{\alpha }B 
-  \frac{1}{8} \delta_{\mu }{}^{\nu } \bar{\nabla}_{\alpha }\bar{R} \bar{\nabla}^{\alpha }B \\
&\qquad+ \bar{R}_{\mu \alpha }{}^{\nu }{}_{\beta } \bar{\nabla}^{\beta }\bar{\nabla}^{\alpha }B 
-  \bar{R}_{\mu \beta }{}^{\nu }{}_{\alpha } \bar{\nabla}^{\beta }\bar{\nabla}^{\alpha }B 
+ \bar{R}^{\nu }{}_{\alpha } \bar{\nabla}_{\mu }A^{\alpha } 
-  \frac{1}{4} \bar{R} \bar{\nabla}_{\mu }A^{\nu } 
+ \frac{1}{2} \bar{R}^{\nu }{}_{\alpha } \bar{\nabla}_{\mu }\bar{\nabla}^{\alpha }B 
+ \bar{R}_{\mu \alpha } \bar{\nabla}^{\nu }A^{\alpha }
-  \bar{S}_{\mu \alpha } \bar{\nabla}^{\nu }A^{\alpha } \\
&\qquad -  \frac{1}{4} \bar{R} \bar{\nabla}^{\nu }A_{\mu } 
+ \frac{1}{2} \bar{R}_{\mu \alpha } \bar{\nabla}^{\nu }\bar{\nabla}^{\alpha }B 
-  \bar{S}_{\mu \alpha } \bar{\nabla}^{\nu }\bar{\nabla}^{\alpha }B 
-  \frac{1}{4} \bar{R} \bar{\nabla}^{\nu }\bar{\nabla}_{\mu }B 
+ \frac{1}{4} \bar{\nabla}^{\nu }\bar{\nabla}_{\mu }\phi \,.
\ea
\ee
(ii) Rewriting the above in terms of irreducible $\bar{R}, \bar{S}_{\mu \nu}, \bar{C}_{\mu \nu \rho \sigma}$ and imposing $\bar{C}_{\mu \nu \rho \sigma} = 0$, we get
\be
\ba
s^{\nu}_{\ \mu}  &= \frac{1}{12} \widehat{h}_{\mu }{}^{\nu } \bar{R} 
-  \frac{1}{4} \delta_{\mu }{}^{\nu } \widehat{h}^{\alpha \beta } \bar{S}_{\alpha \beta } 
+ \widehat{h}_{\mu }{}^{\alpha } \bar{S}^{\nu }{}_{\alpha } 
+ \frac{1}{4} \bar{S}_{\mu }{}^{\nu } \phi 
+ A^{\alpha } \bar{\nabla}_{\alpha }\bar{S}_{\mu }{}^{\nu } 
-  \frac{1}{2} \bar{\square} \widehat{h}_{\mu }{}^{\nu } 
-  \frac{1}{16} \delta_{\mu }{}^{\nu } \bar{\square} \phi \\
&\qquad -  \bar{S}_{\mu \alpha } \bar{\nabla}^{\alpha }A^{\nu } 
+ \frac{1}{2} \bar{\nabla}_{\alpha }\bar{S}_{\mu }{}^{\nu } \bar{\nabla}^{\alpha }B 
+ \bar{S}^{\nu }{}_{\alpha } \bar{\nabla}_{\mu }A^{\alpha } 
+ \frac{1}{2} \bar{S}^{\nu }{}_{\alpha } \bar{\nabla}_{\mu }\bar{\nabla}^{\alpha }B 
-  \frac{1}{2} \bar{S}_{\mu \alpha } \bar{\nabla}^{\nu }\bar{\nabla}^{\alpha }B 
+ \frac{1}{4} \bar{\nabla}^{\nu }\bar{\nabla}_{\mu }\phi\,.
\ea
\ee
(iii) Imposing $\bar{\nabla}_{\alpha} \bar{S}_{\mu \nu} = 0, \partial_{\mu} \bar{R} = 0$ gives
\be
\ba
s^{\nu}_{\ \mu} &= \frac{1}{12} \widehat{h}_{\mu }{}^{\nu } \bar{R} 
-  \frac{1}{4} \delta_{\mu }{}^{\nu } \widehat{h}^{\alpha \beta } \bar{S}_{\alpha \beta } 
+ \widehat{h}_{\mu }{}^{\alpha } \bar{S}^{\nu }{}_{\alpha } 
+ \frac{1}{4} \bar{S}_{\mu }{}^{\nu } \phi 
-  \frac{1}{2} \bar{\square} \widehat{h}_{\mu }{}^{\nu } 
-  \frac{1}{16} \delta_{\mu }{}^{\nu } \bar{\square} \phi 
-  \bar{S}_{\mu \alpha } \bar{\nabla}^{\alpha }A^{\nu } \\
&\qquad + \bar{S}^{\nu }{}_{\alpha } \bar{\nabla}_{\mu }A^{\alpha } 
+ \frac{1}{2} \bar{S}^{\nu }{}_{\alpha } \bar{\nabla}_{\mu }\bar{\nabla}^{\alpha }B 
-  \frac{1}{2} \bar{S}_{\mu \alpha } \bar{\nabla}^{\nu }\bar{\nabla}^{\alpha }B 
+ \frac{1}{4} \bar{\nabla}^{\nu }\bar{\nabla}_{\mu }\phi \,.
\ea
\ee
Let us now compute the SVT decomposition for $(0101)$.
\be
\ba
(0101) = \ s^{\nu}_{\ \mu}  \cF_2(\bar{\square}_{s}) s^{\mu}_{\ \nu} &=  \Bigg[ \frac{1}{12} \widehat{h}_{\mu }{}^{\nu } \bar{R} 
-  \frac{1}{4} \delta_{\mu }{}^{\nu } \widehat{h}^{\alpha \beta } \bar{S}_{\alpha \beta } 
+ \widehat{h}_{\mu }^{\ \alpha } \bar{S}^{\nu }{}_{\alpha } 
+ \frac{1}{4} \bar{S}_{\mu }{}^{\nu } \phi 
-  \frac{1}{2} \bar{\square} \widehat{h}_{\mu }{}^{\nu } 
-  \frac{1}{16} \delta_{\mu }{}^{\nu } \bar{\square} \phi \\
& \qquad -  \bar{S}_{\mu \alpha } \bar{\nabla}^{\alpha }A^{\nu } 
+ \bar{S}^{\nu }{}_{\alpha } \bar{\nabla}_{\mu }A^{\alpha } 
+ \frac{1}{2} \bar{S}^{\nu }{}_{\alpha } \bar{\nabla}_{\mu }\bar{\nabla}^{\alpha }B 
-  \frac{1}{2} \bar{S}_{\mu \alpha } \bar{\nabla}^{\nu }\bar{\nabla}^{\alpha }B 
+ \frac{1}{4} \bar{\nabla}^{\nu }\bar{\nabla}_{\mu }\phi \Bigg] \cF_2(\bar{\square}_{s}) s^{\mu}_{\ \nu}\,, \\
&= \Bigg[ \frac{1}{12} \widehat{h}_{\mu }{}^{\nu } \bar{R}  
+ \widehat{h}_{\mu }{}^{\alpha } \bar{S}^{\nu }{}_{\alpha } 
+ \frac{1}{4} \bar{S}_{\mu }{}^{\nu } \phi 
-  \frac{1}{2} \bar{\square} \widehat{h}_{\mu }{}^{\nu } 
-  \bar{S}_{\mu \alpha } \bar{\nabla}^{\alpha }A^{\nu } 
+ \bar{S}^{\nu }{}_{\alpha } \bar{\nabla}_{\mu }A^{\alpha } \\
& \qquad + \frac{1}{2} \bar{S}^{\nu }{}_{\alpha } \bar{\nabla}_{\mu }\bar{\nabla}^{\alpha }B 
-  \frac{1}{2} \bar{S}_{\mu \alpha } \bar{\nabla}^{\nu }\bar{\nabla}^{\alpha }B 
+ \frac{1}{4} \bar{\nabla}^{\nu }\bar{\nabla}_{\mu }\phi \Bigg] \cF_2(\bar{\square}_{s}) s^{\mu}_{\ \nu} \,,
\ea
\ee
where in the second equality, we canceled the terms to the left (only left, not right) of $\cF_2(\bar{\square}_{s})$ which contain Kronecker deltas $\delta_{\mu }{}^{\nu }$ because they hit $s^{\mu}_{\ \nu}$ to the right of $\cF_2(\bar{\square}_{s})$ and vanish because of tracelessness. Then, we similarly expand $s^{\mu}_{\ \nu}$ on the right, integrate by parts and get the following expressions for purely quadratic in one SVT mode
\be
\ba
(0101)_{\phi \phi} &= \phi \Bigg[ \frac{1}{16}  \cF_2(\bar{\square}_{s}) \bar{S}_{\mu }{}^{\nu }  \bar{S}^{\mu }{}_{\nu }  
+  \frac{1}{8}  \cF_2(\bar{\square}_{s}) \bar{S}_{\mu }{}^{\nu } \bar{\nabla}_{\nu }\bar{\nabla}^{\mu }  
+\frac{1}{16}  \bar{\nabla}_{\mu } \bar{\nabla}^{\nu } \cF_2(\bar{\square}_{s}) \bar{\nabla}_{\nu }\bar{\nabla}^{\mu } 
- \frac{1}{64}  \cF_2(\bar{\square}_{s}) \bar{\square}^{2} 
\Bigg] \phi\,,\\
(0101)_{BB} &= B \Bigg[ 
\frac{1}{2} \bar{\nabla}^{\alpha }  \bar{\nabla}_{\mu }
\cF_2(\bar{\square}_{s})
\bar{S}^{\nu }{}_{\alpha }  \bar{S}^{\mu }{}_{\rho } \bar{\nabla}_{\nu }\bar{\nabla}^{\rho }
-\frac{1}{2} \bar{\nabla}^{\alpha } \bar{\nabla}_{\mu }
\cF_2(\bar{\square}_{s})
\bar{S}^{\nu }{}_{\alpha }  \bar{S}_{\nu \rho } \bar{\nabla}^{\mu }\bar{\nabla}^{\rho } 
\Bigg] B\,,\\
(0101)_{ AA}&= A^{\nu} \Bigg[
-\bar{\nabla}^{\alpha } \cF_2(\bar{\square}_{s})
\bar{S}_{\rho \alpha }  \bar{S}^{\rho }{}_{\mu } \bar{\nabla}_{\nu } 
-  \bar{\nabla}_{\mu } \cF_2(\bar{\square}_{s})
\bar{S}_{\lambda \rho }  \bar{S}^{\lambda }{}_{\nu } \bar{\nabla}^{\rho } 
+  2 \bar{\nabla}^{\alpha } \cF_2(\bar{\square}_{s})
\bar{S}_{\mu \alpha } \bar{S}_{\nu \rho } \bar{\nabla}^{\rho }
\Bigg] A^{\mu}\,,\\
(0101)_{ \widehat{h} \widehat{h}}&=  \widehat{h}_{\mu \nu} \Bigg[
\frac{1}{144}  
\cF_2(\bar{\square}_{s}) \bar{R}^{2} \bar{g}^{\rho \mu} \bar{g}^{\sigma \nu}  
+  \frac{1}{6} 
\cF_2(\bar{\square}_{s}) \bar{R} \bar{S}^{\mu \rho}  \bar{g}^{\nu \sigma} 
-  \frac{1}{12}  
\cF_2(\bar{\square}_{s}) \bar{R} \bar{\square} \bar{g}^{\rho \mu} \bar{g}^{\sigma \nu}  \\
&\qquad \qquad +     \cF_2(\bar{\square}_{s})  \bar{S}^{\sigma \nu }  \bar{S}^{\mu \rho } 
-   \cF_2(\bar{\square}_{s})  \bar{S}^{ \sigma \nu }  \bar{\square} \bar{g}^{\mu \rho} 
+ \frac{1}{4} 
\cF_2(\bar{\square}_{s}) \bar{\square}^{2}  \bar{g}^{\rho \mu} \bar{g}^{\sigma \nu}
- \frac{1}{4}  \cF_2(\bar{\square}_{s}) \bar{S}^{\mu \nu} \bar{S}^{\rho \sigma} 
\Bigg] \widehat{h}_{\rho \sigma}\,,
\ea
\ee
and the mixings are
\be
\ba
(0101)_{ \phi B}&=  0\,,\\
(0101)_{ \phi A}&=  A^{\mu} \left[ \frac{1}{2} \bar{\nabla}^{\alpha } \cF_2(\bar{\square}_{s}) \bar{S}_{\nu \alpha } \bar{\nabla}^{\nu }\bar{\nabla}_{\mu } 
- \frac{1}{2} \bar{\nabla}_{\nu } \cF_2(\bar{\square}_{s})  \bar{S}^{\alpha }{}_{\mu } \bar{\nabla}^{\nu }\bar{\nabla}_{\alpha }
\right] \phi\,,\\
(0101)_{ \phi \widehat{h}}&=  \widehat{h}_{\mu \nu}  \Bigg[ \frac{1}{24}  \bar{R}  \cF_2(\bar{\square}_{s})  \bar{S}^{\mu \nu } 
+ \frac{1}{24}  \bar{R}  \cF_2(\bar{\square}_{s})  \bar{\nabla}^{\nu }\bar{\nabla}^{\mu }
+ \frac{1}{2}  \bar{S}^{\sigma \mu}  \cF_2(\bar{\square}_{s})  \bar{S}_{\sigma }{}^{\nu }
+ \frac{1}{2}  \bar{S}^{ \sigma \mu }  \cF_2(\bar{\square}_{s})  \bar{\nabla}^{\nu }\bar{\nabla}_{\sigma }\\
&\qquad \qquad -  \frac{1}{4} \bar{\square}  \cF_2(\bar{\square}_{s})   \bar{S}^{\mu \nu }
-  \frac{1}{4} \bar{\square}  \cF_2(\bar{\square}_{s})  \bar{\nabla}^{\nu }\bar{\nabla}^{\mu }
- \frac{1}{8}  \bar{S}^{\mu \nu}  \cF_2(\bar{\square}_{s}) \bar{\square} 
\Bigg] \phi\,,\\
(0101)_{ BA}&= A^{\mu}  \Bigg[\bar{\nabla}^{\alpha } \cF_2(\bar{\square}_{s}) \bar{S}_{\nu \alpha }  \bar{S}^{\nu }{}_{\rho } \bar{\nabla}_{\mu }\bar{\nabla}^{\rho }
- 2 \bar{\nabla}^{\alpha } \cF_2(\bar{\square}_{s}) \bar{S}_{\nu \alpha }  \bar{S}_{\mu \rho } \bar{\nabla}^{\nu }\bar{\nabla}^{\rho }
+  \bar{\nabla}_{\nu } \cF_2(\bar{\square}_{s}) \bar{S}^{\sigma }{}_{\mu }\bar{S}_{\sigma \rho } \bar{\nabla}^{\nu }\bar{\nabla}^{\rho }
\Bigg] B\,,\\
(0101)_{ B \widehat{h}}&= \widehat{h}_{\mu \nu} \Bigg[ 
-   \cF_2(\bar{\square}_{s}) \bar{S}^{\sigma \nu } \bar{S}_{\sigma \rho } \bar{\nabla}^{\mu }\bar{\nabla}^{\rho }
+   \cF_2(\bar{\square}_{s}) \bar{S}^{ \sigma \nu  } \bar{S}^{\mu }{}_{\rho } \bar{\nabla}_{\sigma }\bar{\nabla}^{\rho }
\Bigg] B\,,\\
(0101)_{ A \widehat{h}}&=  \widehat{h}_{\mu \nu} \Bigg[ -\frac{1}{6}   \cF_2(\bar{\square}_{s})  \bar{R} \bar{S}^{\nu}_{\ \rho } \bar{\nabla}^{\rho } \delta^{\mu}_{\ \lambda} 
+\frac{1}{6}  \cF_2(\bar{\square}_{s})  \bar{R} \bar{S}^{\mu }{}_{\rho } \bar{\nabla}^{\nu } \delta^{\rho}_{\ \lambda}
- 2  \cF_2(\bar{\square}_{s})  \bar{S}^{ \sigma \nu} \bar{S}_{\sigma \rho } \bar{\nabla}^{\rho } \delta^{\mu}_{\ \lambda} \\
&\qquad \qquad + 2  \cF_2(\bar{\square}_{s})  \bar{S}^{\sigma \nu  }\bar{S}^{\mu }{}_{\rho } \bar{\nabla}_{\sigma } \delta^{\rho}_{\ \lambda}
+    \cF_2(\bar{\square}_{s})   \bar{\square} \bar{S}^{\nu}_{\ \rho } \bar{\nabla}^{\rho } \delta^{\mu}_{\ \lambda}
-    \cF_2(\bar{\square}_{s})   \bar{\square} \bar{S}^{\mu }{}_{\rho } \bar{\nabla}^{\nu } \delta^{\rho}_{\ \lambda}
\Bigg] A^{\lambda}\,.
\ea
\ee
We now consider the perturbation $(1001)$ in SVT form. 
\be
\ba
(1001) = \frac{1}{2} \ \bar{S}^{\nu}_{\ \mu} s^{\mu}_{\ \nu} \cF_2(\bar{\square}_{s}) h &= 
\frac{1}{2} \ \bar{S}^{\nu}_{\ \mu}
\Bigg[
\frac{1}{12} \widehat{h}^{\mu }{}_{\nu } \bar{R} 
+ \widehat{h}_{\nu }{}^{\alpha } \bar{S}^{\mu }{}_{\alpha } 
+ \frac{1}{4} \bar{S}^{\mu }{}_{\nu } \phi 
-  \frac{1}{2} \bar{\square} \widehat{h}^{\mu }{}_{\nu } 
 -  \bar{S}_{\nu \alpha } \bar{\nabla}^{\alpha }A^{\mu } 
-  \frac{1}{2} \bar{S}_{\nu \alpha } \bar{\nabla}^{\mu }\bar{\nabla}^{\alpha }B \\
&\qquad \qquad \qquad + \bar{S}^{\mu }{}_{\alpha } \bar{\nabla}_{\nu }A^{\alpha } 
+ \frac{1}{2} \bar{S}^{\mu }{}_{\alpha } \bar{\nabla}_{\nu }\bar{\nabla}^{\alpha }B 
+ \frac{1}{4} \bar{\nabla}_{\nu }\bar{\nabla}^{\mu }\phi 
\Bigg]
\cF_2 (\bar{\square}_{s} )
\left[ \bar{\square} B - \phi \right]\,,
\ea
\ee
from which we get the terms with purely quadratic in one SVT mode
\be
\ba
(1001)_{\phi \phi} &= \phi
\Bigg[
-\frac{1}{8} \ \bar{S}^{\nu}_{\ \mu} \bar{S}^{\mu }{}_{\nu } \cF_2 (\bar{\square}_{s} )
- \frac{1}{8}  \bar{\nabla}^{\mu } \bar{\nabla}_{\nu } \bar{S}^{\nu}_{\ \mu} \cF_2 (\bar{\square}_{s} )
\Bigg]\phi\,,\\
(1001)_{BB}&= B
\Bigg[
-\frac{1}{4} \bar{S}^{\nu}_{\ \mu} \bar{S}_{\nu \alpha } \bar{\nabla}^{\alpha } \bar{\nabla}^{\mu } \cF_2 (\bar{\square}_{s} )\bar{\square} 
+ \frac{1}{4} \bar{S}^{\nu}_{\ \mu} \bar{S}^{\mu }{}_{\alpha } \bar{\nabla}^{\alpha }  \bar{\nabla}_{\nu } \cF_2 (\bar{\square}_{s} )\bar{\square} 
\Bigg]B\,,
\ea
\ee
and the mixings
\be
\ba
(1001)_{\phi B}&= B
\Bigg[
\frac{1}{8} \bar{S}^{\nu}_{\ \mu}  \bar{S}^{\mu }{}_{\nu } \cF_2 (\bar{\square}_{s} ) \bar{\square} 
+ \frac{1}{8} \bar{S}^{\nu}_{\ \mu}  \cF_2 (\bar{\square}_{s} ) \bar{\square} \bar{\nabla}_{\nu } \bar{\nabla}^{\mu } 
\Bigg]
\phi\,,\\
(1001)_{\phi A}&= 
A^{\mu }
\Bigg[
-  \frac{1}{2} \bar{S}^{\nu}_{\ \mu} \bar{S}_{\nu \alpha } \bar{\nabla}^{\alpha } \cF_2 (\bar{\square}_{s} )
+ \frac{1}{2} \bar{S}^{\nu}_{\ \rho} \bar{S}^{\rho }{}_{\mu } \bar{\nabla}_{\nu }  \cF_2 (\bar{\square}_{s} )
\Bigg]
\phi\,,\\
(1001)_{\phi \widehat{h}} &= \widehat{h}_{\mu \nu }
\Bigg[
-\frac{1}{24} \bar{S}^{\mu \nu}  \bar{R}  \cF_2 (\bar{\square}_{s} )
- \frac{1}{2} \bar{S}^{\nu}_{\ \rho} \bar{S}^{\rho \mu }  \cF_2 (\bar{\square}_{s} )
+ \frac{1}{4} \bar{S}^{\mu \nu}  \cF_2 (\bar{\square}_{s} ) \bar{\square}
\Bigg]
\phi\,,\\
(1001)_{B A} &= A^{\mu }
\Bigg[
 \frac{1}{2} \bar{S}^{\nu}_{\ \mu} \bar{S}_{\nu \alpha }  \bar{\nabla}^{\alpha } \cF_2 (\bar{\square}_{s} ) \bar{\square}
- \frac{1}{2} \bar{S}^{\nu}_{\ \rho} \bar{S}^{\rho }{}_{\mu } \bar{\nabla}_{\nu } \cF_2 (\bar{\square}_{s} ) \bar{\square}
\Bigg]
B\,,\\
(1001)_{B \widehat{h}} &=  \widehat{h}_{\mu  \nu}
\Bigg[
\frac{1}{24} \bar{S}^{\mu \nu}  \bar{R}  \cF_2 (\bar{\square}_{s} ) \bar{\square}
+ \frac{1}{2} \bar{S}^{\nu}_{\ \rho}  \bar{S}^{\rho \mu }  \cF_2 (\bar{\square}_{s} ) \bar{\square}
-  \frac{1}{4} \bar{S}^{\mu  \nu}   \cF_2 (\bar{\square}_{s} ) \bar{\square}^{2}
\Bigg] B\,.
\ea
\ee


\subsection{$ \displaystyle \delta^2 \mathcal{L}_{C^2} =  \delta^2 \left[ \frac{1}{2} \sqrt{-g} \ C^{\rho \sigma}_{\ \ \ \mu \nu}   \Fc_3(\square_{s})   C^{\mu \nu}_{\ \ \ \rho \sigma} \right]$} \label{appendixweylsquare}

We now compute $\delta^2 \left[  \sqrt{-g} \ C^{\rho \sigma}_{\ \ \ \mu \nu}   \Fc_3 (\square_{s})  C^{\mu \nu}_{\ \ \ \rho \sigma} \right]$. We only present the term $(0101) = \sqrt{-\bar{g}}  \  c^{\rho \sigma}_{\ \ \mu \nu}  \mathcal{\bar{F}}_3 (\bar{\square}_{s})  c^{\mu \nu}_{\ \ \rho \sigma}$ which is relevant for our background (see eq.(\ref{weylpert})). Other terms, for an \textit{arbitrary} background, can be computed from Appendix (\ref{appendixperturbations}) but we do not present them here because they are not very enlightening.

(i) Performing SVT decomposition of $c^{\rho \sigma}_{\ \ \mu \nu}$ for an \textit{arbitrary} background, we get
\be
\ba
c^{\rho \sigma}_{\ \ \ \mu \nu (\phi)} ={}&\frac{1}{8} \delta_{\nu }{}^{\sigma } \bar{R}_{\mu }{}^{\rho } \phi
-  \frac{1}{8} \delta_{\nu }{}^{\rho } \bar{R}_{\mu }{}^{\sigma } \phi
-  \frac{1}{8} \delta_{\mu }{}^{\sigma } \bar{R}_{\nu }{}^{\rho } \phi
+ \frac{1}{8} \delta_{\mu }{}^{\rho } \bar{R}_{\nu }{}^{\sigma } \phi
+ \frac{1}{24} \delta_{\mu }{}^{\sigma } \delta_{\nu }{}^{\rho } \bar{R} \phi 
-  \frac{1}{24} \delta_{\mu }{}^{\rho } \delta_{\nu }{}^{\sigma } \bar{R} \phi
-  \frac{1}{4} \bar{R}_{\mu \nu }{}^{\rho \sigma } \phi \,,
\ea
\ee
\be
\ba
c^{\rho \sigma}_{\ \ \ \mu \nu (B)} ={}&- \frac{1}{4} \delta_{\nu }{}^{\sigma } \bar{\nabla}_{\alpha }\bar{R}_{\mu }{}^{\rho } \bar{\nabla}^{\alpha }B
+ \frac{1}{4} \delta_{\nu }{}^{\rho } \bar{\nabla}_{\alpha }\bar{R}_{\mu }{}^{\sigma } \bar{\nabla}^{\alpha }B
+ \frac{1}{4} \delta_{\mu }{}^{\sigma } \bar{\nabla}_{\alpha }\bar{R}_{\nu }{}^{\rho } \bar{\nabla}^{\alpha }B 
-  \frac{1}{4} \delta_{\mu }{}^{\rho } \bar{\nabla}_{\alpha }\bar{R}_{\nu }{}^{\sigma } \bar{\nabla}^{\alpha }B
-  \frac{1}{12} \delta_{\mu }{}^{\sigma } \delta_{\nu }{}^{\rho } \bar{\nabla}_{\alpha }\bar{R} \bar{\nabla}^{\alpha }B\\
& + \frac{1}{12} \delta_{\mu }{}^{\rho } \delta_{\nu }{}^{\sigma } \bar{\nabla}_{\alpha }\bar{R} \bar{\nabla}^{\alpha }B 
-  \frac{1}{2} \delta_{\nu }{}^{\sigma } \bar{R}_{\mu \alpha }{}^{\rho }{}_{\beta } \bar{\nabla}^{\beta }\bar{\nabla}^{\alpha }B
+ \frac{1}{2} \delta_{\nu }{}^{\rho } \bar{R}_{\mu \alpha }{}^{\sigma }{}_{\beta } \bar{\nabla}^{\beta }\bar{\nabla}^{\alpha }B
+ \frac{1}{2} \delta_{\nu }{}^{\sigma } \bar{R}_{\mu \beta }{}^{\rho }{}_{\alpha } \bar{\nabla}^{\beta }\bar{\nabla}^{\alpha }B \\
& -  \frac{1}{2} \delta_{\nu }{}^{\rho } \bar{R}_{\mu \beta }{}^{\sigma }{}_{\alpha } \bar{\nabla}^{\beta }\bar{\nabla}^{\alpha }B
+ \frac{1}{2} \delta_{\mu }{}^{\sigma } \bar{R}_{\nu \alpha }{}^{\rho }{}_{\beta } \bar{\nabla}^{\beta }\bar{\nabla}^{\alpha }B
-  \frac{1}{2} \delta_{\mu }{}^{\rho } \bar{R}_{\nu \alpha }{}^{\sigma }{}_{\beta } \bar{\nabla}^{\beta }\bar{\nabla}^{\alpha }B 
-  \frac{1}{2} \delta_{\mu }{}^{\sigma } \bar{R}_{\nu \beta }{}^{\rho }{}_{\alpha } \bar{\nabla}^{\beta }\bar{\nabla}^{\alpha }B\\
&+ \frac{1}{2} \delta_{\mu }{}^{\rho } \bar{R}_{\nu \beta }{}^{\sigma }{}_{\alpha } \bar{\nabla}^{\beta }\bar{\nabla}^{\alpha }B
-  \frac{1}{4} \delta_{\nu }{}^{\sigma } \bar{R}^{\rho }{}_{\alpha } \bar{\nabla}_{\mu }\bar{\nabla}^{\alpha }B 
+ \frac{1}{4} \delta_{\nu }{}^{\rho } \bar{R}^{\sigma }{}_{\alpha } \bar{\nabla}_{\mu }\bar{\nabla}^{\alpha }B
+ \frac{1}{2} \bar{R}_{\nu }{}^{\rho \sigma }{}_{\alpha } \bar{\nabla}_{\mu }\bar{\nabla}^{\alpha }B
-  \frac{1}{2} \bar{R}_{\nu }{}^{\sigma \rho }{}_{\alpha } \bar{\nabla}_{\mu }\bar{\nabla}^{\alpha }B \\
& + \frac{1}{4} \delta_{\mu }{}^{\sigma } \bar{R}^{\rho }{}_{\alpha } \bar{\nabla}_{\nu }\bar{\nabla}^{\alpha }B
-  \frac{1}{4} \delta_{\mu }{}^{\rho } \bar{R}^{\sigma }{}_{\alpha } \bar{\nabla}_{\nu }\bar{\nabla}^{\alpha }B
-  \frac{1}{2} \bar{R}_{\mu }{}^{\rho \sigma }{}_{\alpha } \bar{\nabla}_{\nu }\bar{\nabla}^{\alpha }B 
+ \frac{1}{2} \bar{R}_{\mu }{}^{\sigma \rho }{}_{\alpha } \bar{\nabla}_{\nu }\bar{\nabla}^{\alpha }B
-  \frac{1}{2} \bar{\nabla}^{\alpha }B \bar{\nabla}^{\rho }\bar{R}_{\mu \nu }{}^{\sigma }{}_{\alpha }\\
& -  \frac{1}{4} \delta_{\nu }{}^{\sigma } \bar{R}_{\mu \alpha } \bar{\nabla}^{\rho }\bar{\nabla}^{\alpha }B 
+ \frac{1}{4} \delta_{\mu }{}^{\sigma } \bar{R}_{\nu \alpha } \bar{\nabla}^{\rho }\bar{\nabla}^{\alpha }B
-  \frac{1}{2} \bar{R}_{\mu \alpha \nu }{}^{\sigma } \bar{\nabla}^{\rho }\bar{\nabla}^{\alpha }B
-  \bar{R}_{\mu \nu }{}^{\sigma }{}_{\alpha } \bar{\nabla}^{\rho }\bar{\nabla}^{\alpha }B 
+ \frac{1}{2} \bar{R}_{\mu }{}^{\sigma }{}_{\nu \alpha } \bar{\nabla}^{\rho }\bar{\nabla}^{\alpha }B\\
& -  \frac{1}{2} \bar{R}_{\nu }{}^{\sigma } \bar{\nabla}^{\rho }\bar{\nabla}_{\mu }B
+ \frac{1}{6} \delta_{\nu }{}^{\sigma } \bar{R} \bar{\nabla}^{\rho }\bar{\nabla}_{\mu }B
+ \frac{1}{2} \bar{R}_{\mu }{}^{\sigma } \bar{\nabla}^{\rho }\bar{\nabla}_{\nu }B 
-  \frac{1}{6} \delta_{\mu }{}^{\sigma } \bar{R} \bar{\nabla}^{\rho }\bar{\nabla}_{\nu }B
+ \frac{1}{2} \bar{\nabla}^{\alpha }B \bar{\nabla}^{\sigma }\bar{R}_{\mu \nu }{}^{\rho }{}_{\alpha }
+ \frac{1}{4} \delta_{\nu }{}^{\rho } \bar{R}_{\mu \alpha } \bar{\nabla}^{\sigma }\bar{\nabla}^{\alpha }B \\
& -  \frac{1}{4} \delta_{\mu }{}^{\rho } \bar{R}_{\nu \alpha } \bar{\nabla}^{\sigma }\bar{\nabla}^{\alpha }B
+ \frac{1}{2} \bar{R}_{\mu \alpha \nu }{}^{\rho } \bar{\nabla}^{\sigma }\bar{\nabla}^{\alpha }B
+ \bar{R}_{\mu \nu }{}^{\rho }{}_{\alpha } \bar{\nabla}^{\sigma }\bar{\nabla}^{\alpha }B 
-  \frac{1}{2} \bar{R}_{\mu }{}^{\rho }{}_{\nu \alpha } \bar{\nabla}^{\sigma }\bar{\nabla}^{\alpha }B
+ \frac{1}{2} \bar{R}_{\nu }{}^{\rho } \bar{\nabla}^{\sigma }\bar{\nabla}_{\mu }B
-  \frac{1}{6} \delta_{\nu }{}^{\rho } \bar{R} \bar{\nabla}^{\sigma }\bar{\nabla}_{\mu }B\\
& -  \frac{1}{2} \bar{R}_{\mu }{}^{\rho } \bar{\nabla}^{\sigma }\bar{\nabla}_{\nu }B 
+ \frac{1}{6} \delta_{\mu }{}^{\rho } \bar{R} \bar{\nabla}^{\sigma }\bar{\nabla}_{\nu }B\,,
\\ \\
c^{\rho \sigma}_{\ \ \ \mu \nu (A)} ={}&- \frac{1}{2} A^{\alpha } \delta_{\nu }{}^{\sigma } \bar{\nabla}_{\alpha }\bar{R}_{\mu }{}^{\rho }
+ \frac{1}{2} A^{\alpha } \delta_{\nu }{}^{\rho } \bar{\nabla}_{\alpha }\bar{R}_{\mu }{}^{\sigma }
+ \frac{1}{2} A^{\alpha } \delta_{\mu }{}^{\sigma } \bar{\nabla}_{\alpha }\bar{R}_{\nu }{}^{\rho } 
-  \frac{1}{2} A^{\alpha } \delta_{\mu }{}^{\rho } \bar{\nabla}_{\alpha }\bar{R}_{\nu }{}^{\sigma }
-  \frac{1}{6} A^{\alpha } \delta_{\mu }{}^{\sigma } \delta_{\nu }{}^{\rho } \bar{\nabla}_{\alpha }\bar{R}\\
& + \frac{1}{6} A^{\alpha } \delta_{\mu }{}^{\rho } \delta_{\nu }{}^{\sigma } \bar{\nabla}_{\alpha }\bar{R} 
+ \frac{1}{2} \bar{R}_{\nu \alpha }{}^{\rho \sigma } \bar{\nabla}^{\alpha }A_{\mu }
+ \frac{1}{2} \bar{R}_{\nu }{}^{\rho \sigma }{}_{\alpha } \bar{\nabla}^{\alpha }A_{\mu }
-  \frac{1}{2} \bar{R}_{\nu }{}^{\sigma \rho }{}_{\alpha } \bar{\nabla}^{\alpha }A_{\mu }
-  \frac{1}{2} \bar{R}_{\mu \alpha }{}^{\rho \sigma } \bar{\nabla}^{\alpha }A_{\nu } 
-  \frac{1}{2} \bar{R}_{\mu }{}^{\rho \sigma }{}_{\alpha } \bar{\nabla}^{\alpha }A_{\nu }\\
&+ \frac{1}{2} \bar{R}_{\mu }{}^{\sigma \rho }{}_{\alpha } \bar{\nabla}^{\alpha }A_{\nu }
-  \frac{1}{2} \bar{R}_{\mu \alpha \nu }{}^{\sigma } \bar{\nabla}^{\alpha }A^{\rho }
-  \frac{1}{2} \bar{R}_{\mu \nu }{}^{\sigma }{}_{\alpha } \bar{\nabla}^{\alpha }A^{\rho }
+ \frac{1}{2} \bar{R}_{\mu }{}^{\sigma }{}_{\nu \alpha } \bar{\nabla}^{\alpha }A^{\rho }
+ \frac{1}{2} \bar{R}_{\mu \alpha \nu }{}^{\rho } \bar{\nabla}^{\alpha }A^{\sigma }
+ \frac{1}{2} \bar{R}_{\mu \nu }{}^{\rho }{}_{\alpha } \bar{\nabla}^{\alpha }A^{\sigma }\\
& -  \frac{1}{2} \bar{R}_{\mu }{}^{\rho }{}_{\nu \alpha } \bar{\nabla}^{\alpha }A^{\sigma } 
-  \frac{1}{2} \delta_{\nu }{}^{\sigma } \bar{R}^{\rho }{}_{\alpha } \bar{\nabla}_{\mu }A^{\alpha }
+ \frac{1}{2} \delta_{\nu }{}^{\rho } \bar{R}^{\sigma }{}_{\alpha } \bar{\nabla}_{\mu }A^{\alpha }
-  \frac{1}{2} \bar{R}_{\nu \alpha }{}^{\rho \sigma } \bar{\nabla}_{\mu }A^{\alpha }
+ \frac{1}{2} \bar{R}_{\nu }{}^{\rho \sigma }{}_{\alpha } \bar{\nabla}_{\mu }A^{\alpha } 
-  \frac{1}{2} \bar{R}_{\nu }{}^{\sigma \rho }{}_{\alpha } \bar{\nabla}_{\mu }A^{\alpha }\\
& -  \frac{1}{2} \bar{R}_{\nu }{}^{\sigma } \bar{\nabla}_{\mu }A^{\rho }
+ \frac{1}{6} \delta_{\nu }{}^{\sigma } \bar{R} \bar{\nabla}_{\mu }A^{\rho }
+ \frac{1}{2} \bar{R}_{\nu }{}^{\rho } \bar{\nabla}_{\mu }A^{\sigma } 
-  \frac{1}{6} \delta_{\nu }{}^{\rho } \bar{R} \bar{\nabla}_{\mu }A^{\sigma }
+ \frac{1}{2} \delta_{\mu }{}^{\sigma } \bar{R}^{\rho }{}_{\alpha } \bar{\nabla}_{\nu }A^{\alpha }
-  \frac{1}{2} \delta_{\mu }{}^{\rho } \bar{R}^{\sigma }{}_{\alpha } \bar{\nabla}_{\nu }A^{\alpha }\\
& + \frac{1}{2} \bar{R}_{\mu \alpha }{}^{\rho \sigma } \bar{\nabla}_{\nu }A^{\alpha } 
-  \frac{1}{2} \bar{R}_{\mu }{}^{\rho \sigma }{}_{\alpha } \bar{\nabla}_{\nu }A^{\alpha }
+ \frac{1}{2} \bar{R}_{\mu }{}^{\sigma \rho }{}_{\alpha } \bar{\nabla}_{\nu }A^{\alpha }
+ \frac{1}{2} \bar{R}_{\mu }{}^{\sigma } \bar{\nabla}_{\nu }A^{\rho }
-  \frac{1}{6} \delta_{\mu }{}^{\sigma } \bar{R} \bar{\nabla}_{\nu }A^{\rho } 
-  \frac{1}{2} \bar{R}_{\mu }{}^{\rho } \bar{\nabla}_{\nu }A^{\sigma }
+ \frac{1}{6} \delta_{\mu }{}^{\rho } \bar{R} \bar{\nabla}_{\nu }A^{\sigma }\\
& -  \frac{1}{2} \delta_{\nu }{}^{\sigma } \bar{R}_{\mu \alpha } \bar{\nabla}^{\rho }A^{\alpha }
+ \frac{1}{2} \delta_{\mu }{}^{\sigma } \bar{R}_{\nu \alpha } \bar{\nabla}^{\rho }A^{\alpha } 
-  \bar{R}_{\mu \nu }{}^{\sigma }{}_{\alpha } \bar{\nabla}^{\rho }A^{\alpha }
-  \frac{1}{2} \bar{R}_{\nu }{}^{\sigma } \bar{\nabla}^{\rho }A_{\mu }
+ \frac{1}{6} \delta_{\nu }{}^{\sigma } \bar{R} \bar{\nabla}^{\rho }A_{\mu }
+ \frac{1}{2} \bar{R}_{\mu }{}^{\sigma } \bar{\nabla}^{\rho }A_{\nu } \\
& -  \frac{1}{6} \delta_{\mu }{}^{\sigma } \bar{R} \bar{\nabla}^{\rho }A_{\nu }
+ \frac{1}{2} A^{\alpha } \bar{\nabla}^{\rho }\bar{R}_{\mu \alpha \nu }{}^{\sigma }
-  \frac{1}{2} A^{\alpha } \bar{\nabla}^{\rho }\bar{R}_{\mu \nu }{}^{\sigma }{}_{\alpha }
-  \frac{1}{2} A^{\alpha } \bar{\nabla}^{\rho }\bar{R}_{\mu }{}^{\sigma }{}_{\nu \alpha } 
+ \frac{1}{2} \delta_{\nu }{}^{\rho } \bar{R}_{\mu \alpha } \bar{\nabla}^{\sigma }A^{\alpha }
-  \frac{1}{2} \delta_{\mu }{}^{\rho } \bar{R}_{\nu \alpha } \bar{\nabla}^{\sigma }A^{\alpha }\\
& + \bar{R}_{\mu \nu }{}^{\rho }{}_{\alpha } \bar{\nabla}^{\sigma }A^{\alpha }
+ \frac{1}{2} \bar{R}_{\nu }{}^{\rho } \bar{\nabla}^{\sigma }A_{\mu } 
-  \frac{1}{6} \delta_{\nu }{}^{\rho } \bar{R} \bar{\nabla}^{\sigma }A_{\mu }
-  \frac{1}{2} \bar{R}_{\mu }{}^{\rho } \bar{\nabla}^{\sigma }A_{\nu }
+ \frac{1}{6} \delta_{\mu }{}^{\rho } \bar{R} \bar{\nabla}^{\sigma }A_{\nu }
-  \frac{1}{2} A^{\alpha } \bar{\nabla}^{\sigma }\bar{R}_{\mu \alpha \nu }{}^{\rho } \\
& + \frac{1}{2} A^{\alpha } \bar{\nabla}^{\sigma }\bar{R}_{\mu \nu }{}^{\rho }{}_{\alpha }
+ \frac{1}{2} A^{\alpha } \bar{\nabla}^{\sigma }\bar{R}_{\mu }{}^{\rho }{}_{\nu \alpha }\,,
\ea
\ee
\be
\ba
c^{\rho \sigma}_{\ \ \ \mu \nu (\widehat{h})} ={}&\frac{1}{6} \delta_{\mu }{}^{\sigma } \delta_{\nu }{}^{\rho } \widehat{h}^{\alpha \beta } \bar{R}_{\alpha \beta }
-  \frac{1}{6} \delta_{\mu }{}^{\rho } \delta_{\nu }{}^{\sigma } \widehat{h}^{\alpha \beta } \bar{R}_{\alpha \beta }
-  \frac{1}{4} \delta_{\nu }{}^{\sigma } \widehat{h}^{\rho \alpha } \bar{R}_{\mu \alpha } 
+ \frac{1}{4} \delta_{\nu }{}^{\rho } \widehat{h}^{\sigma \alpha } \bar{R}_{\mu \alpha }
-  \frac{1}{2} \widehat{h}_{\nu }{}^{\sigma } \bar{R}_{\mu }{}^{\rho }
+ \frac{1}{2} \widehat{h}_{\nu }{}^{\rho } \bar{R}_{\mu }{}^{\sigma }\\
& + \frac{1}{4} \delta_{\mu }{}^{\sigma } \widehat{h}^{\rho \alpha } \bar{R}_{\nu \alpha } 
-  \frac{1}{4} \delta_{\mu }{}^{\rho } \widehat{h}^{\sigma \alpha } \bar{R}_{\nu \alpha }
+ \frac{1}{2} \widehat{h}_{\mu }{}^{\sigma } \bar{R}_{\nu }{}^{\rho }
-  \frac{1}{2} \widehat{h}_{\mu }{}^{\rho } \bar{R}_{\nu }{}^{\sigma }
-  \frac{1}{4} \delta_{\nu }{}^{\sigma } \widehat{h}_{\mu }{}^{\alpha } \bar{R}^{\rho }{}_{\alpha } 
+ \frac{1}{4} \delta_{\mu }{}^{\sigma } \widehat{h}_{\nu }{}^{\alpha } \bar{R}^{\rho }{}_{\alpha }\\
& + \frac{1}{4} \delta_{\nu }{}^{\rho } \widehat{h}_{\mu }{}^{\alpha } \bar{R}^{\sigma }{}_{\alpha }
-  \frac{1}{4} \delta_{\mu }{}^{\rho } \widehat{h}_{\nu }{}^{\alpha } \bar{R}^{\sigma }{}_{\alpha }
+ \frac{1}{6} \delta_{\nu }{}^{\sigma } \widehat{h}_{\mu }{}^{\rho } \bar{R} 
-  \frac{1}{6} \delta_{\nu }{}^{\rho } \widehat{h}_{\mu }{}^{\sigma } \bar{R}
-  \frac{1}{6} \delta_{\mu }{}^{\sigma } \widehat{h}_{\nu }{}^{\rho } \bar{R}
+ \frac{1}{6} \delta_{\mu }{}^{\rho } \widehat{h}_{\nu }{}^{\sigma } \bar{R}\\
&+ \frac{1}{2} \widehat{h}^{\sigma \alpha } \bar{R}_{\mu \alpha \nu }{}^{\rho } 
-  \frac{1}{2} \widehat{h}^{\rho \alpha } \bar{R}_{\mu \alpha \nu }{}^{\sigma }
+ \frac{1}{2} \delta_{\nu }{}^{\sigma } \widehat{h}^{\alpha \beta } \bar{R}_{\mu \alpha }{}^{\rho }{}_{\beta }
-  \frac{1}{2} \delta_{\nu }{}^{\rho } \widehat{h}^{\alpha \beta } \bar{R}_{\mu \alpha }{}^{\sigma }{}_{\beta }
+ \frac{1}{2} \widehat{h}^{\sigma \alpha } \bar{R}_{\mu \nu }{}^{\rho }{}_{\alpha } 
-  \frac{1}{2} \widehat{h}^{\rho \alpha } \bar{R}_{\mu \nu }{}^{\sigma }{}_{\alpha }\\
& -  \frac{1}{2} \widehat{h}^{\sigma \alpha } \bar{R}_{\mu }{}^{\rho }{}_{\nu \alpha }
-  \frac{1}{2} \widehat{h}_{\nu }{}^{\alpha } \bar{R}_{\mu }{}^{\rho \sigma }{}_{\alpha }
+ \frac{1}{2} \widehat{h}^{\rho \alpha } \bar{R}_{\mu }{}^{\sigma }{}_{\nu \alpha } 
+ \frac{1}{2} \widehat{h}_{\nu }{}^{\alpha } \bar{R}_{\mu }{}^{\sigma \rho }{}_{\alpha }
-  \frac{1}{2} \delta_{\mu }{}^{\sigma } \widehat{h}^{\alpha \beta } \bar{R}_{\nu \alpha }{}^{\rho }{}_{\beta }
+ \frac{1}{2} \delta_{\mu }{}^{\rho } \widehat{h}^{\alpha \beta } \bar{R}_{\nu \alpha }{}^{\sigma }{}_{\beta }\\
& + \frac{1}{2} \widehat{h}_{\mu }{}^{\alpha } \bar{R}_{\nu }{}^{\rho \sigma }{}_{\alpha } 
-  \frac{1}{2} \widehat{h}_{\mu }{}^{\alpha } \bar{R}_{\nu }{}^{\sigma \rho }{}_{\alpha }
+ \frac{1}{4} \delta_{\nu }{}^{\sigma } \bar{\square} \widehat{h}_{\mu }{}^{\rho }
-  \frac{1}{4} \delta_{\nu }{}^{\rho } \bar{\square} \widehat{h}_{\mu }{}^{\sigma } 
-  \frac{1}{4} \delta_{\mu }{}^{\sigma } \bar{\square} \widehat{h}_{\nu }{}^{\rho }
+ \frac{1}{4} \delta_{\mu }{}^{\rho } \bar{\square} \widehat{h}_{\nu }{}^{\sigma }
-  \frac{1}{2} \bar{\nabla}^{\rho }\bar{\nabla}_{\mu } \widehat{h}_{\nu }{}^{\sigma }\\
& + \frac{1}{2} \bar{\nabla}^{\rho }\bar{\nabla}_{\nu }\widehat{h}_{\mu }{}^{\sigma } 
+ \frac{1}{2} \bar{\nabla}^{\sigma }\bar{\nabla}_{\mu }\widehat{h}_{\nu }{}^{\rho }
-  \frac{1}{2} \bar{\nabla}^{\sigma }\bar{\nabla}_{\nu }\widehat{h}_{\mu }{}^{\rho }\,.
\ea
\ee

(ii) Rewriting the above in terms of irreducible $\bar{R}, \bar{S}_{\mu \nu}, \bar{C}_{\mu \nu \rho \sigma}$ and imposing $\bar{C}_{\mu \nu \rho \sigma} = 0$, we get
\be
\ba
c^{\rho \sigma}_{\ \ \ \mu \nu (\phi)} ={}&0\,,
\\ \\
c^{\rho \sigma}_{\ \ \ \mu \nu (B)} ={}&\frac{1}{24} \delta_{\mu }{}^{\sigma } \delta_{\nu }{}^{\rho } \bar{\nabla}_{\alpha }\bar{R} \bar{\nabla}^{\alpha }B
-  \frac{1}{24} \delta_{\mu }{}^{\rho } \delta_{\nu }{}^{\sigma } \bar{\nabla}_{\alpha }\bar{R} \bar{\nabla}^{\alpha }B
-  \frac{1}{4} \delta_{\nu }{}^{\sigma } \bar{\nabla}_{\alpha }\bar{S}_{\mu }{}^{\rho } \bar{\nabla}^{\alpha }B 
+ \frac{1}{4} \delta_{\nu }{}^{\rho } \bar{\nabla}_{\alpha }\bar{S}_{\mu }{}^{\sigma } \bar{\nabla}^{\alpha }B
+ \frac{1}{4} \delta_{\mu }{}^{\sigma } \bar{\nabla}_{\alpha }\bar{S}_{\nu }{}^{\rho } \bar{\nabla}^{\alpha }B\\
& -  \frac{1}{4} \delta_{\mu }{}^{\rho } \bar{\nabla}_{\alpha }\bar{S}_{\nu }{}^{\sigma } \bar{\nabla}^{\alpha }B 
+ \frac{1}{24} \delta_{\nu }{}^{\sigma } \bar{\nabla}_{\mu }B \bar{\nabla}^{\rho }\bar{R}
-  \frac{1}{24} \delta_{\mu }{}^{\sigma } \bar{\nabla}_{\nu }B \bar{\nabla}^{\rho }\bar{R}
+ \frac{1}{4} \delta_{\nu }{}^{\sigma } \bar{\nabla}^{\alpha }B \bar{\nabla}^{\rho }\bar{S}_{\mu \alpha } 
-  \frac{1}{4} \bar{\nabla}_{\nu }B \bar{\nabla}^{\rho }\bar{S}_{\mu }{}^{\sigma }\\
& -  \frac{1}{4} \delta_{\mu }{}^{\sigma } \bar{\nabla}^{\alpha }B \bar{\nabla}^{\rho }\bar{S}_{\nu \alpha }
+ \frac{1}{4} \bar{\nabla}_{\mu }B \bar{\nabla}^{\rho }\bar{S}_{\nu }{}^{\sigma }
-  \frac{1}{24} \delta_{\nu }{}^{\rho } \bar{\nabla}_{\mu }B \bar{\nabla}^{\sigma }\bar{R} 
+ \frac{1}{24} \delta_{\mu }{}^{\rho } \bar{\nabla}_{\nu }B \bar{\nabla}^{\sigma }\bar{R}
-  \frac{1}{4} \delta_{\nu }{}^{\rho } \bar{\nabla}^{\alpha }B \bar{\nabla}^{\sigma }\bar{S}_{\mu \alpha }\\
& + \frac{1}{4} \bar{\nabla}_{\nu }B \bar{\nabla}^{\sigma }\bar{S}_{\mu }{}^{\rho } 
+ \frac{1}{4} \delta_{\mu }{}^{\rho } \bar{\nabla}^{\alpha }B \bar{\nabla}^{\sigma }\bar{S}_{\nu \alpha }
-  \frac{1}{4} \bar{\nabla}_{\mu }B \bar{\nabla}^{\sigma }\bar{S}_{\nu }{}^{\rho }\,,
\\ \\
c^{\rho \sigma}_{\ \ \ \mu \nu (A)} ={}&\frac{1}{12} A^{\alpha } \delta_{\mu }{}^{\sigma } \delta_{\nu }{}^{\rho } \bar{\nabla}_{\alpha }\bar{R}
-  \frac{1}{12} A^{\alpha } \delta_{\mu }{}^{\rho } \delta_{\nu }{}^{\sigma } \bar{\nabla}_{\alpha }\bar{R}
-  \frac{1}{2} A^{\alpha } \delta_{\nu }{}^{\sigma } \bar{\nabla}_{\alpha }\bar{S}_{\mu }{}^{\rho } 
+ \frac{1}{2} A^{\alpha } \delta_{\nu }{}^{\rho } \bar{\nabla}_{\alpha }\bar{S}_{\mu }{}^{\sigma }
+ \frac{1}{2} A^{\alpha } \delta_{\mu }{}^{\sigma } \bar{\nabla}_{\alpha }\bar{S}_{\nu }{}^{\rho }\\
& -  \frac{1}{2} A^{\alpha } \delta_{\mu }{}^{\rho } \bar{\nabla}_{\alpha }\bar{S}_{\nu }{}^{\sigma }
-  \frac{1}{12} A_{\nu } \delta_{\mu }{}^{\sigma } \bar{\nabla}^{\rho }\bar{R} 
+ \frac{1}{12} A_{\mu } \delta_{\nu }{}^{\sigma } \bar{\nabla}^{\rho }\bar{R}
+ \frac{1}{2} A^{\alpha } \delta_{\nu }{}^{\sigma } \bar{\nabla}^{\rho }\bar{S}_{\mu \alpha }
-  \frac{1}{2} A_{\nu } \bar{\nabla}^{\rho }\bar{S}_{\mu }{}^{\sigma }\\
& -  \frac{1}{2} A^{\alpha } \delta_{\mu }{}^{\sigma } \bar{\nabla}^{\rho }\bar{S}_{\nu \alpha } 
+ \frac{1}{2} A_{\mu } \bar{\nabla}^{\rho }\bar{S}_{\nu }{}^{\sigma }
+ \frac{1}{12} A_{\nu } \delta_{\mu }{}^{\rho } \bar{\nabla}^{\sigma }\bar{R}
-  \frac{1}{12} A_{\mu } \delta_{\nu }{}^{\rho } \bar{\nabla}^{\sigma }\bar{R}
-  \frac{1}{2} A^{\alpha } \delta_{\nu }{}^{\rho } \bar{\nabla}^{\sigma }\bar{S}_{\mu \alpha } \\
& + \frac{1}{2} A_{\nu } \bar{\nabla}^{\sigma }\bar{S}_{\mu }{}^{\rho }
+ \frac{1}{2} A^{\alpha } \delta_{\mu }{}^{\rho } \bar{\nabla}^{\sigma }\bar{S}_{\nu \alpha }
-  \frac{1}{2} A_{\mu } \bar{\nabla}^{\sigma }\bar{S}_{\nu }{}^{\rho }\,,
\\ \\
c^{\rho \sigma}_{\ \ \ \mu \nu (\widehat{h})} ={}&- \frac{1}{12} \delta_{\nu }{}^{\sigma } \widehat{h}_{\mu }{}^{\rho } \bar{R}
+ \frac{1}{12} \delta_{\nu }{}^{\rho } \widehat{h}_{\mu }{}^{\sigma } \bar{R}
+ \frac{1}{12} \delta_{\mu }{}^{\sigma } \widehat{h}_{\nu }{}^{\rho } \bar{R}
-  \frac{1}{12} \delta_{\mu }{}^{\rho } \widehat{h}_{\nu }{}^{\sigma } \bar{R} 
-  \frac{1}{3} \delta_{\mu }{}^{\sigma } \delta_{\nu }{}^{\rho } \widehat{h}^{\alpha \beta } \bar{S}_{\alpha \beta }
+ \frac{1}{3} \delta_{\mu }{}^{\rho } \delta_{\nu }{}^{\sigma } \widehat{h}^{\alpha \beta } \bar{S}_{\alpha \beta }\\
& -  \frac{1}{2} \delta_{\nu }{}^{\sigma } \widehat{h}^{\rho \alpha } \bar{S}_{\mu \alpha } 
+ \frac{1}{2} \delta_{\nu }{}^{\rho } \widehat{h}^{\sigma \alpha } \bar{S}_{\mu \alpha }
-  \frac{1}{4} \widehat{h}_{\nu }{}^{\sigma } \bar{S}_{\mu }{}^{\rho }
+ \frac{1}{4} \widehat{h}_{\nu }{}^{\rho } \bar{S}_{\mu }{}^{\sigma }
+ \frac{1}{2} \delta_{\mu }{}^{\sigma } \widehat{h}^{\rho \alpha } \bar{S}_{\nu \alpha } 
-  \frac{1}{2} \delta_{\mu }{}^{\rho } \widehat{h}^{\sigma \alpha } \bar{S}_{\nu \alpha }\\
& + \frac{1}{4} \widehat{h}_{\mu }{}^{\sigma } \bar{S}_{\nu }{}^{\rho }
-  \frac{1}{4} \widehat{h}_{\mu }{}^{\rho } \bar{S}_{\nu }{}^{\sigma }
-  \frac{1}{4} \delta_{\nu }{}^{\sigma } \widehat{h}_{\mu }{}^{\alpha } \bar{S}^{\rho }{}_{\alpha } 
+ \frac{1}{4} \delta_{\mu }{}^{\sigma } \widehat{h}_{\nu }{}^{\alpha } \bar{S}^{\rho }{}_{\alpha }
+ \frac{1}{4} \delta_{\nu }{}^{\rho } \widehat{h}_{\mu }{}^{\alpha } \bar{S}^{\sigma }{}_{\alpha }
-  \frac{1}{4} \delta_{\mu }{}^{\rho } \widehat{h}_{\nu }{}^{\alpha } \bar{S}^{\sigma }{}_{\alpha }\\
& + \frac{1}{4} \delta_{\nu }{}^{\sigma } \bar{\square} \widehat{h}_{\mu }{}^{\rho } 
-  \frac{1}{4} \delta_{\nu }{}^{\rho } \bar{\square} \widehat{h}_{\mu }{}^{\sigma }
-  \frac{1}{4} \delta_{\mu }{}^{\sigma } \bar{\square} \widehat{h}_{\nu }{}^{\rho }
+ \frac{1}{4} \delta_{\mu }{}^{\rho } \bar{\square} \widehat{h}_{\nu }{}^{\sigma } 
-  \frac{1}{2} \bar{\nabla}^{\rho }\bar{\nabla}_{\mu }\widehat{h}_{\nu }{}^{\sigma }
+ \frac{1}{2} \bar{\nabla}^{\rho }\bar{\nabla}_{\nu }\widehat{h}_{\mu }{}^{\sigma }\\
& + \frac{1}{2} \bar{\nabla}^{\sigma }\bar{\nabla}_{\mu }\widehat{h}_{\nu }{}^{\rho }
-  \frac{1}{2} \bar{\nabla}^{\sigma }\bar{\nabla}_{\nu }\widehat{h}_{\mu }{}^{\rho }\,.
\ea
\ee

(iii) Imposing $\bar{\nabla}_{\alpha} \bar{S}_{\mu \nu} = 0, ~~\partial_{\mu} \bar{R} = 0$ to above gives
\be
\ba
c^{\rho \sigma}_{\ \ \ \mu \nu (\phi)} ={}&0 \,,\qquad \qquad
c^{\rho \sigma}_{\ \ \ \mu \nu (B)} ={}&0 \,, \qquad \qquad
c^{\rho \sigma}_{\ \ \ \mu \nu (A)} ={}&0 \,.
\ea
\ee
\be
\ba
c^{\rho \sigma}_{\ \ \ \mu \nu (\widehat{h})} ={}&- \frac{1}{12} \delta_{\nu }{}^{\sigma } \widehat{h}_{\mu }{}^{\rho } \bar{R}
 + \frac{1}{12} \delta_{\nu }{}^{\rho } \widehat{h}_{\mu }{}^{\sigma } \bar{R}
 + \frac{1}{12} \delta_{\mu }{}^{\sigma } \widehat{h}_{\nu }{}^{\rho } \bar{R}
 -  \frac{1}{12} \delta_{\mu }{}^{\rho } \widehat{h}_{\nu }{}^{\sigma } \bar{R} 
 -  \frac{1}{3} \delta_{\mu }{}^{\sigma } \delta_{\nu }{}^{\rho } \widehat{h}^{\alpha \beta } \bar{S}_{\alpha \beta }
 + \frac{1}{3} \delta_{\mu }{}^{\rho } \delta_{\nu }{}^{\sigma } \widehat{h}^{\alpha \beta } \bar{S}_{\alpha \beta }\\
& -  \frac{1}{2} \delta_{\nu }{}^{\sigma } \widehat{h}^{\rho \alpha } \bar{S}_{\mu \alpha } 
 + \frac{1}{2} \delta_{\nu }{}^{\rho } \widehat{h}^{\sigma \alpha } \bar{S}_{\mu \alpha }
 -  \frac{1}{4} \widehat{h}_{\nu }{}^{\sigma } \bar{S}_{\mu }{}^{\rho }
 + \frac{1}{4} \widehat{h}_{\nu }{}^{\rho } \bar{S}_{\mu }{}^{\sigma }
 + \frac{1}{2} \delta_{\mu }{}^{\sigma } \widehat{h}^{\rho \alpha } \bar{S}_{\nu \alpha } 
 -  \frac{1}{2} \delta_{\mu }{}^{\rho } \widehat{h}^{\sigma \alpha } \bar{S}_{\nu \alpha }
 + \frac{1}{4} \widehat{h}_{\mu }{}^{\sigma } \bar{S}_{\nu }{}^{\rho }\\
& -  \frac{1}{4} \widehat{h}_{\mu }{}^{\rho } \bar{S}_{\nu }{}^{\sigma }
 -  \frac{1}{4} \delta_{\nu }{}^{\sigma } \widehat{h}_{\mu }{}^{\alpha } \bar{S}^{\rho }{}_{\alpha } 
 + \frac{1}{4} \delta_{\mu }{}^{\sigma } \widehat{h}_{\nu }{}^{\alpha } \bar{S}^{\rho }{}_{\alpha }
 + \frac{1}{4} \delta_{\nu }{}^{\rho } \widehat{h}_{\mu }{}^{\alpha } \bar{S}^{\sigma }{}_{\alpha }
 -  \frac{1}{4} \delta_{\mu }{}^{\rho } \widehat{h}_{\nu }{}^{\alpha } \bar{S}^{\sigma }{}_{\alpha }
 + \frac{1}{4} \delta_{\nu }{}^{\sigma } \bar{\square} \widehat{h}_{\mu }{}^{\rho } 
 -  \frac{1}{4} \delta_{\nu }{}^{\rho } \bar{\square} \widehat{h}_{\mu }{}^{\sigma }\\
& -  \frac{1}{4} \delta_{\mu }{}^{\sigma } \bar{\square} \widehat{h}_{\nu }{}^{\rho }
 + \frac{1}{4} \delta_{\mu }{}^{\rho } \bar{\square} \widehat{h}_{\nu }{}^{\sigma } 
-  \frac{1}{2} \bar{\nabla}^{\rho }\bar{\nabla}_{\mu }\widehat{h}_{\nu }{}^{\sigma }
 + \frac{1}{2} \bar{\nabla}^{\rho }\bar{\nabla}_{\nu }\widehat{h}_{\mu }{}^{\sigma }
 + \frac{1}{2} \bar{\nabla}^{\sigma }\bar{\nabla}_{\mu }\widehat{h}_{\nu }{}^{\rho }
 -  \frac{1}{2} \bar{\nabla}^{\sigma }\bar{\nabla}_{\nu }\widehat{h}_{\mu }{}^{\rho }\,.
\ea
\ee
Now we find the quadratic in $\widehat{h}_{\mu \nu}$ part. There are many Kronecker delta terms in $c^{\rho \sigma}_{\ \ \ \mu \nu (\widehat{h})}$ with free indices. Since the totally tracelessness property of the Weyl tensor is preserved at each order of perturbation, all the Kronecker delta terms in $c^{\rho \sigma}_{\ \ \ \mu \nu (\widehat{h})}$ will hit $c^{\mu \nu}_{\ \ \rho \sigma (\widehat{h})}$ and go to zero. We are then left with only a few terms in $c^{\rho \sigma}_{\ \ \ \mu \nu (\widehat{h})}$. Assuming that it will be multiplied to $\mathcal{\bar{F}}_3 (\bar{\square}_s) c^{\mu \nu}_{\ \ \rho \sigma (\widehat{h})}$ later, the simplified expression is
\be
\ba
c^{\rho \sigma}_{\ \ \ \mu \nu (\widehat{h})} = {}&- \frac{1}{4} \widehat{h}_{\nu }{}^{\sigma } \bar{S}_{\mu }{}^{\rho }
 + \frac{1}{4} \widehat{h}_{\nu }{}^{\rho } \bar{S}_{\mu }{}^{\sigma }
 + \frac{1}{4} \widehat{h}_{\mu }{}^{\sigma } \bar{S}_{\nu }{}^{\rho }
 -  \frac{1}{4} \widehat{h}_{\mu }{}^{\rho } \bar{S}_{\nu }{}^{\sigma }
 -  \frac{1}{2} \bar{\nabla}^{\rho }\bar{\nabla}_{\mu }\widehat{h}_{\nu }{}^{\sigma }
 + \frac{1}{2} \bar{\nabla}^{\rho }\bar{\nabla}_{\nu }\widehat{h}_{\mu }{}^{\sigma }
 \\ &+ \frac{1}{2} \bar{\nabla}^{\sigma }\bar{\nabla}_{\mu }\widehat{h}_{\nu }{}^{\rho }
 -  \frac{1}{2} \bar{\nabla}^{\sigma }\bar{\nabla}_{\nu }\widehat{h}_{\mu }{}^{\rho }\,.
\ea
\ee
Further simplifications beyond this happen when the above simplified $c^{\rho \sigma}_{\ \ \ \mu \nu (\widehat{h})}$ is multiplied with $\mathcal{\bar{F}}_3 (\bar{\square}_{s}) c^{\mu \nu}_{\ \ \rho \sigma (\widehat{h})}$, and the index symmetries of $c^{\mu \nu}_{\ \ \rho \sigma (\widehat{h})}$ are used. We ultimately get
\be \label{weylquadpert}
c^{\rho \sigma}_{\ \ \ \mu \nu (\widehat{h})}  \mathcal{\bar{F}}_3 (\bar{\square}_{s}) c^{\mu \nu}_{\ \ \rho \sigma (\widehat{h})} = \left[ \widehat{h}_{\nu }{}^{\rho } \bar{S}_{\mu }{}^{\sigma } + 2 \bar{\nabla}^{\rho }\bar{\nabla}_{\nu }\widehat{h}_{\mu }{}^{\sigma } \right] \mathcal{\bar{F}}_3 (\bar{\square}_{s}) \left[ c^{\mu \nu}_{\ \ \rho \sigma (\widehat{h})} \right]\,,
\ee
where
\be
\ba
c^{\mu \nu}_{\ \ \rho \sigma (\widehat{h})} ={}&- \frac{1}{12} \delta^{\nu }{}_{\sigma } \widehat{h}^{\mu }{}_{\rho } \bar{R}
 + \frac{1}{12} \delta^{\nu }{}_{\rho } \widehat{h}^{\mu }{}_{\sigma } \bar{R}
 + \frac{1}{12} \delta^{\mu }{}_{\sigma } \widehat{h}^{\nu }{}_{\rho } \bar{R}
 -  \frac{1}{12} \delta^{\mu }{}_{\rho } \widehat{h}^{\nu }{}_{\sigma } \bar{R} 
 -  \frac{1}{3} \delta^{\mu }{}_{\sigma } \delta^{\nu }{}_{\rho } \widehat{h}^{\alpha \beta } \bar{S}_{\alpha \beta }
 + \frac{1}{3} \delta^{\mu }{}_{\rho } \delta^{\nu }{}_{\sigma } \widehat{h}^{\alpha \beta } \bar{S}_{\alpha \beta }\\
 &-  \frac{1}{2} \delta^{\nu }{}_{\sigma } \widehat{h}_{\rho }{}^{\alpha } \bar{S}^{\mu }{}_{\alpha } 
 + \frac{1}{2} \delta^{\nu }{}_{\rho } \widehat{h}_{\sigma }{}^{\alpha } \bar{S}^{\mu }{}_{\alpha }
 -  \frac{1}{4} \widehat{h}^{\nu }{}_{\sigma } \bar{S}^{\mu }{}_{\rho }
 + \frac{1}{4} \widehat{h}^{\nu }{}_{\rho } \bar{S}^{\mu }{}_{\sigma }
 + \frac{1}{2} \delta^{\mu }{}_{\sigma } \widehat{h}_{\rho }{}^{\alpha } \bar{S}^{\nu }{}_{\alpha } 
 -  \frac{1}{2} \delta^{\mu }{}_{\rho } \widehat{h}_{\sigma }{}^{ \alpha } \bar{S}^{\nu }{}_{\alpha }
 + \frac{1}{4} \widehat{h}^{\mu }{}_{\sigma } \bar{S}^{\nu }{}_{\rho }\\
 &-  \frac{1}{4} \widehat{h}^{\mu }{}_{\rho } \bar{S}^{\nu }{}_{\sigma }
 -  \frac{1}{4} \delta^{\nu }{}_{\sigma } \widehat{h}^{\mu \alpha } \bar{S}_{\rho \alpha } 
 + \frac{1}{4} \delta^{\mu }{}_{\sigma } \widehat{h}^{\nu \alpha } \bar{S}_{\rho \alpha }
 + \frac{1}{4} \delta^{\nu }{}_{\rho } \widehat{h}^{\mu \alpha } \bar{S}_{\sigma \alpha }
 -  \frac{1}{4} \delta^{\mu }{}_{\rho } \widehat{h}^{\nu \alpha } \bar{S}_{\sigma \alpha }
 + \frac{1}{4} \delta^{\nu }{}_{\sigma } \bar{\square} \widehat{h}^{\mu }{}_{\rho } 
 -  \frac{1}{4} \delta^{\nu }{}_{\rho } \bar{\square} \widehat{h}^{\mu }{}_{\sigma }\\
& -  \frac{1}{4} \delta^{\mu }{}_{\sigma } \bar{\square} \widehat{h}^{\nu }{}_{\rho }
 + \frac{1}{4} \delta^{\mu }{}_{\rho } \bar{\square} \widehat{h}^{\nu }{}_{\sigma } 
 -  \frac{1}{2} \bar{\nabla}_{\rho }\bar{\nabla}^{\mu }\widehat{h}^{\nu }{}_{\sigma }
 + \frac{1}{2} \bar{\nabla}_{\rho }\bar{\nabla}^{\nu }\widehat{h}^{\mu }{}_{\sigma }
 + \frac{1}{2} \bar{\nabla}_{\sigma }\bar{\nabla}^{\mu }\widehat{h}^{\nu }{}_{\rho }
 -  \frac{1}{2} \bar{\nabla}_{\sigma }\bar{\nabla}^{\nu }\widehat{h}^{\mu }{}_{\rho }\,.
\ea
\ee
It is easy to determine SVT expressions for $c^{\mu \nu}_{\ \ \rho \sigma}$ from $c^{\rho \sigma}_{\ \ \ \mu \nu}$ since $ c^{\mu \nu}_{\ \ \rho \sigma} =  c_{\alpha \beta}^{\ \ \gamma \lambda}  \bar{g}^{\alpha \mu} \bar{g}^{\beta \nu} \bar{g}_{\gamma \rho} \bar{g}_{\lambda \sigma} + \bar{C}_{\alpha \beta}^{\ \ \gamma \lambda} \delta ( g^{\alpha \mu} g^{\beta \nu} g_{\gamma \rho} g_{\lambda \sigma} )$ where the second term is zero for our conformally-flat background. So expressions for $c^{\mu \nu}_{\ \ \rho \sigma}$ are obtained from their $c^{\rho \sigma}_{\ \ \ \mu \nu}$ counterparts with lower (upper) indices raised (lowered).


\section{Perturbed equations of motion at the linear level}

Perturbations at the quadratic level of the action are equivalent to linear perturbation of the equations of motion and sometimes it is easier to study the latter. The variation of trace equations of motion eq.(\ref{traceequations of motion}) under the usual conditions eq.(\ref{mss20}) is
\be \label{firstvariationtraceeq}
\ba
-M_P^2 \delta R &= -6 \bar{\square} \Fc_1(\bar{\square}_s) \delta R -( \delta {L_1} +2 \delta \bar{L}_{1})
- 2 \delta ({\bar{\nabla}_\mu\bar{\nabla}_\nu })\Fc_2(\bar{\square}_s){\bar{S}}^{\mu\nu} - 2\bar{\nabla}_\mu\bar{\nabla}_\nu \delta  ({\Fc_2(\square_{s})}){\bar{S}}^{\mu\nu}
- 2\bar{\nabla}_\mu\bar{\nabla}_\nu \Fc_2 (\bar{\square}_s) \delta {\bar{S}}^{\mu\nu} \\
& \qquad -( \delta {L_2} +2 \delta \bar{L}_{2})+2 \delta \widetilde\Delta\,.
\ea
\ee
$ \delta L_1 = \delta L_2 = 0$, because each term in their respective infinite sums have $2$ background curvatures, each of which has at least one derivative acting on it. Taking a first variation always results in a derivative acting on these background curvatures (from eq.(\ref{mss21})), resulting in zero. We have
\be
\ba
\delta \bar{L}_1 = \bar{R} \left( \Fc_1 (\bar{\square}_s) - f_{1,0} \right) \delta R \qquad \qquad 
\delta \bar{L}_2 = \bar{S}^{\alpha \beta} \left( \Fc_2 (\bar{\square}_s) - f_{2,0} \right) s_{\alpha \beta}\,,
\ea
\ee
\be
\delta \widetilde\Delta = \frac{1}{M_s^2} \bar{S}^\beta_{\ \gamma} \bar{\nabla}_\beta \bar{\nabla}^\mu \Fc_4(\bar{\square}_s) s^\gamma_{\ \mu} -  \frac{1}{M_s^2} \bar{S}^\gamma_{\ \mu} \bar{\nabla}_\beta \bar{\nabla}^\mu \Fc_4(\bar{\square}_s) s^\beta_{\ \gamma}\,,
\ee
\be
\ba
- 2 \delta ({\bar{\nabla}_\mu\bar{\nabla}_\nu })\Fc_2(\bar{\square}_s){\bar{S}}^{\mu\nu} &= 2 f_{2,0} \bar{\nabla}_\mu \bar{\nabla}_\nu s^{\mu \nu}\\
- 2\bar{\nabla}_\mu\bar{\nabla}_\nu \delta  ({\Fc_2(\square_{s})}){\bar{S}}^{\mu\nu} &= -2 \bar{\nabla}_\mu\bar{\nabla}_\nu \left[ \frac{ \Fc_2 (\bar{\square}_s) - f_{2,0} }{\bar{\square}_s} \right] \delta(\square_{s}) \bar{S}^{\mu \nu}\,.
\ea
\ee
Collecting all terms, first variation of the trace equations of motion eq.(\ref{firstvariationtraceeq}) becomes
\be \label{traceequations of motionpert}
\ba
-M_P^2 \delta R &= -6 \bar{\square} \Fc_1(\bar{\square}_s) \delta R - 2 \bar{R} \Fc_4(\bar{\square}_s) \bar{\square}_s \delta R - 2 \bar{S}^{\alpha \beta} \Fc_5 (\bar{\square}_s) \bar{\square}_s s_{\alpha \beta} 
+ 2 f_{2,0} \bar{\nabla}_\mu \bar{\nabla}_\nu s^{\mu \nu}\\
& \qquad -2 \bar{\nabla}_\mu \bar{\nabla}_\nu \Fc_5 (\bar{\square}_s) \delta(\square_{s}) \bar{S}^{\mu \nu}
- 2 \bar{\nabla}_\mu \bar{\nabla}_\nu \bar{\Fc_2} (\bar{\square}_s) \delta {S}^{\mu\nu}\\
& \qquad + \frac{1}{M_s^2} 2 \bar{S}^\beta_{\ \gamma} \bar{\nabla}_\beta \bar{\nabla}^\mu \Fc_5 (\bar{\square}_s) s^\gamma_{\ \mu} - \frac{1}{M_s^2} 2  \bar{S}^\gamma_{\ \mu} \bar{\nabla}_\beta \bar{\nabla}^\mu \Fc_5 (\bar{\square}_s) s^\beta_{\ \gamma}\,.
\ea
\ee



\section{Commutation relations}
\label{comrelsec}
In  higher derivative gravity like IDG, there are many covariant derivatives, some of which are to the left of form factors $\Fc_i\LF  \square_{s} \RF$ eq.(\ref{formfactor}). In order to contract them with covariant derivatives lying to the right of $\Fc_i\LF  \square_{s} \RF$, we need to perform many commutations. These commutations, in a theory like IDG with infinite number of box operators in $\Fc_i\LF  \square_{s} \RF$, lead to generation of an infinite number of background curvatures. When the background is MSS, it is possible in some cases to obtain compact expressions after the many commutations. It is apparently not possible to obtain compact expressions when the background is arbitrary. Nevertheless, it is useful to have formulas for such commutation relations to understand the general structure of perturbative expansions of a gravity action. In this section, we will generalize the commutation relations derived in Appendix B of \cite{Biswas:2016egy} where the background was MSS, to those when the background is \textit{arbitrary}.

For an arbitrary scalar $\phi$, we have
\be
\ba
\bar{\nabla}_\mu \bar{\nabla}_\nu \phi &= \bar{\nabla}_\nu \bar{\nabla}_\mu \phi\,, \qquad \text{provided torsion $T_{\mu \nu}^\lambda = 0$}\,,\\
\bar{\nabla}_\mu \bar{\nabla}_\alpha \bar{\nabla}_\beta \phi &= \bar{\nabla}_\alpha \bar{\nabla}_\beta \bar{\nabla}_\mu \phi - \bar{R}^\lambda_{\ \beta \mu \alpha} \bar{\nabla}_\lambda \phi\,,\\
\bar{\nabla}_\mu \bar{\square} \phi &= \bar{\square} \bar{\nabla}_\mu \phi - \bar{R}_{\mu \nu} \bar{\nabla}^\nu \phi\,. \label{f3}
\ea
\ee
For a vector $t^\mu$, we have
\be
\ba
\bar{\nabla}_\mu \bar{\nabla}_\alpha t^\mu &= \bar{\nabla}_\alpha \bar{\nabla}_\mu t^\mu + \bar{R}_{\alpha \mu} t^\mu\,,\\
\bar{\nabla}_\mu  \bar{\nabla}_\alpha \bar{\nabla}_\beta t^\mu &= \bar{\nabla}_\alpha \bar{\nabla}_\beta \bar{\nabla}_\mu t^\mu + \bar{\nabla}_\alpha (\bar{R}_{\beta \mu} t^\mu) - \bar{R}^\lambda_{\ \beta \mu \alpha} \bar{\nabla}_\lambda t^\mu + \bar{R}_{\lambda \alpha} \bar{\nabla}_\beta t^\lambda\,,\\
\bar{\nabla}_\mu \bar{\square} t^\mu &= \bar{\square} \bar{\nabla}_\mu t^\mu + \bar{\nabla}^\mu (\bar{R}_{\mu \nu} t^\nu) \label{f7} \,,\\
\bar{\nabla}_\nu \bar{\nabla}^\mu \bar{\nabla}_\rho t^\sigma &= \bar{\nabla}_\rho \bar{\nabla}^\mu \bar{\nabla}_\nu t^\sigma - \bar{\nabla}_\lambda t^\sigma (\bar{R}^{\lambda \ \mu}_{\ \nu \ \rho} + \bar{R}^{\lambda \ \ \mu}_{\ \rho \nu}) + \bar{R}^{\sigma \ \mu}_{\ \lambda \ \rho} \bar{\nabla}_\nu t^\lambda + \bar{R}^{\sigma \ \ \mu}_{\ \lambda \nu} \bar{\nabla}_\rho t^\lambda + \bar{\nabla}^\mu (\bar{R}^\sigma_{\ \lambda \nu \rho} t^\lambda)\,.
\ea
\ee
If $t^\mu$ was a transverse vector $A^\mu$ satisfying $\bar{\nabla}_\mu A^\mu =0$, the above expressions simplify to
\bea
\bar{\nabla}_\mu \bar{\nabla}_\alpha A^\mu &=& \bar{R}_{\alpha \mu} A^\mu\,,\\
\bar{\nabla}_\mu  \bar{\nabla}_\alpha \bar{\nabla}_\beta A^\mu &=& \bar{\nabla}_\alpha (\bar{R}_{\beta \mu} A^\mu) - \bar{R}^\lambda_{\ \beta \mu \alpha} \bar{\nabla}_\lambda A^\mu + \bar{R}_{\lambda \alpha} \bar{\nabla}_\beta A^\lambda\,,\\
\bar{\nabla}_\mu \bar{\square} A^\mu &=& \bar{\nabla}^\mu (\bar{R}_{\mu \nu} A^\nu)\,.
\eea
The last expression tells us that given a transverse vector $A^\mu$, $\bar{\square} A^\mu$ is not a transverse vector.
For symmetric (and traceless) tensor $t^{\mu \nu}$, we have
\be
\ba
\bar{\nabla}_\mu \bar{\nabla}_\alpha t^{\mu \nu} &= \bar{\nabla}_\alpha \bar{\nabla}_\mu t^{\mu \nu} + \bar{R}_{\lambda \alpha} t^{\lambda \nu} + \bar{R}^\nu_{\ \lambda \mu \alpha} t^{\mu \lambda}\,,\\
\bar{\nabla}_\mu \bar{\nabla}_\alpha \bar{\nabla}_\beta t^{\mu \nu} &= \bar{\nabla}_\alpha \bar{\nabla}_\beta \bar{\nabla}_\mu t^{\mu \nu} + \bar{\nabla}_\alpha (\bar{R}_{\lambda \beta} t^{\lambda \nu} + \bar{R}^\nu_{\ \lambda \mu \beta} t^{\mu \lambda}) - \bar{R}^\lambda_{\ \beta \mu \alpha} \bar{\nabla}_\lambda t^{\mu \nu} + \bar{R}_{\lambda \alpha} \bar{\nabla}_\beta t^{\lambda \nu} + \bar{R}^\nu_{\ \lambda \mu \alpha} \bar{\nabla}_\beta t^{\mu \lambda}\,,\\
\bar{\nabla}_\mu \bar{\square} t^{\mu \nu} &= \bar{\square} \bar{\nabla}_\mu t^{\mu \nu} + \bar{R}^\nu_{\ \lambda \mu \rho} \bar{\nabla}^\rho t^{\mu \lambda} + \bar{\nabla}^\sigma (\bar{R}_{\lambda \sigma} t^{\lambda \nu} + \bar{R}^\nu_{\ \lambda \mu \sigma} t^{\mu \lambda})\,,\\
\bar{\nabla}^\sigma \bar{\nabla}_\rho \bar{\nabla}_\sigma t_{\mu \nu} &= \bar{\nabla}_\rho \bar{\square} t_{\mu \nu} + \bar{R}_{\lambda \rho} \bar{\nabla}^\lambda t_{\mu \nu} - \bar{R}^\lambda_{\ \mu \alpha \rho} \bar{\nabla}^\alpha t_{\lambda \nu} - \bar{R}^\lambda_{\ \nu \alpha \rho} \bar{\nabla}^\alpha t_{\mu \lambda}\,.
\ea
\ee
For a symmetric, transverse (and traceless) tensor $T^{\mu \nu}$ satisfying $\bar{\nabla}_\mu T^{\mu \nu} =0$, we have
\be
\ba
\bar{\nabla}_\mu \bar{\nabla}_\alpha T^{\mu \nu} &= \bar{R}_{\lambda \alpha} T^{\lambda \nu} + \bar{R}^\nu_{\ \lambda \mu \alpha} T^{\mu \lambda}\,,\\
\bar{\nabla}_\mu \bar{\nabla}_\alpha \bar{\nabla}_\beta T^{\mu \nu} &= \bar{\nabla}_\alpha (\bar{R}_{\lambda \beta} T^{\lambda \nu} + \bar{R}^\nu_{\ \lambda \mu \beta} T^{\mu \lambda}) - \bar{R}^\lambda_{\ \beta \mu \alpha} \bar{\nabla}_\lambda T^{\mu \nu} + \bar{R}_{\lambda \alpha} \bar{\nabla}_\beta T^{\lambda \nu} + \bar{R}^\nu_{\ \lambda \mu \alpha} \bar{\nabla}_\beta T^{\mu \lambda}\,,\\
\bar{\nabla}_\mu \bar{\square} T^{\mu \nu} &= \bar{R}^\nu_{\ \lambda \mu \rho} \bar{\nabla}^\rho T^{\mu \lambda} + \bar{\nabla}^\sigma (\bar{R}_{\lambda \sigma} T^{\lambda \nu} + \bar{R}^\nu_{\ \lambda \mu \sigma} T^{\mu \lambda})\,,\\
\bar{\nabla}^\sigma \bar{\nabla}_\rho \bar{\nabla}_\sigma T_{\mu \nu} &= \bar{\nabla}_\rho \bar{\square} T_{\mu \nu} + \bar{R}_{\lambda \rho} \bar{\nabla}^\lambda T_{\mu \nu} - \bar{R}^\lambda_{\ \mu \alpha \rho} \bar{\nabla}^\alpha T_{\lambda \nu} - \bar{R}^\lambda_{\ \nu \alpha \rho} \bar{\nabla}^\alpha T_{\mu \lambda}\,.
\ea
\ee
For a tensor $t_{\mu \nu}$, we have
\be
\bar{\nabla}_\rho \bar{\square} t_{\mu \nu} = \bar{\square} \bar{\nabla}_\rho t_{\mu \nu} - \bar{\nabla}^\alpha (\bar{R}^\lambda_{\ \mu \rho \alpha} t_{\lambda \nu} + \bar{R}^\lambda_{\ \nu \rho \alpha} t_{\mu \lambda}) - \bar{R}_{\lambda \sigma} \bar{\nabla}^\lambda t_{\mu \nu} - \bar{R}^\lambda_{\ \mu \rho \sigma} \bar{\nabla}^\sigma t_{\lambda \nu} - \bar{R}^\lambda_{\ \nu \rho \sigma} \bar{\nabla}^\sigma t_{\mu \lambda}\,.
\ee
Now, when the background is MSS, some of these commutations relations simplify into a compact recursion formulas. For example, we have
\be
\bar{\nabla}_\mu \bar{\square} \phi = \left( \bar{\square} - \frac{\bar{R}}{4} \right) \bar{\nabla}_\mu \phi \implies \bar{\nabla}_\mu \bar{\square}^n \phi = \left( \bar{\square} - \frac{\bar{R}}{4} \right)^n \bar{\nabla}_\mu \phi\,.
\ee
In eq.(\ref{f3}), in order to get a recursion relation, we must express $\bar{R}_{\mu \nu} \bar{\nabla}^\nu \phi$ in terms of $\bar{\nabla}_\mu \phi$. To that end, suppose
\be
\bar{\nabla}_\mu \bar{\square} \phi = \bar{\square} \bar{\nabla}_\mu \phi - \bar{R}_{\mu \nu} \bar{\nabla}^\nu \phi = \bar{\square} \bar{\nabla}_\mu \phi - x \bar{\nabla}_{\mu} \phi\,,
\ee
where $x$ is some scalar to be determined. Rewriting this twice, we get
\bea
\bar{R}_\mu^{\ \rho} \bar{\nabla}_\rho \phi &=& x \bar{\nabla}_\mu \phi\,, \\
\bar{R}_\nu^{\ \alpha} \bar{\nabla}_\alpha \phi &=& x \bar{\nabla}_\nu \phi\,.
\eea
Multiplying both expressions above, contracting with $\bar{g}^{\mu \nu}$ and rearranging gives
\be
x = \sqrt{ \frac{\bar{R}_{\mu \nu} \bar{R}^{\mu \nu}}{4}}\,,
\ee
so that now, we have a recursion relation for a general metric:
\be \label{recursiongeneral1}
\bar{\nabla}_\mu \bar{\square}^n \phi = (\bar{\square} - x)^n \bar{\nabla}_{ \mu} \phi  = \left( \bar{\square} - \sqrt{ \frac{\bar{R}_{\mu \nu} \bar{R}^{\mu \nu}}{4}} \right)^{n} \bar{\nabla}_\mu \phi\,.
\ee
For MSS, $x = \bar{R}/4$, as expected.

Similarly, for a vector $t^\mu$ in eq.(\ref{f7}), we have
\be
\bar{\nabla}_\mu \bar{\square} t^\mu = \bar{\square} \bar{\nabla}_\mu t^\mu + \bar{\nabla}^\mu (\bar{R}_{\mu \nu} t^\nu) = \bar{\square} \bar{\nabla}_\mu t^\mu + y \bar{\nabla}_\mu t^\mu = (\bar{\square} + y)\bar{\nabla}_\mu t^\mu\,,
\ee
where the scalar $y$ is given by
\be
y (t^\mu) = \frac{\bar{\nabla}^\rho(\bar{R}_{\rho \sigma}t^\sigma)}{\bar{\nabla}_\alpha t^\alpha}\,,
\ee
such that we get the recursion relation
\be \label{recursiongeneral2}
\bar{\nabla}_\mu \bar{\square}^n t^\mu =  (\bar{\square} + y)^n \bar{\nabla}_\mu t^\mu\,, \qquad \qquad \text{where } y = \frac{\bar{R}}{4} + \frac{(\bar{\nabla}^{\rho} \bar{R}_{\rho \sigma}) t^{\sigma}}{(\bar{\nabla}^{\rho} t^{\sigma}) \bar{g}_{\rho \sigma}}\,.
\ee
For MSS, $y =\bar{R}/4$, as expected.

Given a tensor $t^{\beta \mu \alpha}$, we have
\be
\ba
\bar{\nabla}_\alpha \bar{\square} t^{\beta \mu \alpha} &= \bar{\square} \bar{\nabla}_\alpha t^{\beta \mu \alpha} +  \bar{R}^\beta_{\ \lambda \alpha \rho} \bar{\nabla}^\rho t^{\lambda \mu \alpha} -\bar{R}_{\lambda \alpha} \bar{\nabla}^\lambda t^{\beta \mu \alpha}  + \bar{R}^\mu_{\ \lambda \alpha \rho} \bar{\nabla}^\rho t^{\beta \lambda \alpha} + \bar{R}_{\lambda \rho} \bar{\nabla}^\rho t^{\beta \mu \lambda} \\
& \qquad + \bar{\nabla}^\rho \left(\bar{R}^\beta_{\ \lambda \alpha \rho} t^{\lambda \mu \alpha} + \bar{R}^\mu_{\ \lambda \alpha \rho} t^{\beta \lambda \alpha} + \bar{R}_{\lambda \rho} t^{\beta \mu \lambda} \right)\,.
\ea
\ee
Given a tensor $t^{\mu \alpha \nu \beta}$, we have
\be
\ba
\bar{\nabla}_\mu \bar{\square} t^{\mu \alpha \nu \beta} &= \bar{\square} \bar{\nabla}_\mu t^{\mu \alpha \nu \beta}
+ \bar{R}_{\lambda \rho} \bar{\nabla}^\rho t^{\lambda \alpha \nu \beta} -\bar{R}_{\lambda \mu} \bar{\nabla}^\lambda t^{\mu \alpha \nu \beta} + \bar{R}^\alpha_{\ \lambda \mu \rho} \bar{\nabla}^\rho t^{\mu \lambda \nu \beta} + \bar{R}^\nu_{\ \lambda \mu \rho} \bar{\nabla}^\rho t^{\mu \alpha \lambda \beta} + \bar{R}^\beta_{\ \lambda \mu \rho} \bar{\nabla}^\rho t^{\mu \alpha \nu \lambda}  \\
& \qquad + \bar{\nabla}^\rho \left( \bar{R}_{\lambda \rho} t^{\lambda \alpha \nu \beta} + \bar{R}^\alpha_{\ \lambda \mu \rho} t^{\mu \lambda \nu \beta} + \bar{R}^\nu_{\ \lambda \mu \rho} t^{\mu \alpha \lambda \beta} + \bar{R}^\beta_{\ \lambda \mu \rho} t^{\mu \alpha \nu \lambda} \right)
\,.
\ea
\ee

\bibliographystyle{utphys}
\bibliography{ssa}

\end{document}